\documentclass[twocolumn,table,xcolor]{aastex62}
\usepackage{xcolor}
\usepackage{dirtytalk}
\usepackage{amsmath}
\usepackage{array}
\usepackage{float}
\usepackage{enumitem}
\usepackage{CJKutf8}
\newcommand{\chinesename}{{\begin{CJK}{UTF8}{gbsn}(王加冕)\end{CJK}}}
\newcolumntype{L}[1]{>{\raggedright\arraybackslash}p{#1}}

\begin{document}

\title{\textbf{Performance Verification of the EXtreme PREcision Spectrograph}}

\correspondingauthor{Ryan T. Blackman} 
\author[0000-0002-0303-3276]{Ryan T. Blackman} 
\author[0000-0003-2221-0861]{Debra A. Fischer}
\affiliation{Department of Astronomy, Yale University, 52 Hillhouse Ave, New Haven, CT 06511, USA}

\author[0000-0002-5018-7761]{Colby A. Jurgenson}
\affiliation{Department of Astronomy, The Ohio State University, 4055 McPherson Laboratory, 140 West 18th Avenue, Columbus, OH 43210, USA}

\author[0000-0001-8483-4667]{David Sawyer}
\affiliation{Lowell Observatory, 1400 Mars Hill Rd, Flagstaff, AZ 86001, USA}

\author{Tyler M. McCracken}
\affiliation{Ball Aerospace and Technologies Corporation, 1600 Commerce St, Boulder, CO 80301, USA}

\author[0000-0002-4974-687X]{Andrew E. Szymkowiak}
\author[0000-0003-2168-0191]{Ryan R. Petersburg}
\affil{Department of Physics, Yale University, 217 Prospect St, New Haven, CT 06511, USA}

\author[0000-0001-7664-648X]{J. M. Joel Ong \chinesename}
\affiliation{Department of Astronomy, Yale University, 52 Hillhouse Ave, New Haven, CT 06511, USA}

\author[0000-0002-9873-1471]{John M. Brewer}
\affiliation{Department of Physics and Astronomy, San Francisco State University, 1600 Holloway Ave, San Francisco, CA 94132, USA}

\author[0000-0002-3852-3590]{Lily L. Zhao}
\author[0000-0003-2369-0481]{Christopher Leet}
\affiliation{Department of Astronomy, Yale University, 52 Hillhouse Ave, New Haven, CT 06511, USA}

\author[0000-0003-1605-5666]{Lars A. Buchhave}
\author[0000-0003-1001-0707]{René Tronsgaard}
\affiliation{DTU Space, National Space Institute, Technical University of Denmark, Elektrovej 328, DK-2800 Kgs. Lyngby, Denmark}

\author[0000-0003-4450-0368]{Joe Llama}
\affiliation{Lowell Observatory, 1400 Mars Hill Rd, Flagstaff, AZ 86001, USA}

\author{Travis Sawyer}
\affiliation{College of Optical Sciences, University of Arizona, 1630 E University Blvd, Tucson, AZ 85719, USA}

\author[0000-0002-5070-8395]{Allen B. Davis}
\author[0000-0001-9749-6150]{Samuel H. C. Cabot}
\affiliation{Department of Astronomy, Yale University, 52 Hillhouse Ave, New Haven, CT 06511, USA}

\author{Michael Shao}
\author[0000-0003-0159-8668]{Russell Trahan}
\affiliation{Jet Propulsion Laboratory, California Institute of Technology, 4800 Oak Grove Drive, Pasadena, CA 91109, USA}

\author{Bijan Nemati}
\affiliation{Center for Applied Optics, University of Alabama in Huntsville, 301 Sparkman Drive, Huntsville, AL 35899, USA}

\author[0000-0001-7966-8182]{Matteo Genoni}
\author[0000-0003-3724-7667]{Giorgio Pariani}
\author[0000-0003-1580-6966]{Marco Riva}
\affiliation{INAF - Osservatorio Astronomico di Brera, Via Emilio Bianchi 46, 23807 Merate, Italy}

\author{Rafael A. Probst}
\affiliation{Menlo Systems GmbH, Martinsried, Germany}

\author{Ronald Holzwarth}
\affiliation{Menlo Systems GmbH, Martinsried, Germany} 
\affiliation{Max-Planck-Institut für Quantenoptik, Garching, Germany}

\author{Tilo Steinmetz}
\affiliation{Menlo Systems GmbH, Martinsried, Germany}

\author{Paul Fournier}
\author{Rafal Pawluczyk}
\affiliation{FiberTech Optica Inc, 330 Gage Avenue, Suite 1, Kitchener, ON, N2M 5C6 Canada}


\begin{abstract}
The EXtreme PREcision Spectrograph (EXPRES) is a new Doppler spectrograph designed to reach a radial velocity measurement precision sufficient to detect Earth-like exoplanets orbiting nearby, bright stars. We report on extensive laboratory testing and on-sky observations to quantitatively assess the instrumental radial velocity measurement precision of EXPRES, with a focused discussion of individual terms in the instrument error budget. We find that EXPRES can reach a single-measurement instrument calibration precision better than 10 cm s$^{-1}$, not including photon noise from stellar observations. We also report on the performance of the various environmental, mechanical, and optical subsystems of EXPRES, assessing any contributions to radial velocity error. For atmospheric and telescope related effects, this includes the fast tip-tilt guiding system, atmospheric dispersion compensation, and the chromatic exposure meter. For instrument calibration, this includes the laser frequency comb (LFC), flat-field light source, CCD detector, and effects in the optical fibers. Modal noise is mitigated to a negligible level via a chaotic fiber agitator, which is especially important for wavelength calibration with the LFC. Regarding detector effects, we empirically assess the impact on radial velocity precision due to pixel-position non-uniformities (PPNU) and charge transfer inefficiency (CTI). EXPRES has begun its science survey to discover exoplanets orbiting G-dwarf and K-dwarf stars, in addition to transit spectroscopy and measurements of the Rossiter-McLaughlin effect. 
\end{abstract}

\section{Introduction}
\label{sec:intro}

The discovery of exoplanets was first enabled by Doppler spectroscopy \citep{mayor1995}, which detects the reflex radial velocity of stars orbiting a common center of mass with a planetary companion. Since higher mass planets in short period orbits produce larger reflex stellar velocities, the distribution of Doppler-detected exoplanets reflects this observational bias. Small rocky planets in Earth-like orbits have eluded detection because they induce sub-m s$^{-1}$ reflex velocities that have historically been buried in systematic errors from instruments, the analysis, and stellar photospheric velocities.

Design specifications for the newest generation of spectrographs leverage both technological advancements and detailed analysis of previous instruments. Systems engineering methodology with detailed error budgets attempts to identify and mitigate known sources of instrumental error \citep[e.g.,][]{podgorski2014, halverson2016}. These studies inform all aspects of the instrument and optical design, including the materials, fibers, wavelength calibration sources, and the choice of detectors. In addition, studies of the effects of stellar activity also inform design choices regarding instrument resolution and the desired signal-to-noise ratio (S/N) of stellar spectra \citep{davis2017}. 

The EXtreme PREcision Spectrograph (EXPRES) is a new Doppler spectrograph that has been commissioned at Lowell Observatory's 4.3-m  Lowell Discovery Telescope \cite[LDT, formerly known as the Discovery Channel Telescope,][]{levine2016} in Happy Jack, AZ, USA \citep{jurgenson2016}. EXPRES completed its commissioning period in February 2019. Science operations are now underway, though minor software tools are still being developed to improve instrument control.

The primary design driver for EXPRES was the goal of exploiting high spectral resolution to mitigate the effects of stellar activity and enhance the Doppler signature of orbiting exoplanets. The science program for EXPRES is a radial velocity survey of nearby, bright G-dwarf and K-dwarf stars to search for rocky exoplanets. To accomplish this, stringent requirements were placed on instrument performance, so that the radial velocity error contribution from instrumental effects would be significantly lower than the errors induced by stellar activity. Additional science goals include the study of hot Jupiter atmospheres during transit events, measurements of the Rossiter-McLaughlin Effect, and follow-up mass measurements of transiting exoplanets. 

This paper evaluates the performance of the EXPRES instrument and a companion paper, \cite{petersburg2019}, presents the reduction pipeline and the first on-sky radial velocity measurements. This paper is organized as follows. In Section \ref{sec:design}, we summarize the final design and current status of EXPRES following the commissioning period, with updates since the initial design described in \citet{jurgenson2016}. A detailed breakdown of known sources of radial velocity error in EXPRES is presented in Section \ref{sec:errors}, along with a discussion of each error source. Section \ref{sec:throughput} details the measured throughput of the instrument and characterizes the detected S/N for stars as a function of brightness, and  the measured S/N of the calibration sources with nominal exposure lengths. We then explore specific aspects of the instrument in more detail, quantifying the expected contributions to radial velocity measurement error. In section \ref{sec:environment}, we examine the thermo-mechanical, pressure, and vibrational stability of the instrument. Illumination stability is explored in Section \ref{sec:illumination}. The CCD detector is discussed in Section \ref{sec:detector}. Our treatment of stray light and cosmic ray removal is summarized in Section \ref{sec:stray_light}. A discussion of sky and Moonlight contamination is presented in Section \ref{sec:sky}. The performance of the chromatic exposure meter is detailed in Section \ref{sec:expm}. The results of our lab tests for instrument calibration precision are shown in Section \ref{sec:lab_tests}. In Section \ref{sec:discussion}, we discuss recommendations for instrument development for radial velocity work. We hope that this paper illustrates the type of spectrograph evaluation tests that could be further developed and shared for all new Doppler spectrographs \cite[e.g.,][]{wright2017}.

\section{Final Instrument Design}
\label{sec:design}
\subsection{Summary of Instrument Hardware}
\begin{figure*}[t]
\begin{center}
\includegraphics{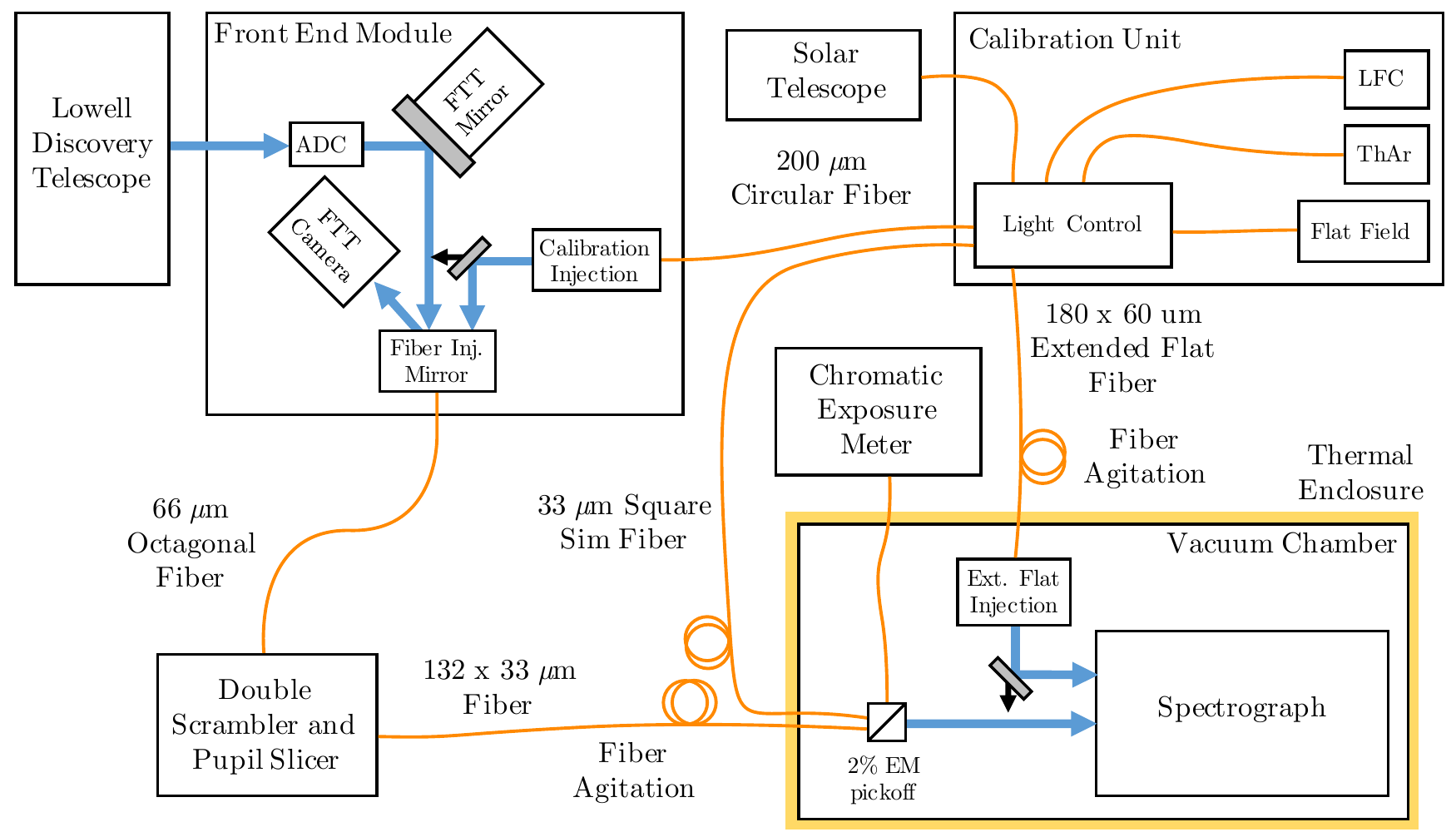} 
\end{center}
\caption{A high-level view of the EXPRES optomechanical sub-systems that follows the light path from the telescope to the spectrograph. Orange lines are optical fibers and blue lines are light in open-air.}
\label{fig:arch}
\end{figure*}

A detailed description of the EXPRES design was presented in \cite{jurgenson2016}. Here, we summarize that design and highlight changes that have been made in the time since. A simplified schematic of the EXPRES hardware architecture is shown in Figure \ref{fig:arch}.

\begin{figure}
\begin{center}
\includegraphics[scale=0.47]{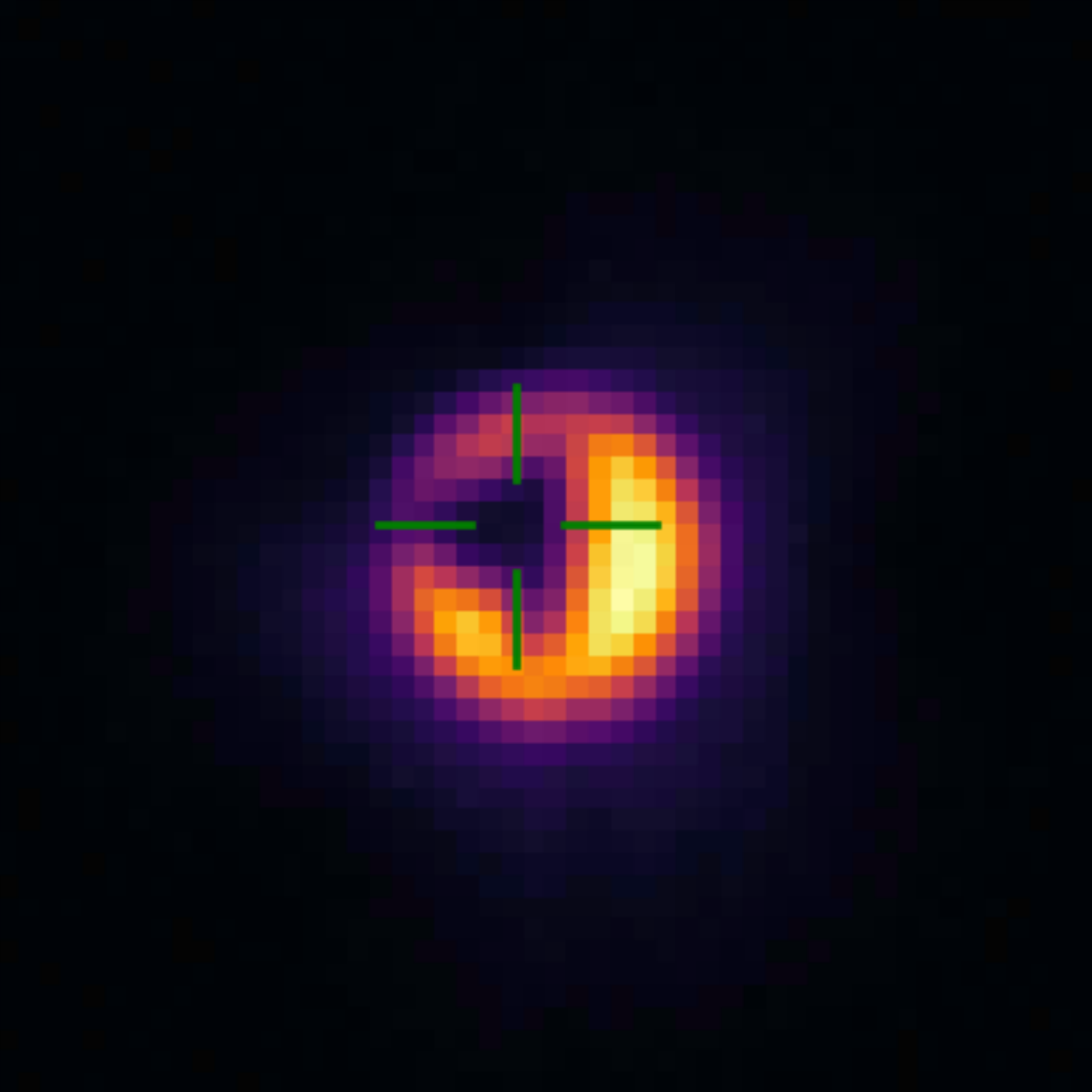} 
\end{center}
\caption{An image of of the FIM as seen by the FTT camera during a ThAr calibration exposure. The spot of calibration light illuminates the dark triangle that contains the core extension of the octagonal science fiber.}
\label{fig:triangle}
\end{figure}

Stellar light comes to a focus in the EXPRES front-end module (FEM) via the primary, secondary, and tertiary mirrors of the LDT \citep{macfarlane2004}. The FEM occupies one of the five ports of the LDT instrument cube. In the FEM, the beam is collimated and then atmospheric dispersion compensation (ADC) is performed with a prism-pair system \citep[e.g.,][]{wynne1986}. This corrected light is then reimaged and guided into a $66\;\micron$ core octagonal science fiber via a fast tip-tilt (FTT) system. A cylindrical core is fused to the end of the science fiber, extending it out of the cladding, and is embedded directly into a V-groove in the fiber injection mirror (FIM). Target acquisition is performed by centering the image of the star onto cross-hairs that are aligned with the dark triangular shadow from the V-groove that supports the extension of the science fiber core. The FTT detector is an Andor iXon 897 electron-multiplying charge-coupled device (EMCCD). An image of a calibration source being injected into the octagonal science fiber, as seen by the FTT detector, is shown in Figure \ref{fig:triangle}. Spill-over light from the image of the star on the fiber is reflected to the FTT detector during observations. When the servo loop on the FTT camera is closed, the FTT system samples the image at rates up to 600 Hz and performs corrections at rates up to 100 Hz. 

The octagonal science fiber runs through the cable wrap of the telescope and the 65 m length is fed through the core of the telescope pier down to the basement level in the stabilized EXPRES instrument room. In this room, light from the octagonal fiber passes through a pupil slicer and double scrambler module that contains the EXPRES shutter. The pupil slicer divides the beam into two half-moon images that are stacked and injected into the rectangular science fiber. This rectangular fiber is half the width and twice the height of the octagonal fiber with core dimensions of $132\times 33$ $\mu$m. The pupil slicer and double scrambler inverts the near and far field of the octagonal fiber output and effectively doubles the resolution of the spectrograph with only modest losses from reflections on the associated optics, alignment errors, and the injection of light into the rectangular fiber. The 5 m rectangular fiber enters the vacuum chamber and serves as the slit for EXPRES. 

\begin{figure}
\begin{center}
\includegraphics{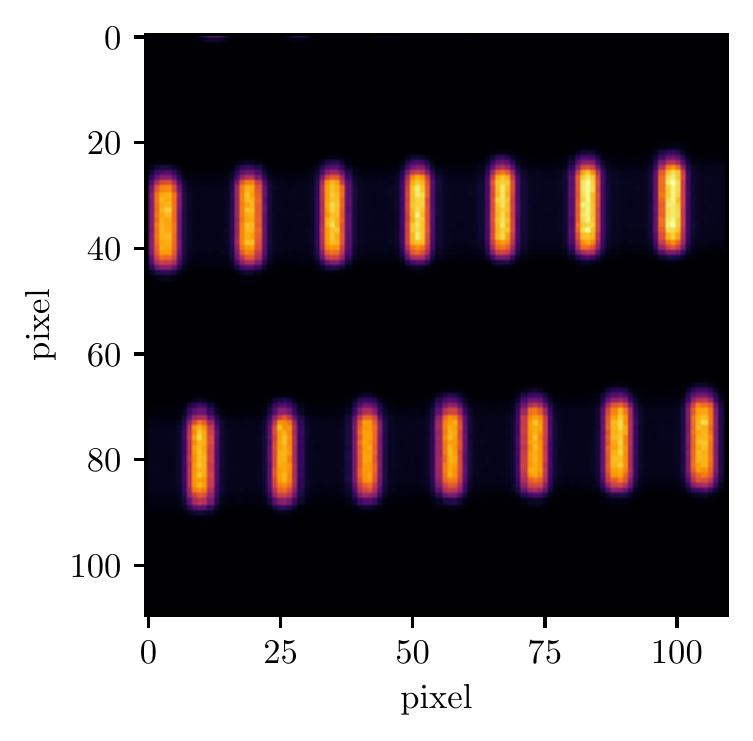} 
\end{center}
\caption{A small region of the 2D LFC spectrum in two adjacent spectral orders, showing the rectangular PSF of the instrument.}
\label{fig:slit}
\end{figure}

The optical design of EXPRES was presented in \cite{jurgenson2016}. Light injected into the spectrograph undergoes focal ratio conversion from f/3 to f/8.5 before light is collimated. An R4 dual mosaic echelle grating is etched into a single piece of Zerodur with 30 lines/mm for the primary dispersing element. Light reflects back to the main collimator and the beam comes to a focus behind a Mangin mirror, which is used to correct cylindrical field curvature. A transfer collimator is then used to re-collimate light before it passes through two cross-dispersing prisms. An 8-element camera is used to focus the beam with a highly stable line-spread-function. The detector is an STA1600 CCD with 10.6k $\times$ 10.6k 9 $\mu$m pixels built by Semiconductor Technology Associates, Inc (STA). The rectangular science fiber produces a rectangular point spread function (PSF) on the detector, shown in Figure \ref{fig:slit}. This image is of a small region of the LFC spectrum as detected by EXPRES, where the extremely narrow emission lines of the LFC are broadened in the dispersion direction into a rectangular shape by the instrument optics. 

Precise wavelength calibration is accomplished with a Menlo Systems LFC \citep[e.g.,][]{wilken2012,molaro2013,probst2014}, and a thorium-argon (ThAr) lamp is used for the initial, coarse wavelength solutions. Calibration light may be injected into the octagonal science fiber via a retractable fold-mirror in the FEM. This enables the ThAr lamp, LFC, and flat-field light source to be injected into the same fiber as the science light. Each calibration source can also be injected into a square $33 \times 33$ $\mu$m core simultaneous fiber with spectral orders offset from the science orders by about 10 pixels, if such a calibration is desired.

The EXPRES flat-field light source is a custom, LED-based device that can feed both the science fiber and an oversized, or extended, fiber for accurate flat-fielding of the edges of the science orders in the cross-dispersion direction. A large set of extended flats is typically taken every few months and added to a master, 2D flat for that epoch. Flats through the science fiber are taken every night, and are used for optimal extraction and normalization of stellar spectra. The spectral characteristics of both flat-field modes are discussed in Section \ref{sec:throughput}. The light source is composed of 25 LEDs positioned on a compact chip. Emission from the chip is coupled to the various fibers via a 4 inch diameter integrating sphere. The total power emitted from the chip reaches 12.5 W, ensuring that enough light is coupled to the fibers despite poor efficiency from the integrating sphere to the fibers. Different frequency LEDs were chosen with power that approximately matches the inverse of the EXPRES instrumental throughput. The power output from each LED has a small range of adjustability to help ensure that a relatively smooth spectrum can be obtained. This light source also includes a more traditional quartz lamp that can be optionally injected into the fibers with or without the LEDs for an additional calibration option in red wavelengths.

\begin{figure}
\begin{center}
\includegraphics{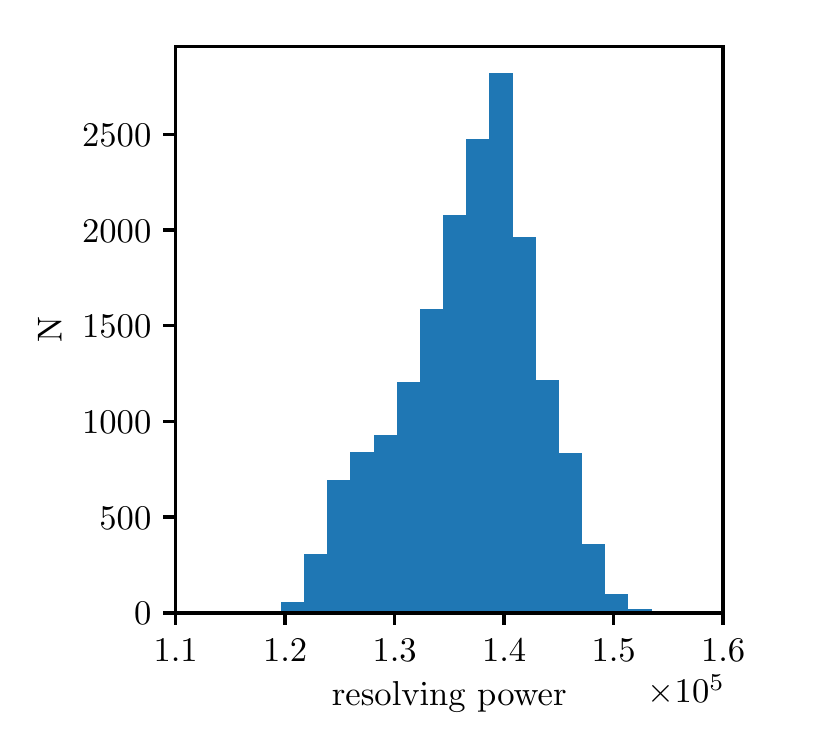} 
\end{center}
\caption{The distribution of resolving powers measured for EXPRES across the wavelength range of the LFC. The median resolving power is 137,500 with a standard deviation of 6,100.}
\label{fig:resolution}
\end{figure}

EXPRES spans a wavelength range of 3800-7800 \AA with a median resolving power of R=137,500 $\pm$ 6,100 and a sampling of four pixels. However, the wavelength range used for radial velocity analysis is approximately 4850-7150 \AA. This is set primarily by the range of the LFC, however, some extrapolation from the LFC wavelength solution can be made with ThAr wavelength solutions. The resolution has been empirically measured across the spectral format using an LFC spectrum. Each emission line of the LFC is fit with a gaussian profile, and the resolution is computed from the full-width half-max (FWHM) of each line in frequency space, at the central frequency of that line. The resolution is converted to resolving power via $R=f_{\mathrm{FWHM}}/f$. The distribution of the measured resolving powers are plotted in Figure \ref{fig:resolution}. The resolution is not constant across each order, as the blue sides of the orders tend to be lower, leading to the asymmetry in Figure \ref{fig:resolution} around R=130,000. 

\subsection{Revised Calibration Unit}
The design of the calibration unit from \cite{jurgenson2016} has been revised to address two issues that arose during commissioning. First, the flip mirrors that were used to feed multiple calibration sources to the calibration fiber occasionally failed. It was difficult for the observer to troubleshoot this issue because the calibration unit is enclosed in a light-tight box in the spectrograph room of the LDT, which should generally not be entered to maintain thermal stability. Second, alignment of the different fibers was challenging and optimal efficiency was not achieved. The alignment state of some sources and fibers were effectively dependent on other alignment states. This resulted in a very small tolerance for the position of other fiber mounting posts to achieve maximum efficiency. While this design theoretically provided the best coupling efficiency of the calibration sources to the proper fibers, in practice we found that this was not easily achievable. 

\begin{figure*}[t]
\begin{center}
\includegraphics{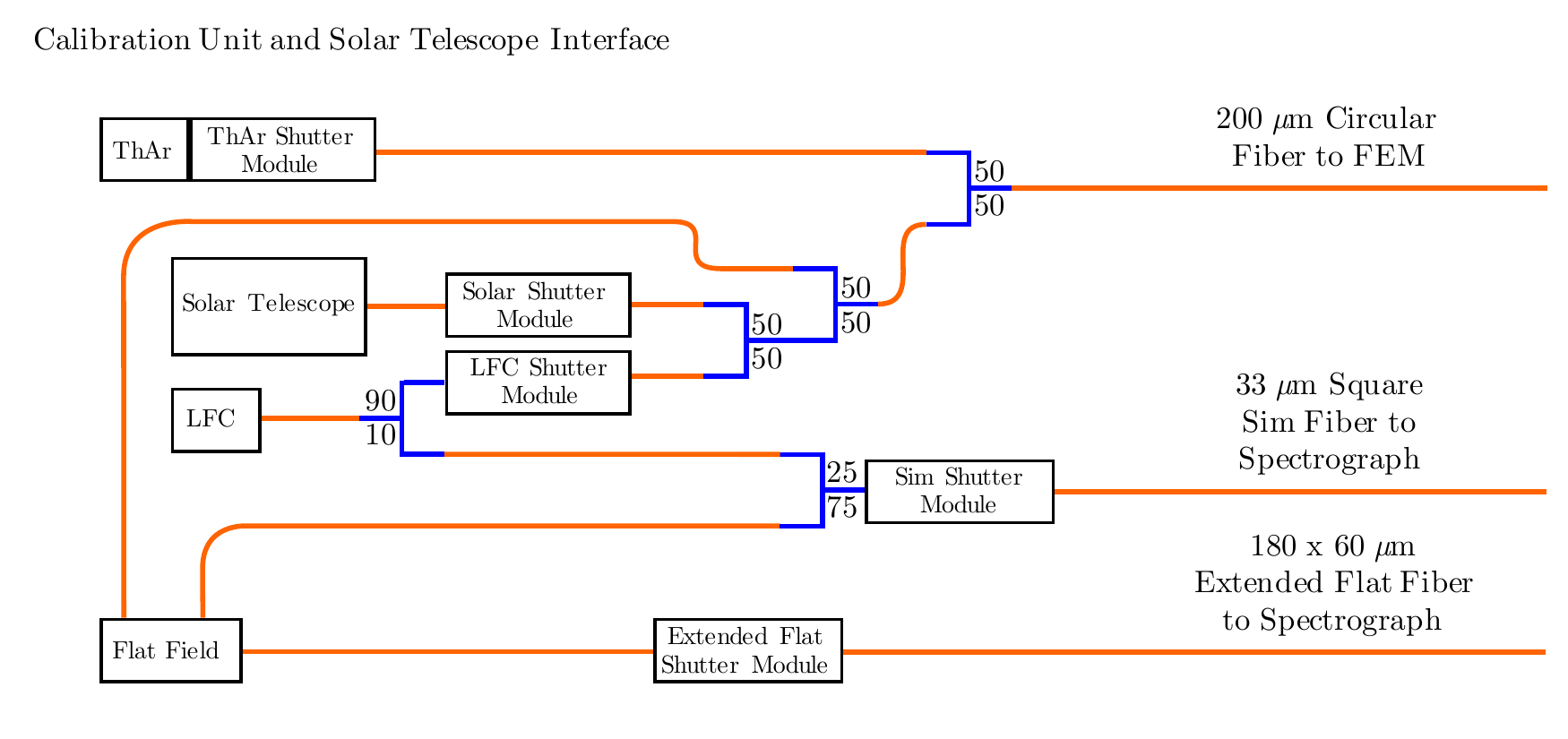} 
\end{center}
\caption{Schematic of the revised calibration unit of EXPRES, including the planned solar telescope implementation for the calibration unit. In this figure, fibers are represented in orange, and blue represents the splitters/combiners.}
\label{fig:cal_unit}
\end{figure*}

To address these issues, we designed a new calibration unit that eliminated the use of flip mirrors and made the alignment of each source and fiber independent. This was accomplished with the use of commercially available $2\times1$ fiber couplers which can be used as both splitters and combiners. In this setup, light from different calibration sources can be combined for injection into the calibration fiber. Light from the LFC is split so that it can be injected into both the calibration fiber and the simultaneous fiber. Different splitting/combining ratios were chosen to make the exposure times appropriate for each source. While these devices are inherently inefficient, it was much easier to reach the maximum possible efficiency through alignment with this system, resulting in shorter exposure times of the calibration sources being attainable. In addition, the calibration sources were bright enough such that sacrifice in efficiency was possible without a corresponding increase in exposure time, even if no gains were realized. For example, the LFC is naturally several orders of magnitude brighter than what would be appropriate for 10 second exposures with EXPRES. Even with a more inefficient system, use of a neutral density filter would still be required to make the source dim enough to be exposed on for the appropriate duration. Each source in the new calibration unit has its own shutter that is controlled by the instrument software. These shutters are opened at the same time as the EXPRES shutter in the double scrambler when a hardware signal from the CCD controller is received. A final benefit to this design is that it is highly modular, so new calibration sources can easily be added later without a large impact on other sources. A schematic of this setup is shown in Figure \ref{fig:cal_unit}. The light path from each of the calibration sources, as well as a planned solar telescope, are shown to each of the three fibers of EXPRES. These are the three fibers exiting the calibration unit in Figure \ref{fig:arch}. The splitting/combining ratios are shown for each splitter/combiner.

\subsection{Solar Telescope}
A solar telescope for daytime observations of the Sun is also being implemented for EXPRES. This telescope is based on the solar telescope for HARPS-N \citep{dumusque2015b,phillips2016,milbourne2019}. The telescope itself consists of a 400 mm focal length, and a 3-inch aperture lens that focuses light into a 2-inch diameter integrating sphere. The output from the integrating sphere is coupled via an optical fiber to the calibration unit of EXPRES. The telescope tracker is an off-the-shelf guider specifically designed for solar observing, and the entire assembly sits on an equatorial motorized mount. The solar telescope is housed under a fixed, acrylic dome situated on an auxiliary building to the LDT. The telescope will be completely automated, observing on every clear day while the Sun is above 30 degrees in elevation.

Combining EXPRES solar observations with simultaneous spacecraft data from missions such as NASA’s Solar Dynamics Observatory (SDO) will enable us to determine methods to mitigate signals from stellar activity \citep[e.g.,][]{haywood2016}. Additionally, Lowell Observatory has been actively involved in measuring the activity of the Sun and sun-like stars for over 25 years with the Solar-Stellar Spectrograph \citep[SSS;][]{hall1995}. The EXPRES solar telescope will replace the solar element of the SSS, enabling the continuous monitoring of solar activity into the next cycle.

\section{Error Breakdown}
\label{sec:errors}
\subsection{Table of Terms}
In this section, we synthesize everything we have learned about each known source of radial velocity error, based on our measurements during the commissioning of EXPRES and the literature regarding similar instruments. In Table \ref{tab:table_errors}, we list each effect relevant for EXPRES, note the mitigation method used if any, state the uncorrected magnitude of the error, the magnitude of the residual error after mitigation/calibration, and the source of each numerical estimate that we provide. Most of the instrumental error sources are discussed throughout this paper. In many cases, errors have been constrained to upper or lower limits, and some errors are small enough that we have not pursued more exact numbers. Some errors are merely estimates due to the difficulty in isolating them. Errors that are not calibratable are listed with the same uncorrected magnitude and residual error. Many of the constraints come from our measured instrument calibration precision, which we define as the expected radial velocity error contributed by instrumentation effects. This has been measured to be well-under 10 cm s$^{-1}$ by cross-correlating LFC exposures to one reference exposure, and is discussed further in Section \ref{sec:lab_tests}. This test effectively demonstrates the amount of instrumental drift expected between calibration and science exposures. It does not include all of the on-sky effects listed in Table \ref{tab:table_errors}, vibrations from the telescope dome and slewing, calibration injection repeatability errors, in addition to CTI errors caused by mismatched S/N of stellar spectra.

\subsection{Definitions of Terms}
Here, we define each term listed in Table \ref{tab:table_errors}, along with some brief discussion of the effects and references for further explanations. This is intended as a brief summary, as many of these effects warrant entire papers of their own or are discussed in more detail throughout this paper. This section is intended to be the most concise reference to the material in Table \ref{tab:table_errors}. \\

\noindent
\underline{Image motion on fiber} - this is the motion of the star on the fiber during observations, which induces radial velocity error with finite scrambling. It is usually reported as the RMS or standard deviation of the motion in arcseconds or milliarcseconds (mas). With EXPRES, this has been empirically measured at 30 mas on-sky, when guiding on the fiber, and the radial velocity error calculation has been done with the methods of \cite{halverson2015}. This is discussed more in Section \ref{sec:ftt}. \\

\noindent
\underline{Atmospheric dispersion} - the effect of chromatic dispersion when starlight enters Earth's atmosphere, which induces radial velocity error via chromatic coupling efficiency. This is partially mitigated with a chromatic exposure meter, though it can mimic guiding errors with a chromatic dependence, which will also depend on scrambling gain in the fibers. Atmospheric dispersion compensation (ADC) is used to mitigate this effect, but is limited up to a certain zenith distance \citep[see][also discussed in Section \ref{sec:adc}]{halverson2016}. \\

\noindent
\underline{Barycentric correction} - the residual error from the barycentric correction comes from several sources. The accuracy of the correction algorithm, accuracy in the stellar and observatory coordinates, accuracy of the reported shutter times, and accuracy of the chromatic weights measured by the exposure meter all impact the fidelity of the correction \citep[discussed further in][]{blackman2019}. The correction is performed with the weighting scheme of \cite{tronsgaard2019}, which improves over the weighted midpoint method commonly used in the past. The barycentric correction and performance of the exposure meter is discussed in more detail in Section \ref{sec:expm}. \\

\noindent
\underline{Photon noise} - the number of photons detected from the star limits the precision of a radial velocity measurement due to Poisson noise. This results in degraded absorption line profiles. Photon noise can be improved with longer exposures to reach higher S/N values, and also averaged over by binning consecutive or phase-folded radial velocity measurements. The radial velocity error from photon noise can be assessed with on-sky data. The cross-correlation function (CCF) method of solving for radial velocity returns a formal error that is derived primarily from photon noise, which has been measured as a function of S/N with EXPRES. Further analysis and discussion of this result is presented in \cite{petersburg2019}. \\  

\noindent
\underline{Stellar activity} - effects from the surfaces of stars such as spots, plages, faculae \citep{davis2017}, p-modes \citep{chaplin2019}, granulation and supergranulation \citep{meunier2019}, and magnetic activity \citep{milbourne2019} can induce radial velocity error. For example, absorption line asymmetries or large-scale flows can manifest as spurious radial velocity signals, and different absorption lines may be impacted differently. Further discussion of stellar activity is outside the scope of this paper. \\

\noindent
\underline{Telluric contamination} - absorption lines caused by Earth's atmosphere, which vary in strength from just a couple percent to full saturation, contaminate the recorded stellar spectra. These lines are imprinted on the stellar spectrum and move with respect to it due to the barycentric motion of the Earth. Modeling and proper weighting of identified tellurics with the methods of \cite{leet2019} have led to between a few cm s$^{-1}$ up to 15 cm s$^{-1}$ RMS radial velocity improvement on-sky with EXPRES, as discussed in Section \ref{sec:sky}. However, we do not have an estimate for the absolute error due to telluric contamination before or after this process, as the effects are essentially degenerate with radial velocity errors from stellar activity. We currently estimate approximately 25 cm s$^{-1}$ of uncorrected error contribution from telluric contamination, based on the results in \cite{cunha2014}. However, this exact error depends on many factors, such as spectral type, air mass, systemic radial velocity, and the method used for solving for radial velocity. \\ 

\noindent
\underline{Sky/Moon Contamination} - sunlight reflected from the Moon and scattered in the atmosphere makes its way into the fiber, contaminating the observed stellar spectrum with a fainter, reddened spectrum of the Sun. The impact is more significant for fainter stars, as discussed in \cite{halverson2016}, and later in Section \ref{sec:sky} of this paper. \\

\noindent
\underline{Analysis errors} - errors coming from the way we treat the data and determine the radial velocity. One of our radial velocity analysis pipelines uses the CCF method, which may have some drawbacks, such as losing information content when using only a fraction of the available stellar lines, errors in line positions, and using different lines over time due to barycentric motion. \\

\noindent
\underline{Calibration Modal noise} - interference of spatial propagation modes in optical fibers leads to radial velocity errors due to the induced speckle patter in illumination. This is more of an issue for the highly coherent laser frequency comb calibration source. This is mostly mitigated by 1) fiber agitation and 2) longer exposure times of the calibration source, achieved by using a neutral density filter to dim the light source. The improvement with agitation is shown in Section \ref{sec:modal_noise}, and is discussed in more detail in \cite{mahadevan2014} and \cite{petersburg2018}. \\

\noindent
\underline{Calibration refresh rate beat frequency} - the spatial light modulator (SLM) of the LFC effectively causes this source to flicker, and if this frequency is near vibrational frequencies in the instrument, a beat pattern may manifest as spurious radial velocity shifts, as discussed in Section \ref{sec:detector}. \\

\noindent
\underline{Calibration photon noise} - photon noise will fundamentally impact the accuracy of the wavelength calibration, as with stellar observations. The LFC is brighter and has far more emission lines compared to the number of absorption lines in G-type and K-type stars, leading to a much smaller contribution to radial velocity error. \\

\noindent
\underline{Calibration accuracy} - fundamental limit of accuracy from the LFC emission lines, discussed in Section \ref{sec:cal_sources}. This value has been shown to be close to 1 cm s$^{-1}$ in the literature \citep{milakovic2020,probst2020}. \\

\noindent
\underline{Calibration Injection Repeatability} - the positional repeatability of the calibration light injection into the science fiber will lead to reduced measurement precision, as any shifts throughout the nightly calibration sequences will lead to uncalibratable errors in the wavelength solution. Instrument drifts when moving the calibration injection mirror in and out are not significantly larger than test sequences with the mirror in a static position, so this error is small, as discussed further in \ref{sec:cal_sources}. \\

\noindent
\underline{Calibration background noise} - the background of the LFC emission is variable over time and degrades as the photonic crystal fiber (PCF) ages. Much of this background comes from incoherent light that is amplified in the broadening process. Additionally, as the PCF degrades over time, the desired peaks in the LFC become fainter, the structure of the broadened optical spectrum becomes more unstable, and as a result, the spectral flattener performs sub-optimally. This may introduce a requirement for more complicated fitting algorithms when it occurs. This error could be absorbed in the calibration algorithms term, but it seems distinct enough to include it separately. This is discussed in more detail in Section \ref{sec:cal_sources}. \\

\noindent
\underline{Calibration algorithms} - the method used to fit the calibration emission lines and treatment of the background from the source will limit the accuracy of the wavelength assigned to each pixel, which is also limited by the PSF of the instrument. This may only become a significant issue when the PCF degrades and the LFC background becomes larger and more variable, as discussed in Section \ref{sec:cal_sources}. \\

\noindent
\underline{Detector temperature changes} - temperature changes in the CCD, for example those caused by readout, will deform the chip, leading to radial velocity error. A firmware update to keep power dissipation constant was implemented in the EXPRES CCD during commissioning, greatly improving thermal stability, as discussed in section \ref{sec:detector}. \\

\noindent
\underline{Detector electronics noise} - read noise from the detector and imperfect bias removal and gain correction will lead to error, although read noise is included in our S/N estimates along with photon noise, as discussed in Section \ref{sec:detector}. \\

\noindent
\underline{Detector pixel position non-uniformity} - small pixel position errors or non-uniform quantum efficiency across individual pixels will lead to incorrect wavelengths being assigned to pixels. This term has been measured for EXPRES, as discussed in Section \ref{sec:detector}. \\

\noindent
\underline{Detector charge transfer inefficiency} - some amount of accumulated charge is lost with each pixel transfer during readout, the effect being worse for pixels farther from the location of readout, leading to degraded line symmetry. Absorption lines farther from readout will be shifted by greater velocities. This effect has been studied in \cite{goudfrooij2006,bouchy2009,blake2017}, and measured for EXPRES in Section \ref{sec:detector}. This can be mitigated by matching the S/N of different exposures of the same star or by restoring counts in pixels based on their measured signal and location on the chip. \\

\noindent
\underline{Detector brighter-fatter effect} - the brighter-fatter effect introduces a flux-dependence in the instrument PSF detected by the CCD. Due to lateral charge diffusion, a brighter light source imaged by the CCD will exhibit a larger PSF than a fainter source \citep{antilogus2014}. This effect is minimized with thinned CCDs such as the one used in EXPRES (30 $\mu$m) compared to thick devices ($\sim 200$ $\mu$m) that are optimized for near-infrared wavelengths \citep{coulton2018}. In the case of a spectrograph, the brighter-fatter effect can impact the height of each diffraction order as well as the width of the PSF in the dispersion direction. This may lead to a slight differences in resolution. Radial velocity errors stemming from brighter-fatter can be minimized by matching the S/N of different exposures of the same star, as we normally do to mitigate effects from CTI as well. To absolutely minimize this effect, the S/N of the flat-field exposures could be matched to each other as well as the science data. However, this is difficult to achieve in practice. Observations of different stars may have different S/N requirements, which would necessitate different sets of flat-fields at different S/N. In addition, there is benefit to maximizing S/N of the flat-field exposures, as higher S/N flat-fields will produce a higher quality extraction \citep[see][for more details about this process]{petersburg2019}. \\

\noindent
\underline{Detector imperfect flat-field} - this calibration is done to account for quantum efficiency variations between different pixels on the CCD. Any residual errors will lead to degraded line profiles, however, this effect is expected to mostly average out given the large number of pixels used in the spectral format, and is included as a noise term in the EXPRES S/N, as discussed in Section \ref{sec:cal_sources}. \\

\noindent
\underline{Detector fringing} - due to the thinning of the CCD, interference may occur in red wavelength orders from photons reflecting off of different layers in the CCD. This mostly occurs outside of the region used for radial velocity analysis (approximately 4850-7150 \AA), and is generally calibratable with a flat-field and accurate continuum normalization, as presented in \cite{xu2019} and discussed in Section \ref{sec:detector}. \\

\noindent
\underline{Detector stray light and cosmic rays} - unwanted reflected light from various instrument surfaces may hit the detector in specific regions, leading to asymmetries in spectral line profiles. This is not strictly an effect of the detector itself, but can be mitigated with specific techniques in the data reduction step. It is difficult to constrain a specific value to the error term, as the extraction techniques developed for EXPRES naturally account for it. Cosmic ray removal may also introduce some amount of noise. The effects of these sources are included in the EXPRES noise model, as discussed in Section \ref{sec:stray_light}. \\

\noindent
\underline{Instrument temperature changes} - temperature changes in the chamber lead to calibratable errors. For example, the length of optical elements will change depending on their coefficients of thermal expansion (CTE), and movement of the optical bench will lead to centroid shifts. This can be minimized by thermal control of the instrument, use of materials with low CTE materials such as Invar and Zerodur, and frequent wavelength calibration, as discussed in Section \ref{sec:environment}.  \\

\noindent
\underline{Instrument pressure changes} - pressure changes in the instrument vacuum chamber between calibrations will cause a change in the index of refraction of the medium, this changes the wavelength of photons entering the chamber, introducing spurious velocity shifts. The error contribution from this effect is negligible with continuous pumping of the chamber, as described in Section \ref{sec:environment}. We have not observed any negative impact from leaving the pumps on while observing, as discussed in Section \ref{sec:lab_tests}. \\

\noindent
\underline{Instrument vibrational stability} - vibrations in the instrument due to pumps, coolers, and the observatory dome and telescope will cause a mechanical drift between calibration and science frames. The effect of the vacuum pumps has been measured to be negligible, as discussed in Section \ref{sec:lab_tests}, which we expect to be one of the largest sources of vibrational instability. \\

\noindent
\underline{Mechanical creep} - mechanical creep in the instrument due to stresses of the optical bench holding heavy optical elements may lead to drift, as well as growth in materials such as Invar and Zerodur over time \citep{bayer-helms1987,steele1992}. These drifts are very slow in time, thus are easily calibratable, as discussed in Section \ref{sec:environment}. \\

\noindent
\underline{Zerodur phase change} - Zerodur is the grating substrate of EXPRES, as it ages, it slowly undergoes a phase change that effectively changes the groove spacing, leading to a calibratable velocity drift \citep{halverson2016}.

\startlongtable
\begin{deluxetable*}{p{30mm}p{30mm}p{20mm}p{20mm}p{40mm}}
\centering
\tablecaption{Radial velocity error sources identified in EXPRES. \label{tab:table_errors}}
\startdata
\\
\hline
\rowcolor[HTML]{CBCEFB}[\tabcolsep]
\multicolumn{1}{|p{30mm}|}{Error term}        
    & \multicolumn{1}{p{30mm}|}{Mitigation method}    
        & \multicolumn{1}{p{20mm}|}{Uncorrected \newline Magnitude} 
            & \multicolumn{1}{p{25mm}|}{Residual \newline Error}   
                & \multicolumn{1}{p{40mm}|}{Source of Estimate}         
                    \\ \hline
                        \\
On-sky Effects                                               
    &       
        &       
            &                                                                                   
                &        
                    \\ \hline
\rowcolor[HTML]{EFEFEF}[\tabcolsep]
\multicolumn{5}{|l|}{Instrument}     
    \\ \hline    
\multicolumn{1}{|p{30mm}|}{\raggedright Image motion on fiber}                     
    & \multicolumn{1}{p{30mm}|}{FTT, double scrambler} 
        & \multicolumn{1}{l|}{$> 10$ cm s$^{-1}$}           
            & \multicolumn{1}{l|}{2.0 cm s$^{-1}$}                                      
                & \multicolumn{1}{p{40mm}|}{\raggedright Measured motion on sky, error calculated theoretically}      
                    \\ \hline
\multicolumn{1}{|p{30mm}|}{\raggedright Atmospheric dispersion}                    
    & \multicolumn{1}{p{30mm}|}{ADC, chromatic \newline exposure meter}                     
        & \multicolumn{1}{l|}{$\sim$10 cm s$^{-1}$} 
            & \multicolumn{1}{l|}{2.0 cm s$^{-1}$} 
                & \multicolumn{1}{p{40mm}|}{\raggedright Estimate based on ADC design, FTT images}   
                    \\ \hline
\multicolumn{1}{|p{30mm}|}{\raggedright Barycentric correction}                    
    & \multicolumn{1}{p{30mm}|}{\raggedright chromatic exposure meter, proper weighting method} 
        & \multicolumn{1}{l|}{$\sim$10 cm s$^{-1}$}                                 
            & \multicolumn{1}{l|}{1.0 cm s$^{-1}$}                                                      
                & \multicolumn{1}{p{40mm}|}{\raggedright Calculated from knowns (algorithms, shutter timing, coordinates)} 
                    \\ \hline
\rowcolor[HTML]{EFEFEF}[\tabcolsep]
\multicolumn{5}{|l|}{Non-instrument}     
    \\ \hline
\multicolumn{1}{|p{30mm}|}{Photon noise}
    & \multicolumn{1}{p{30mm}|}{\raggedright Increase exposure time, radial velocity binning}    
        & \multicolumn{1}{p{20mm}|}{30 cm s$^{-1}$ \newline (S/N = 250 \newline per pixel at \newline 578 nm)}    
            & \multicolumn{1}{p{20mm}|}{30 cm s$^{-1}$ \newline (S/N = 250 \newline per pixel at \newline 578 nm)} 
                & \multicolumn{1}{p{40mm}|}{\raggedright Formal error returned from CCF pipeline} 
                    \\ \hline
\multicolumn{1}{|p{30mm}|}{Stellar Activity}
    & \multicolumn{1}{p{30mm}|}{\raggedright Statistics, strategic observing}  
        & \multicolumn{1}{l|}{$> 50$ cm s$^{-1}$}  
            & \multicolumn{1}{l|}{unknown}
                & \multicolumn{1}{p{30mm}|}{Literature}                
                    \\ \hline
\multicolumn{1}{|p{30mm}|}{\raggedright Telluric Contamination}                          
    & \multicolumn{1}{p{30mm}|}{\raggedright Modeling and division}  
        & \multicolumn{1}{l|}{$\sim$25 cm s$^{-1}$}  
            & \multicolumn{1}{l|}{$\sim$10-25 cm s$^{-1}$}
                & \multicolumn{1}{p{30mm}|}{\raggedright On-sky improvements}
                    \\ \hline
\multicolumn{1}{|p{30mm}|}{Sky/Moon Contamination}       
    & \multicolumn{1}{p{30mm}|}{\raggedright Observe bright stars, avoid Moon}
        & \multicolumn{1}{l|}{$\sim$1 m s$^{-1}$}  
            & \multicolumn{1}{l|}{$<$10 cm s$^{-1}$}
                & \multicolumn{1}{p{30mm}|}{\raggedright Sky brightness calculation}
                    \\ \hline
\multicolumn{1}{|p{30mm}|}{Analysis Errors}
    & \multicolumn{1}{p{30mm}|}{-} 
        & \multicolumn{1}{l|}{2.0 cm s$^{-1}$}
           & \multicolumn{1}{l|}{2.0 cm s$^{-1}$}  
               & \multicolumn{1}{p{40mm}|}{\raggedright Analysis of data reduction}           
                   \\ \hline
                       \\
\multicolumn{5}{l}{Calibration Source (LFC)}     
    \\ \hline
\multicolumn{1}{|p{30mm}|}{Modal noise}
    & \multicolumn{1}{p{30mm}|}{\raggedright Fiber agitation}   
        & \multicolumn{1}{l|}{30 cm s$^{-1}$}  
            & \multicolumn{1}{p{20mm}|}{$<5.0$ cm s$^{-1}$} 
                & \multicolumn{1}{p{40mm}|}{\raggedright Test with agitator on/off}
                    \\ \hline
\multicolumn{1}{|p{30mm}|}{\raggedright Refresh rate beat frequency}                              
    & \multicolumn{1}{p{30mm}|}{\raggedright Set SLM frequency away from any vibrational frequencies}   
        & \multicolumn{1}{l|}{up to 5 m s$^{-1}$}  
            & \multicolumn{1}{p{20mm}|}{$<5.0$ cm s$^{-1}$} 
                & \multicolumn{1}{p{40mm}|}{\raggedright Measured with different LFC SLM frequencies}
                    \\ \hline
\multicolumn{1}{|p{30mm}|}{Photon noise}
    & \multicolumn{1}{p{30mm}|}{High S/N exposures}     
        & \multicolumn{1}{l|}{$2.0$ cm s$^{-1}$}    
            & \multicolumn{1}{l|}{$2.0$ cm s$^{-1}$}
                & \multicolumn{1}{p{40mm}|}{\raggedright Measured S/N, formal error returned from CCF}           
                    \\ \hline
\multicolumn{1}{|p{30mm}|}{Calibration Accuracy}
    & \multicolumn{1}{p{30mm}|}{-}          
        & \multicolumn{1}{l|}{2.0 cm s$^{-1}$}  
           & \multicolumn{1}{l|}{2.0 cm s$^{-1}$}               
               & \multicolumn{1}{p{30mm}|}{Literature}      
                   \\ \hline
\multicolumn{1}{|p{30mm}|}{\raggedright Calibration Injection Repeatability}
    & \multicolumn{1}{p{30mm}|}{-}          
        & \multicolumn{1}{l|}{$<$ 1.0 cm s$^{-1}$}  
           & \multicolumn{1}{l|}{$<$ 1.0 cm s$^{-1}$}               
               & \multicolumn{1}{p{30mm}|}{Lab tests}      
                   \\ \hline
\multicolumn{1}{|p{30mm}|}{\raggedright Calibration background noise}                    
    & \multicolumn{1}{p{30mm}|}{Monitor PCF health and replace}
        & \multicolumn{1}{l|}{$\sim$30 cm s$^{-1}$} 
            & \multicolumn{1}{l|}{$<5.0$ cm s$^{-1}$}   
                & \multicolumn{1}{p{30mm}|}{Lab tests} 
                    \\ \hline
\multicolumn{1}{|p{30mm}|}{\raggedright Calibration algorithms}          
    & \multicolumn{1}{p{30mm}|}{\raggedright Must account for variable background}               
        & \multicolumn{1}{l|}{$>5.0$ cm s$^{-1}$}
            & \multicolumn{1}{l|}{$<5.0$ cm s$^{-1}$}   
                & \multicolumn{1}{p{30mm}|}{Data analysis}   
                    \\ \hline
                        \\
\multicolumn{5}{l}{Detector Effects}     
    \\ \hline
\multicolumn{1}{|p{30mm}|}{Temperature Changes}
    & \multicolumn{1}{p{30mm}|}{\raggedright Constant power during integration and readout}   
        & \multicolumn{1}{l|}{$\sim$10 cm s$^{-1}$}  
            & \multicolumn{1}{p{20mm}|}{$<1.0$ cm s$^{-1}$} 
                & \multicolumn{1}{p{30mm}|}{\raggedright Environmental monitoring}
                    \\ \hline
\multicolumn{1}{|p{30mm}|}{Electronics Noise}                          
    & \multicolumn{1}{p{30mm}|}{\raggedright Readout rate important} 
        & \multicolumn{1}{l|}{$\sim$10 cm s$^{-1}$}    
            & \multicolumn{1}{l|}{$<5.0$ cm s$^{-1}$}
                & \multicolumn{1}{p{30mm}|}{Measured S/N}
                    \\ \hline
\multicolumn{1}{|p{30mm}|}{\raggedright Pixel Position Non-uniformity}                           
    & \multicolumn{1}{p{30mm}|}{Sub-pixel CCD characterization}
        & \multicolumn{1}{l|}{5.0 cm s$^{-1}$}
           & \multicolumn{1}{l|}{5.0 cm s$^{-1}$}
               & \multicolumn{1}{p{30mm}|}{Lab tests}          
                   \\ \hline
\multicolumn{1}{|p{30mm}|}{Charge Transfer \textbf{Inefficiency (CTI)}} 
    & \multicolumn{1}{p{30mm}|}{\raggedright matching high S/N exposures, correction to restore charge}       
        & \multicolumn{1}{l|}{up to 5 m s$^{-1}$} 
            & \multicolumn{1}{l|}{$<10$ cm s$^{-1}$}
                & \multicolumn{1}{p{30mm}|}{Lab tests} 
                    \\ \hline
\multicolumn{1}{|p{30mm}|}{Brighter-fatter} 
    & \multicolumn{1}{p{30mm}|}{Match exposure S/N, model PSF}       
        & \multicolumn{1}{l|}{$<5.0$ cm s$^{-1}$} 
            & \multicolumn{1}{l|}{$<5.0$ cm s$^{-1}$}
                & \multicolumn{1}{p{30mm}|}{Lab tests} 
                    \\ \hline
\multicolumn{1}{|p{30mm}|}{Imperfect Flat-field}
    & \multicolumn{1}{p{30mm}|}{\raggedright Flat-field correction with many exposures}
        & \multicolumn{1}{l|}{$>5.0$ cm s$^{-1}$}    
            & \multicolumn{1}{l|}{$<5.0$ cm s$^{-1}$}
                & \multicolumn{1}{p{30mm}|}{Data analysis}    
                    \\ \hline
\multicolumn{1}{|p{30mm}|}{Fringing}                    
    & \multicolumn{1}{p{30mm}|}{\raggedright Flat-field correction, continuum removal}
        & \multicolumn{1}{l|}{$<5.0$ cm s$^{-1}$}    
            & \multicolumn{1}{l|}{$<5.0$ cm s$^{-1}$}     
                & \multicolumn{1}{p{30mm}|}{Data analysis}    
                    \\ \hline
\multicolumn{1}{|p{30mm}|}{\raggedright Stray Light and Cosmic Rays}                    
    & \multicolumn{1}{p{30mm}|}{\raggedright Extraction techniques}
        & \multicolumn{1}{l|}{$<5.0$ cm s$^{-1}$}    
            & \multicolumn{1}{l|}{$<5.0$ cm s$^{-1}$}
                & \multicolumn{1}{p{30mm}|}{Data analysis}
                    \\ \hline
                        \\
\multicolumn{5}{l}{Environmental Stability}     
    \\ \hline
\multicolumn{1}{|p{30mm}|}{Temperature Changes}
    & \multicolumn{1}{p{30mm}|}{\raggedright Temperature controlled room/instrument, low-CTE materials}   
        & \multicolumn{1}{l|}{$>10$ cm s$^{-1}$}  
            & \multicolumn{1}{p{20mm}|}{$<5.0$ cm s$^{-1}$} 
                & \multicolumn{1}{p{30mm}|}{\raggedright Environmental monitoring}
                    \\ \hline
\multicolumn{1}{|p{30mm}|}{Pressure Changes}                          
    & \multicolumn{1}{p{30mm}|}{\raggedright Continuous pumping, ion getters} 
        & \multicolumn{1}{l|}{$4.0$ cm s$^{-1}$}    
            & \multicolumn{1}{l|}{$<1.0$ cm s$^{-1}$}
                & \multicolumn{1}{p{30mm}|}{\raggedright Environmental monitoring}
                    \\ \hline
\multicolumn{1}{|p{30mm}|}{\raggedright Vibrational stability (pumps, coolers, observatory)}                           
    & \multicolumn{1}{p{30mm}|}{Mechanical isolation}          
        & \multicolumn{1}{l|}{$>5.0$ cm s$^{-1}$}
           & \multicolumn{1}{l|}{$<5.0$ cm s$^{-1}$}
               & \multicolumn{1}{p{30mm}|}{Lab tests}                                
                   \\ \hline
\multicolumn{1}{|p{30mm}|}{Mechanical Creep}                           
    & \multicolumn{1}{p{30mm}|}{Frequent calibration}          
        & \multicolumn{1}{l|}{$>5.0$ cm s$^{-1}$}
           & \multicolumn{1}{l|}{$<1.0$ cm s$^{-1}$}
               & \multicolumn{1}{p{30mm}|}{Literature, lab tests}                                
                   \\ \hline
\multicolumn{1}{|p{30mm}|}{\raggedright Zerodur Phase Change (grating)}                           
    & \multicolumn{1}{p{30mm}|}{\raggedright Frequenct calibration, aged Zerodur}          
        & \multicolumn{1}{l|}{$5.0$ cm s$^{-1}$}
           & \multicolumn{1}{l|}{$<1.0$ cm s$^{-1}$}
               & \multicolumn{1}{p{30mm}|}{Literature}                                
                   \\ \hline
                       \\
\hline
\rowcolor[HTML]{FFB1B1}[\tabcolsep]
\multicolumn{5}{|l|}{Summary of Errors}     
    \\ \hline
\rowcolor[HTML]{FFB1B1}[\tabcolsep]
\multicolumn{1}{|p{30mm}|}{Component}
    & \multicolumn{1}{p{30mm}|}{Error}
        & \multicolumn{3}{l|}{Justification}
            \\ \hline
\multicolumn{1}{|p{30mm}|}{\raggedright Instrument Errors}
    & \multicolumn{1}{p{30mm}|}{\raggedright $<10$ cm s$^{-1}$}   
        & \multicolumn{3}{p{80mm}|}{\raggedright Quadrature sum of error terms in the on-sky, calibration, detector, and environmental categories, constrained to be under 10 cm s$^{-1}$ by the LFC stability tests}  
            \\ \hline
\multicolumn{1}{|p{30mm}|}{Photon Noise}                          
    & \multicolumn{1}{p{30mm}|}{30 cm s$^{-1}$ \newline (@ S/N = 250 \newline per pixel)} 
        & \multicolumn{3}{p{80mm}|}{\raggedright Error term from photon shot noise, confirmed by the formal error returned from the CCF of on-sky radial velocity measurements}
            \\ \hline
\multicolumn{1}{|p{30mm}|}{\raggedright Single-Measurement Precision}   
    & \multicolumn{1}{p{30mm}|}{$<32$ cm s$^{-1}$} 
        & \multicolumn{3}{p{80mm}|}{Quadrature sum of instrument errors and photon noise terms}    
                    \\ \hline
\enddata
\end{deluxetable*}

\subsection{Summary of Errors}
At the end of Table \ref{tab:table_errors}, we note the final estimation for the magnitude of instrument errors, combined with the photon noise, to obtain a single-measurement precision below 32 cm s$^{-1}$. In the following sections, justification for the estimate of total instrumental error is provided. The photon noise error has been confirmed with many on-sky measurements, presented in \citep{petersburg2019}. The remaining terms contributing to measurement error are telluric contamination and stellar activity. Constraining the magnitude of these terms and mitigating them is an active research area in Doppler spectroscopy. Further discussion of them is beyond the instrumental scope of this paper.

\section{Instrument Efficiency}
\label{sec:throughput}
\subsection{Throughput of Optical Subsystems}
Maintaining high throughput is challenging for any instrument with many optomechanical subsystems, yet is a requirement for obtaining high S/N with reasonable exposure lengths on stars. We have empirically measured the throughput for as many of these subsystems as possible. In Table \ref{tab:throughput}, we show the throughput for the different subsystems for stellar light, assuming two different seeing values of 0.7 arcseconds and 1.5 arcseconds. The values we present come from different sources. Where possible, we experimentally measured throughput in the lab. This included measurements of the FEM, the various fibers, and the pupil slicer/double scrambler module. The method used was to inject lasers or LEDs of different wavelengths into the various subsystems, and compare the power output with the measured intrinsic power of the sources, while accounting for background noise. The pupil slicer/double scrambler was only measured at one wavelength due to the high power requirements of the light source in order to get significant signal at this point in the instrument. Based on the optical design, we assume that losses are achromatic, and so we have applied the measured throughput to all wavelengths. In the case of the spectrograph optics and CCD, we take the specifications from the manufacturers of the optical components. The efficiency of the telescope was provided by the observatory. Those values are likely upper limits, as the true efficiencies will be lower in practice. Given that there is considerable uncertainty when making these measurements, and possibly not every source of coupling loss has been taken into account, the total throughput values provided here are estimates for the upper limits as well.

\begin{table}
\caption{Throughput values for the optical subsystems of EXPRES. The values were obtained through either measurement, theory, and specifications provided by the manufacturers. Two different values of seeing are included for throughput of the fiber injection mirror (FIM) and the total. Only atmospheric losses from seeing are included, general attenuation is not.}
\label{tab:throughput}
\centering
\begin{tabular}{l|lll}
Component          &  455 nm &  530 nm &  625 nm \\ \hline
Telescope          & 71.20\%                                               & 71.20\%                                               & 69.60\%                                               \\
FEM                & 72.06\%                                               & 79.95\%                                               & 81.01\%                                               \\
FIM (0.7")         & 86.30\%                                               & 86.30\%                                               & 86.30\%                                               \\
FIM (1.5")         & 51.30\%                                               & 51.30\%                                               & 51.30\%                                               \\
65 m Science Fiber & 65.23\%                                               & 78.16\%                                               & 81.88\%                                               \\
Slicer / Scrambler & \multicolumn{1}{c}{85.51\%}                           & \multicolumn{1}{c}{85.51\%}                           & \multicolumn{1}{c}{85.51\%}                           \\
Rectangular Fiber  & 89.63\%                                               & 90.62\%                                               & 90.92\%                                               \\
Spectrograph + CCD       & 45.80\%                                               & 49.80\%                                               & 49.80\%                                               \\ \hline
Total (0.7")       & 9.88\%                                                & 13.87\%                                               & 15.03\%                                               \\
Total (1.5")       & 5.87\%                                                & 8.24\%                                                & 8.93\%                                               
\end{tabular}
\end{table}

\subsection{Science Light S/N}
\begin{figure}
\begin{center}
\includegraphics{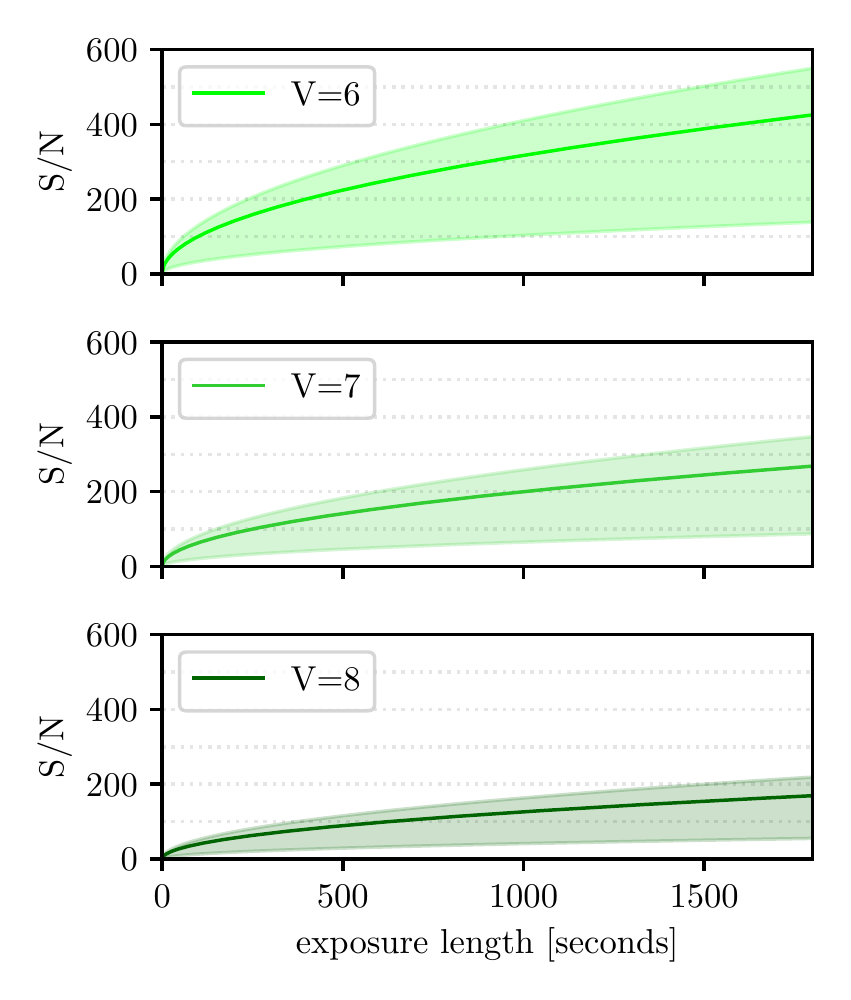} 
\end{center}
\caption{The S/N per pixel values obtained with EXPRES under different seeing conditions at a wavelength of 578 nm, extrapolated for stars of different V-band magnitudes. The solid line represents the median S/N values, while the region around it denotes the best and worse seeing conditions we have encountered.}
\label{fig:stellar_snr}
\end{figure}

\begin{figure*}
\begin{center}
\includegraphics{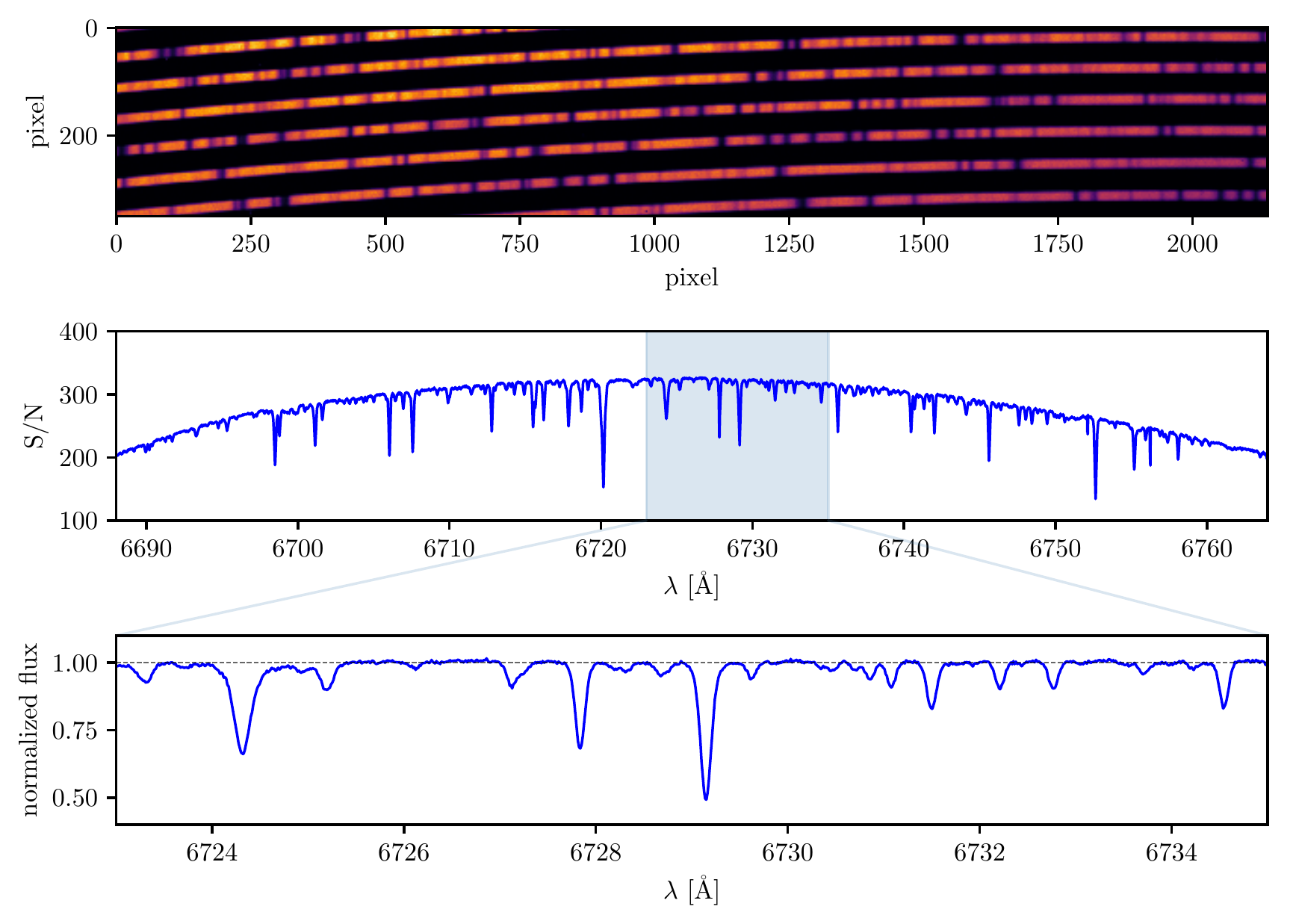} 
\end{center}
\caption{\textit{Top}: Small region of a 2D spectrum of 55 Cnc from a 600 second exposure. \textit{Middle}: S/N per pixel in the 1D extraction of one spectral order from the same exposure. \textit{Bottom}: Continuum normalized region within the same order as above.}
\label{fig:55Cnc_snr}
\end{figure*}

The S/N obtained on stars is dependent on both magnitude of the star and the atmospheric seeing, as a large spot size on the fiber will inherently decrease efficiency of the instrument. The angular size of the fiber on the sky is fixed at 0.9 arcseconds. Figure \ref{fig:stellar_snr} shows the range of S/N values per pixel obtained for stars of different magnitudes, extrapolated from extensive observations of a single star under a range of atmospheric conditions. This S/N is taken at the peak of the order containing a wavelength of 578 nm. With a sampling of four pixels, the S/N per resolution element is roughly double these values. The range of S/N at a given exposure length is driven by variable seeing conditions on different nights. The solid lines denote the median S/N, and the region around it denotes the best and worse seeing conditions we have observed in. The best seeing conditions we have observed in are around 0.7 arcseconds. The median seeing has been close to 1 arcsecond, and the maximum seeing has been several arcseconds. In Figure \ref{fig:55Cnc_snr}, we show a small region of the 2D spectrum of 55 Cnc (top panel), S/N plotted per pixel for a full spectral order (middle panel), and a continuum normalized region within that order (bottom panel). This exposure length was 600 seconds, which yielded a peak S/N in this red order of 327 per pixel. At this level, radial velocity error from photon noise is reduced to around 20 cm s$^{-1}$, as confirmed by the formal error returned from the CCF method of solving for radial velocity. The details of this CCF and other reduction and extraction methods used for EXPRES are presented in \cite{petersburg2019}. 

\subsection{Calibration S/N}
\begin{figure*}
\begin{center}
\includegraphics{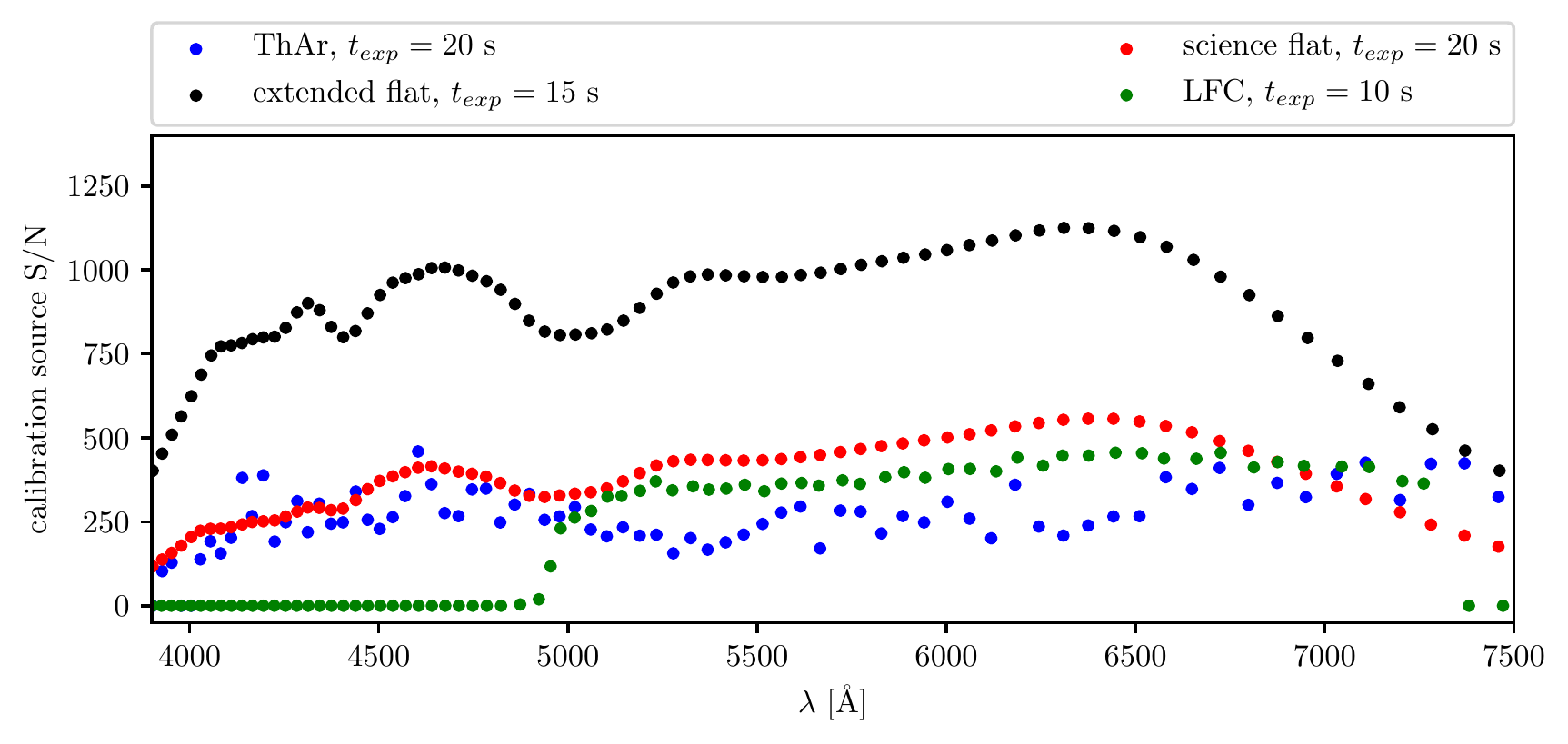} 
\end{center}
\caption{The S/N per pixel in each order for each of the calibration sources, including the extended flat. The length of the exposures is noted in the legend. The peak S/N per pixel in each order is shown for the LFC and flat-field light source, while the mean line S/N per pixel for lines above a S/N of 100 in each order is shown for the ThAr lamp.}
\label{fig:cal_snrs}
\end{figure*}

In Figure \ref{fig:cal_snrs}, we show the S/N per pixel across the wavelength range of the instrument for each calibration mode for the typical exposure length used in normal operation. In this figure, the science flat, ThAr, and LFC calibrations are taken from the science fiber, and the flat is shown for the extended fiber as well. The LFC exposures need to be long enough for residual modal noise to average out, discussed more in Section \ref{sec:illumination}. The efficiency from the calibration unit to the spectrograph is very low $(<1\%)$, but the light sources are very bright, enabling reasonable exposure times. In the case of the ThAr spectra, there is considerable variation in the brightness of different emission lines. The brightest lines exhibit the best S/N, and may even be outside the linear regime of the CCD. In Figure \ref{fig:cal_snrs}, we show the mean S/N of the peak of emission lines in the ThAr lamp above a S/N of 100. The mean of all lines is therefore somewhat lower. In the rest of the calibration modes, we show the peak S/N in each order, this is much more uniform for the LFC and flat-field light source, owing to the considerable variability in line strength in ThAr spectra.

\section{Environmental Stability}
\label{sec:environment}
\subsection{Room and Chamber Temperature Stability}
Minimizing temperature deviations is of high importance for instrument stability, as thermal expansion and contraction of optical components will inevitably lead to spurious shifts that may only be partly calibratable. Additionally, the LFC requires temperature stability of $\pm 1$ K to operate effectively. EXPRES is contained in a vacuum chamber within a temperature-controlled room that is isolated from the rest of the observatory, but the chamber itself is not actively temperature-controlled. Therefore, any temperature changes in the room may propagate to the chamber. Such changes are minimized by a thermal enclosure surrounding the vacuum chamber, two layers of radiation shielding covering the inside walls of the chamber, and G10 thermal isolating blocks between the spectrograph optical bench and chamber, and between the chamber and the ground. The optical bench and mirror mounts were crafted from Invar, a nickel-iron alloy with extremely low coefficient of thermal expansion at room temperature \cite{steele1992}. 

\begin{figure*}[t]
\begin{center}
\includegraphics{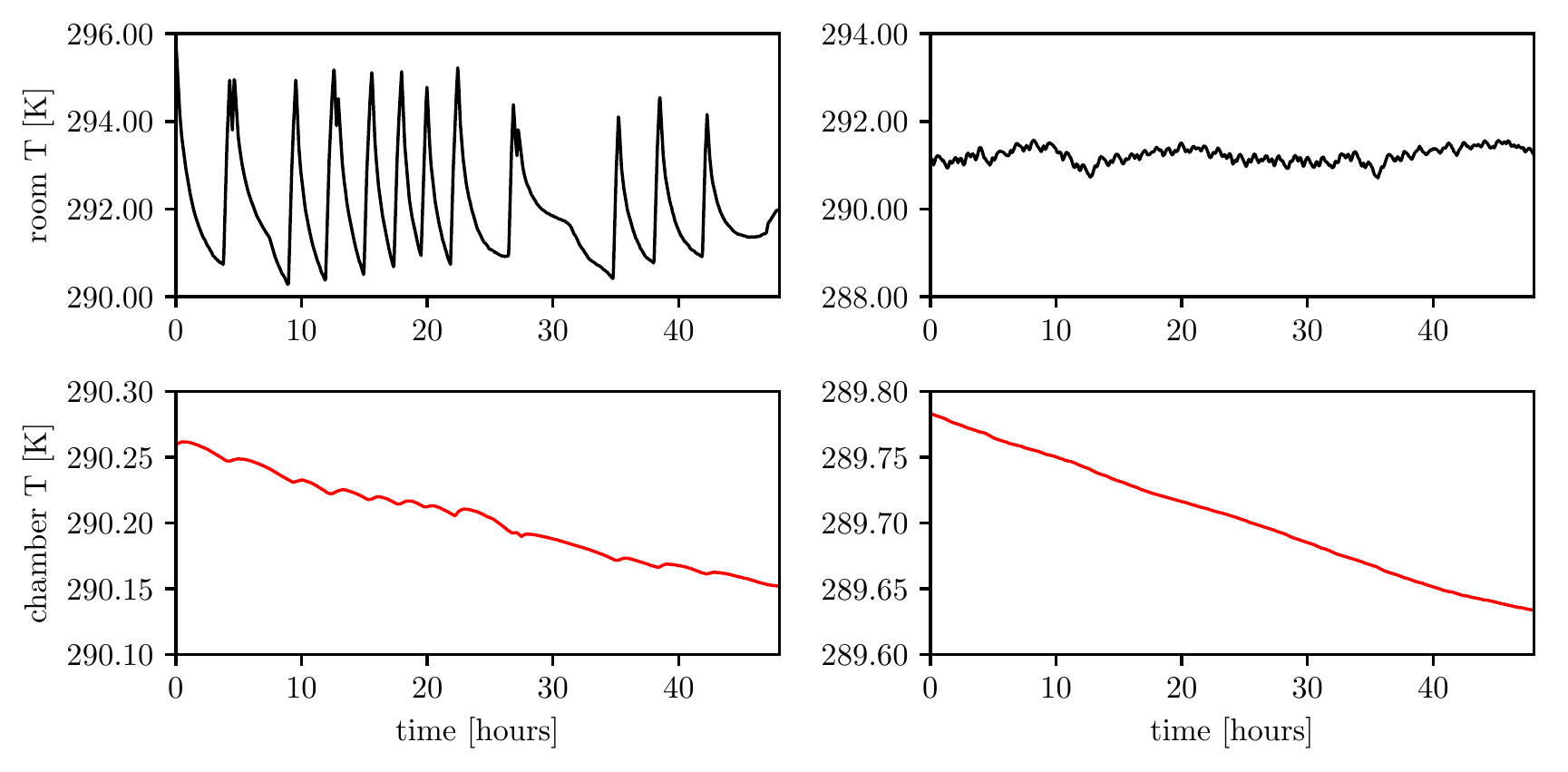} 
\end{center}
\caption{Temperature drifts in the spectrograph room (top) and vacuum chamber (bottom) before the the HVAC system was upgraded (left) and after (right) over 48-hour periods. Sudden spikes in temperature in the spectrograph room resulted in very small spikes in the vacuum chamber temperature, but even after this was fixed, a temperature drift of 0.075 K per day persists in the chamber. }
\label{fig:rc_temps}
\end{figure*}

The heating, ventilation, and air conditiong (HVAC) system of the spectrograph room was rebuilt after the initial design did not meet the goal of minimizing temperature variations to $\pm 0.5$ K per day. The initial problem was probably caused by a combination of stratification of air layers in the room due to insufficient circulation, a single-stage heating element that caused rapid temperature spikes when turned on, and the drawing in of outside air caused a rapid decrease in temperature when the heater was off. This effect was mitigated by moving the return duct to floor-level in the room, adding a multi-stage heating element that allows for gradual temperature changes, and closing off the outside air mix. The original acceptance tests were performed in an empty room with no equipment to restrict airflow or add heat. In Figure \ref{fig:rc_temps}, we show the measured temperatures in the spectrograph room and chamber, before and after the fix was implemented, for a 48-hour period. Before the additional ventilation was added, there were consistent, somewhat-periodic dips and spikes in temperature of $\pm 2.5$ K in the room. Following the upgrade, this variation was reduced to less than $0.5$ K per day, meeting the design specifications.

The rapid temperature spikes in the room shown in Figure \ref{fig:rc_temps} propagated to the spectrograph chamber, illustrated by the small spikes in the chamber temperature occurring at the same times. Both temperature spikes were eliminated by the upgraded HVAC system, however, a temperature drift in the chamber remains. This is not likely caused by temperature instability in the room, as the room temperature is level over time. The long-term temperature of the chamber has been observed to correlate weakly with the ground temperature, most likely due to residual thermal conductivity between the chamber legs and the slab on which it is mounted, despite isolation with G10 blocks. This drift is fairly linear and slow over time, so regular wavelength calibration of the instrument mitigates the impact of thermal contraction or expansion of the optical components. 

\subsection{Chamber Pressure Stability}
Pressure variations in the spectrograph chamber will impact the refractive indices of optical components, and change the wavelength of light, inducing spurious velocity shifts. Results from \citet{wilson2012,hearty2014,halverson2016} indicate that pressure variations of $<0.01$ $\mu$torr translate to calibratable velocity shifts of 0.05 cm s$^{-1}$. Large changes in pressure between calibrations frames and science exposures will result in uncalibratable radial velocity errors.

\begin{figure}
\begin{center}
\includegraphics{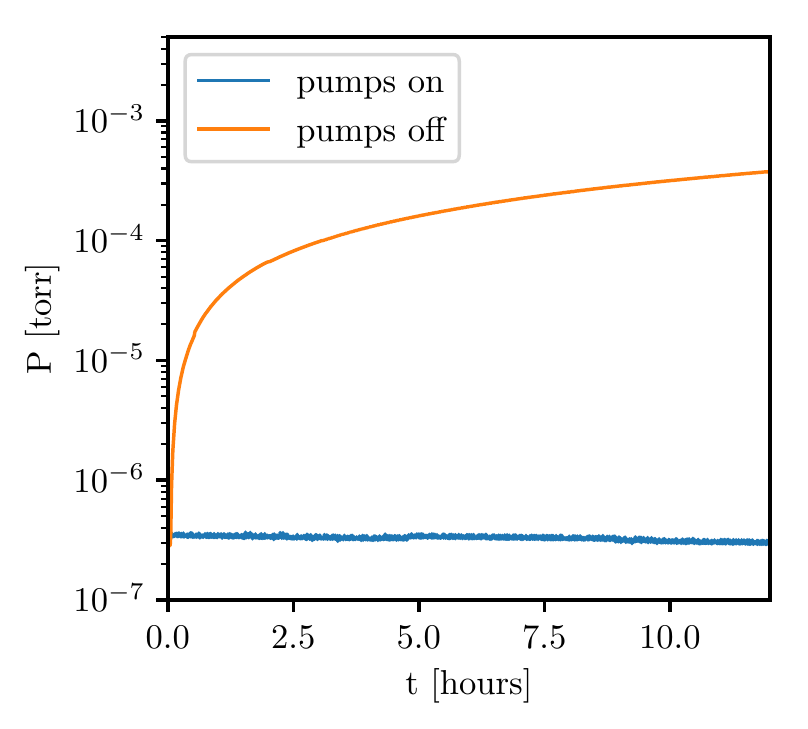} 
\end{center}
\caption{Pressure change in the spectrograph chamber over time, with the vacuum pumps on and with the vacuum pumps off. With the pumps on, pressure stability is achieved at a level of $2\times 10^{-8}$ torr.}
\label{fig:ror}
\end{figure}

The EXPRES vaccuum chamber features two vacuum pumps, which may or may not be used during operation. After extensive pumping, the pressure in the chamber reaches a low of about $1\times 10^{-7}$ torr. When the pumps are turned off, the chamber experiences a steady increase in pressure. This rise is shown in Figure \ref{fig:ror} for a twelve hour period, along with pressure over the same time span with the pumps on. The rate-of-rise in pressure is consistently about $3\times 10^{-5}$ torr/hour. The pressure rises to a maximum of $3.5\times 10^{-4}$ torr in twelve hours. When the pumps are left on, the pressure is stable, with a standard deviation of $2\times 10^{-8}$ torr. 

The radial velocity error for a given pressure change between calibration and science exposures can be obtained from the Edl\'{e}n equation \citep{edlen1966}, revised in \cite{birch1993}, which gives the index of refraction of air at different pressures and temperatures. This index of refraction at a given temperature and pressure is given by
\begin{equation}
\begin{aligned}
n_\mathrm{tp}-1  = & \frac{(p/\mathrm{torr})(n_\mathrm{s}-1)}{720.777} \\ 
                     & \times \frac{1+10^{-6}(0.801-0.01296[T/^\circ\mathrm{C}])(P/\mathrm{torr})}{1+0.0036610(T/^\circ\mathrm{C})},
\end{aligned}
\end{equation}
where $P$ is the pressure in torr and $T$ is the temperature in C$^\circ$, which we assume to be the typical 20 $^\circ$C. $(n_\mathrm{s}-1)$ is the standard index refraction of air, given by
\begin{equation}
\begin{aligned}
n_\mathrm{s}-1 = & 8343.05 \\
                   & + 2406294\times[130-(\sigma/\mu\mathrm{m}^{-1})^2]^{-1} \\
                   & + 15999\times[38.9-(\sigma/\mu\mathrm{m}^{-1})^2]^{-1},
\end{aligned}
\end{equation}
where $\sigma$ is the vacuum wavenumber of the light. The shifted wavelength $\lambda_\mathrm{s}$ is then
\begin{equation}
\lambda_\mathrm{s} = \lambda_\mathrm{0}/n_{\mathrm{tp}},
\end{equation}
where $\lambda_0$ is the initial, vacuum wavelength and $n_\mathrm{tp}$ is the index of refraction of the medium. We have implicitly assumed that the shifted wavelength is from a vacuum to some medium with index of refraction $n_\mathrm{tp}$, however, what we are really interested in is the shift in wavelength from the medium at the time of calibration to the medium at the time of science exposures. Any difference in index of refraction between calibration and science exposures changes the wavelength of light passing through the spectrograph, which is manifested as a spurious radial velocity shift of the star. For practical purposes, at pressures less than 1 atm, this true shift in wavelength can be approximately calculated from just the change in pressure, $\Delta \mathrm{P}$, regardless of what the initial pressure is, even though we implicitly assume vacuum as the reference for the calculation. Therefore, we can express the radial velocity error due to pressure changes as
\begin{equation}
\mathrm{RV}_{\mathrm{error}}/\mathrm{cm\:s}^{-1} = 1.07\times 10^4 (\Delta \mathrm{P/torr})
\end{equation}
at the relevant pressures.

\begin{figure}
\begin{center}
\includegraphics{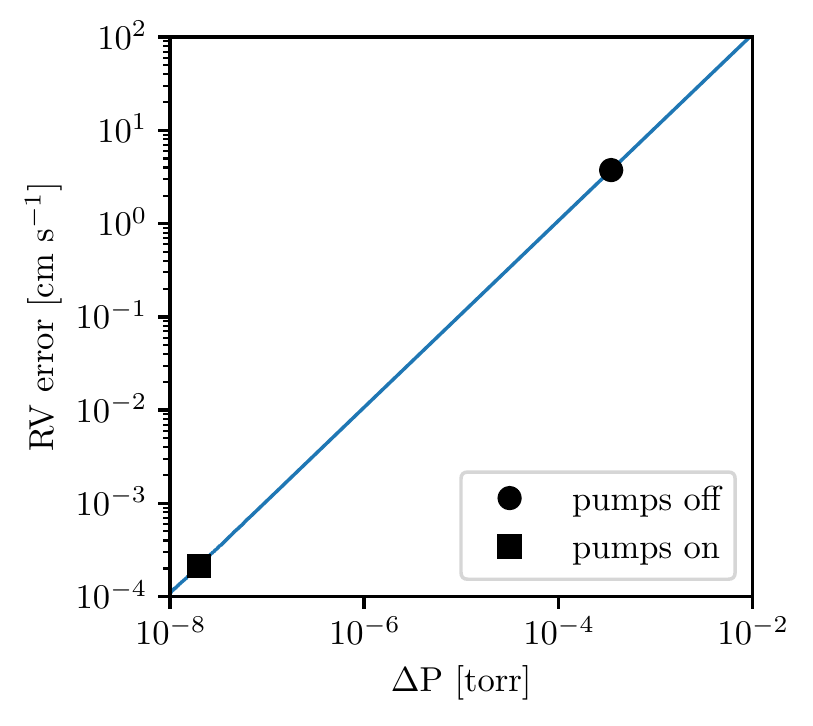} 
\end{center}
\caption{The radial velocity error incurred at different changes in pressure within the spectrograph chamber, from the Edl\'{e}n equation. The error indicated for the case of vacuum pumps off assumes 12 hours of unpumped operation.}
\label{fig:edlen}
\end{figure}

With the current rate-of-rise in pressure from Figure \ref{fig:ror}, the change in pressure from the beginning to end of the night equates to a relative velocity shift of 3.75 cm s$^{-1}$. However, with wavelength solutions being obtained at least once per hour, the maximum uncalibrated error would be 0.3 cm s$^{-1}$. In Figure \ref{fig:edlen}, we show the theoretical radial velocity shifts incurred at different pressure changes in the spectrograph. Points are marked for the expected radial velocity error when the pumps are running and the pressure change after no pumping for 12 hours. However, an additional potentially undesirable impact of changing pressure is change in focus of the instrument, which is generally not checked throughout the night. A second consequence may be that changing pressure facilitates temperature changes within the chamber. Considering this, we generally choose to leave the pumps on during operation. The stability of the pressure with the pumps on is at the level of $2 \times10^{-8}$ torr, which equates to a negligible radial velocity error of $2.1 \times 10^{-4}$ cm s$^{-1}$. The impact of additional vibration from the vacuum pumps on spectrograph stability is explored in section \ref{sec:lab_tests}. 

\subsection{Vibrational Stability}
Several layers of vibration isolation have been implemented to prevent vibrations from both geological and observatory sources. The EXPRES vacuum chamber is mounted on an isolated concrete footing that is decoupled from the rest of the observatory structure. Furthermore, vibration dampening springs are placed between the chamber and this footing, and the optical bench and the chamber. Two additional sources of vibration are from the instrument itself: the vacuum pumps and the CCD cooler. One adverse effect from the CCD cooler pump has already been observed and mitigated; vibrations of the CCD coupled with the refresh rate of the LFC spatial light modulator induced a beat frequency that propagated to a periodic Doppler shift in the wavelength solutions. 

\subsection{Mechanical Creep}
The stability of optical components in 3D space is constrained by the mechanical structures that support them. In addition to the effects of thermal expansion and contraction discussed previously, one more component of instability in the instrument is mechanical creep. Creep can be caused by mechanical stresses, such as the optical bench bearing the weight of large mirrors, prisms, and the grating. Additionally, even at constant temperature, Invar grows due to the movement of carbon atoms within the nickel-iron matrix, at an initial rate of 8-12 ppm yr$^{-1}$ and a long-term slower rate of 2-6 ppm yr$^{-1}$ \citep{steele1992}. This effect depends on several factors, such as time since final heat treatment, method of heat treatment, carbon content, and ambient temperature. The Invar optical bench of EXPRES was never measured to know these exact effects, but they occur on timescales much longer than the nightly calibration sequences, and therefore do not have a significant impact on radial velocity measurement precision.

\subsection{Zerodur Phase Change}
The effective groove spacing of the R4 grating changes slightly over time due to the phase change of the Zerodur substrate \citep{bayer-helms1987}. \citet{halverson2016} note that this induces a calibratable velocity drift of 5 cm s$^{-1}$ day$^{-1}$, measured with the \textit{High Accuracy Radial Velocity Planet Searcher} \citep[HARPS,][]{mayor2003}, but depends on the exact Zerodur being used and its age. This effect is too small to isolate and measure with EXPRES, but it does contribute to the daily drift of the wavelength solution seen in the calibration exposures. This slow change is nearly completely calibratable with the frequent wavelength solutions routinely obtained with EXPRES.

\section{Illumination Stability}
\label{sec:illumination}

\subsection{Atmospheric Dispersion Compensation}
\label{sec:adc}
Atmospheric dispersion compensation (ADC) is used prior to guiding light into the science fiber in order to prevent the chromatic elongation of stars due to refraction in Earth's atmosphere. The EXPRES ADC consists of four prisms, with pairs of prisms mounted on independently rotatable stages. Each pair consists of materials whose refractive indices are chosen so that the central wavelength of the corrected range pass through the pair with no deviation in angle. During an observation, the LDT telescope control system (TCS) sends a packet of current pointing values to the EXPRES control system once every second. From these packets, the current elevation is used to solve for the relative angle required between the two prism pair axes, and the current parallactic angle is used to determine the absolute rotation angles of the two stages. The requested prism rotations are translated into step positions, which are transmitted to the motor controllers, that then move until the commanded configuration is obtained.

\begin{figure*}
\begin{center}
\includegraphics[scale=1.15]{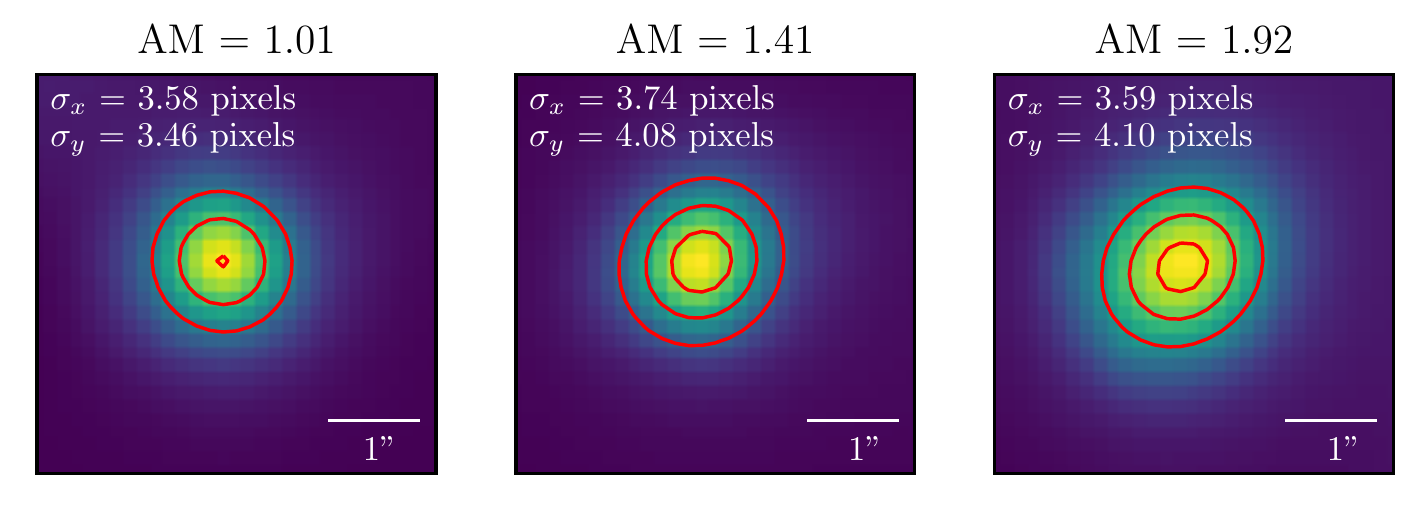} 
\end{center}
\caption{Stellar images at three different air masses on the FTT camera. Atmospheric dispersion increases the ellipticity of the stars as air mass increases, indicating sub-optimal performance of the ADC.}
\label{fig:ftt_am}
\end{figure*}

Several problems may arise if this dispersion is left uncorrected. Throughput would decrease with a chromatic dependence as the spot would be larger on the fiber. Furthermore, a chromatic flux dependence may manifest as chromatic error in the barycentric correction if guiding moves to different chromatic images of the star, although this would be mitigated by the chromatic exposure meter discussed in Section \ref{sec:expm}. This may also cause a radial velocity error due to color differences between different observations, although this can be mitigated by repeatable continuum normalization and color index corrections \citep{halverson2016,wehbe2019}. The EXPRES ADC is designed to work effectively up to zenith distance of $65^\circ$, or 2 air masses, at which point the expected dispersion of the star is expected to be limited to 0.029 arcseconds. In practice, it is rare for us to observe stars at such high air mass in our survey; however, it is occasionally necessary to observe stars at higher air masses, either during transit follow-up or to extend the seasonal time baseline in our data sets. In Figure \ref{fig:ftt_am}, we show example stellar images on the FTT camera at three different air masses, away from the fiber. Many FTT exposures of each star are summed together to get a low-noise image of each of these stars. A two-dimensional Gaussian is fitted to determine the widths of the spots, reported as one standard deviation in each dimension. From an air mass of 1 to an air mass of 2, the stellar image increases in ellipticity, indicating sub-optimal performance of the ADC. The amount of dispersion observed is better than expected from the atmosphere without correction, except in the lowest air mass case, indicating that the ADC is improving the image quality but not yet at the expected level. Remaining issues could be poor image quality delivered by the telescope, or non-optimized rotational angles of the prism pairs for the associated air masses at the LDT site. Future work will include optimizing the rotation angles for a given target elevation, to minimize the ellipticity of the stars. \cite{wehbe2019} indicate that residual dispersion of several hundred mas will contribute a radial velocity error on the order of cm s$^{-1}$ or less.

\subsection{Fast Tip/Tilt Guiding}
\label{sec:ftt}
Image stability on the fiber is important to both maintain high throughput and eliminate radial velocity errors due to image motion, as discussed further in the next subsection. The guiding of starlight on the fiber is performed with the FTT system. A  proportional-integral-derivative (PID) loop is used to rapidly issue updated mirror positions to keep the star centered on the fiber, the position of which is determined via a ``center-of-mass" of the detector counts. The Andor EMCCD is used as the guide camera, as discussed in Section \ref{sec:design}. The advantage of an EMCCD is that extremely fast readout rates are possible with sub-frames, and that read noise is low even in the regime of low photon counts. The camera can image at a rate of up to 666 Hz (1.5 ms exposures) with a $32 \times 32$ pixel sub-frame. In this mode, the FTT mirror can perform corrective motions at a rate of up to 100 Hz. When guiding a star on the fiber, only the halo of the star is visible on the camera, as most light is entering the fiber. Therefore, it is of high interest to understand how stable the illumination is in this setup.

\begin{figure*}
\begin{center}
\includegraphics{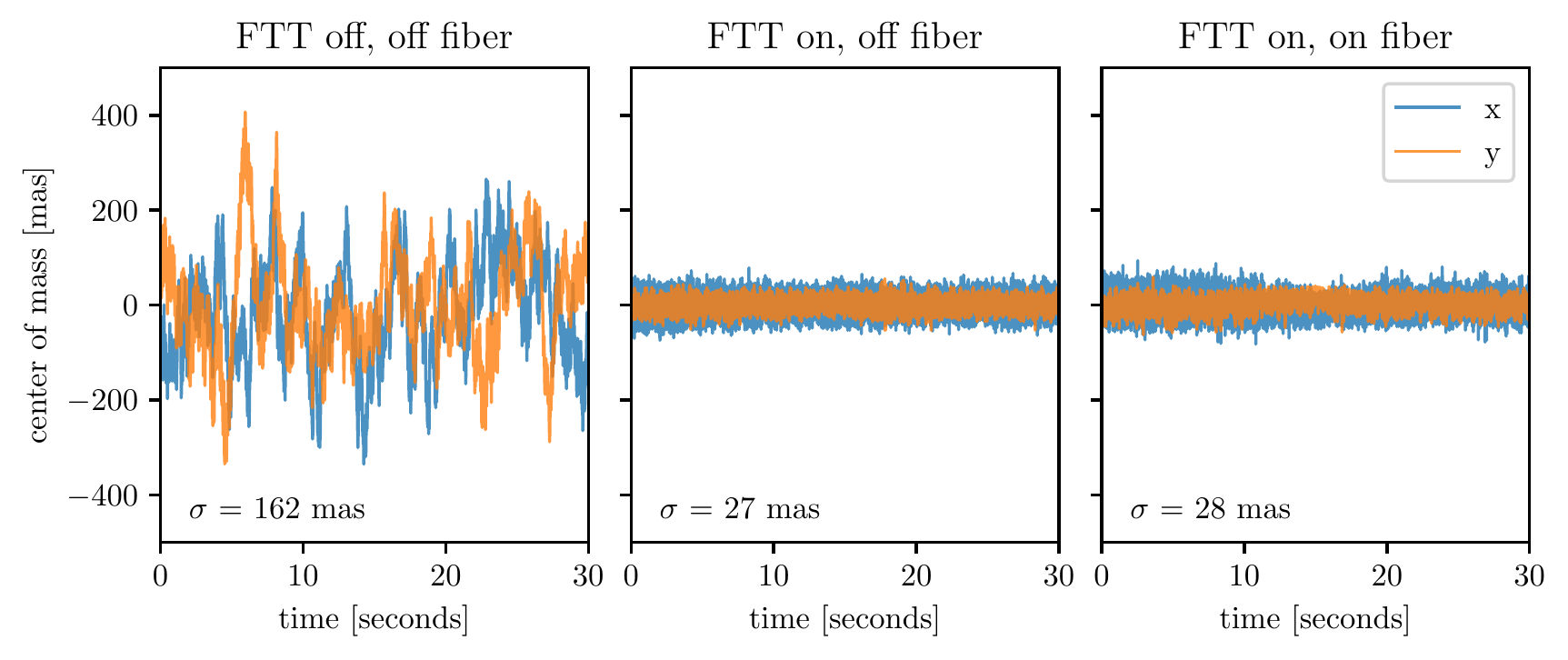} 
\end{center}
\caption{\textit{Left}: Centroid positions of the observed star over time without FTT corrections. \textit{Middle}: FTT guiding on same star, showing factor of six improvement in centroid position. \textit{Right}: FTT guiding active, with the star on the fiber. The image stability is almost as good as with the star on the fiber compared to} off of the fiber.
\label{fig:ftt}
\end{figure*}

To assess the image stability with the FTT guide system, we observed a faint star in several test modes. First, the star was imaged on the camera without the FTT system active, away from the fiber on the mirror, to gauge the natural image motion due to seeing, guiding errors from the telescope, and windshake. Next, the FTT system was activated with the star still away from the fiber, to gauge the improvement with a clean image. Finally, the star was moved onto the fiber, so that guiding with the FTT was actively performed on the residual light reflected from the FIM. The results from these tests are shown in Figure \ref{fig:ftt}. The image centroid is assessed in the horizontal and vertical dimensions via the ``center-of-mass" of counts on the detector. To estimate the total variability in spot position, the standard deviations of the two centroids are added in quadrature. For the test with no active FTT guiding, the star position was stable to 162 mas. With the FTT active, guiding the star to a spot away from the fiber, the image stability improved by a factor of six to 27 mas. When guiding the star onto the fiber using only the residual reflect light from beside the fiber, the image stability was only slightly worse at 28 mas. This is the mode in which science spectra are taken. The ``center-of-mass" algorithm is used due to the speed requirements, but it is also possible to fit two-dimensional Gaussians to all of the sub-frames in post-processing to determine image stability. We found slightly worse stability in the image centroids in this manner, but the performance was still under 50 mas, meeting the original EXPRES specifications for guiding errors. Similar image stability tests were also repeated with the telescope and image rotator in various orientations, and confirmed that flexure of the instrumentation did not have an impact on performance.

\begin{figure*}
\begin{center}
\includegraphics{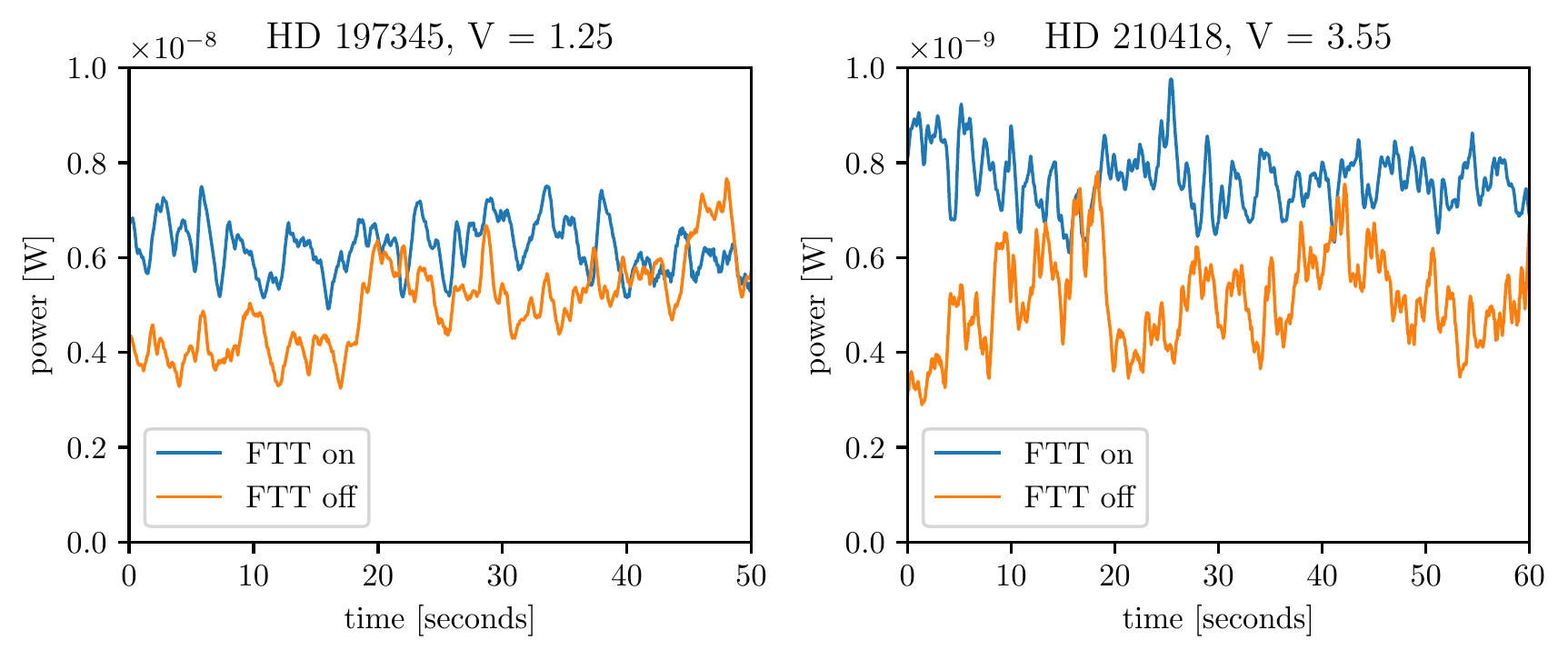} 
\end{center}
\caption{Power measurements at the output of the octagonal science fiber on two different stars, with the FTT on and the FTT off. A consistent improvement in measured power, as well as less variability in measured power, was observed in both cases.}
\label{fig:power_meter_tests}
\end{figure*}

Stability tests with the FTT were also performed with a power meter at the output of the octagonal science fiber, before the slicer and spectrograph were installed. The results of these tests, on two different stars, as a function of time, is shown in Figure \ref{fig:power_meter_tests}. With the FTT on and guiding starlight to the fiber, a consistent improvement in power output from the science fiber was observed. The mean, median, and standard deviation in measured power for both stars is shown in Table \ref{tab:power_meter}. On HD 197345, the mean improvement in power with the FTT on was 25\%, while the standard deviation in measured power decreased by 59\%. On HD 210418, the mean power increased by 50\% with the FTT on, while the standard deviation in power decreased by 54\%. From these measurements, use of the FTT is not only important for image stability, but for improving throughput as well. The remaining variability in these measurements comes primarily from transmission changes in Earth's atmosphere.

\begin{table*}
\caption{Results from power meter tests with the FTT on and off. With the FTT on, measured power improved with less variability.}
\label{tab:power_meter}
\centering
\begin{tabular}{l|llll}
star      & FTT status & mean P {[}W{]}         & median P {[}W{]}       & $\sigma_\mathrm{P}$ {[}W{]} \\ \hline
HD 197345 & on         & $6.21 \times 10^{-9}$  & $6.22 \times 10^{-9}$  & $5.48 \times 10^{-10}$      \\
HD 197345 & off        & $5.01 \times 10^{-9}$  & $5.04 \times 10^{-9}$  & $9.36 \times 10^{-10}$      \\ \hline
HD 210418 & on         & $7.66 \times 10^{-10}$ & $7.67 \times 10^{-10}$ & $6.24 \times 10^{-11}$      \\
HD 210418 & off        & $5.11 \times 10^{-10}$ & $5.07 \times 10^{-10}$ & $9.64 \times 10^{-11}$     
\end{tabular}
\end{table*}

\subsection{Scrambling}
EXPRES has implemented a Bowen-Walraven pupil slicer with a double scrambler. A high scrambling gain, defined as the ratio of displacements between the weighted center of the starlight spot on the input face of the fiber and the weighted center of light at the output of the fiber, is important for minimizing radial velocity error due to inhomogeneous illumination of the spectrograph \citep{perruchot2011}. Light that exits the 66 $\mu$m octagonal science fiber is sliced and stacked in the pupil plane, while the near and far field outputs are inverted. This process results in an estimated scrambling gain of $5\times 10^{3}$ being achieved, in addition to the natural scrambling of multimode fibers. This scrambling gain is a theoretical estimate, however, measurements on prototype devices yielded results of the same magnitude. Double scrambler designs presented in \cite{halverson2015} and \cite{podgorski2014} with near and far field inversion were estimated to have scrambling gains of $10^{4}$. Based on this, the scrambling gain of the EXPRES double scrambler is conservatively estimated at $5\times 10^{3}$. Inverting the near and far fields smooths out non-uniformities in the pupil plane, producing a PSF that is more uniform and constant in time. Light from the pupil slicer is injected into the rectangular science fiber with core dimensions $132\times33$ $\mu$m and a 5-meter length. The octagonal and rectangular fiber geometries also provide additional scrambling compared to that of circular fibers \citep{sturmer2016}.

\begin{figure}
\begin{center}
\includegraphics{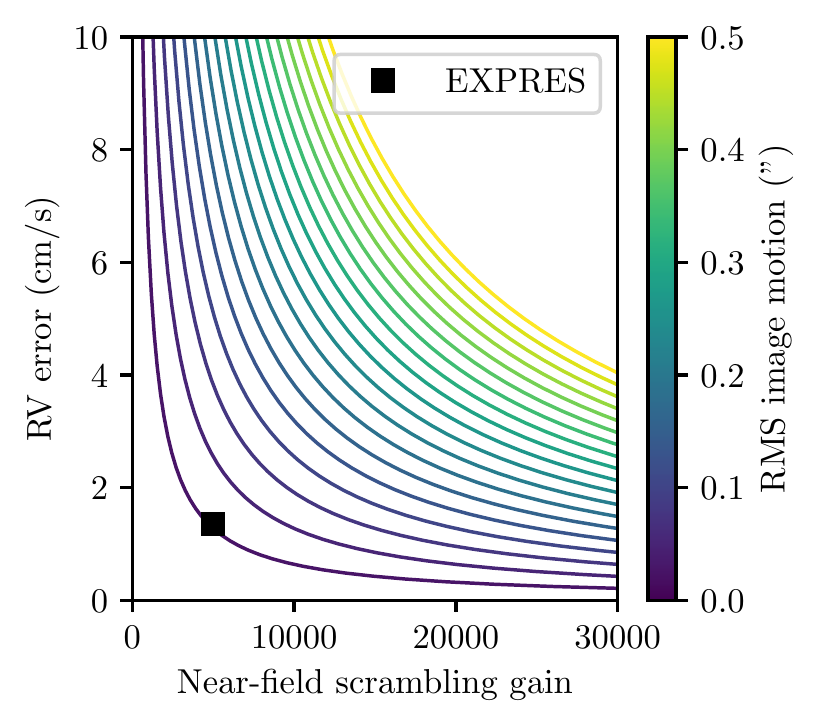} 
\end{center}
\caption{Family of curves indicating the radial velocity error for a given amount of image motion on the fiber and different scrambling gains.}
\label{fig:scrambling_error}
\end{figure}

Throughput tests of the pupil slicer and double scrambler showed an efficiency of about $85\%$. This enables the high resolution of EXPRES without the same loss of light that would come with using a slit to achieve the same resolution. Figure \ref{fig:scrambling_error} shows the expected radial velocity error for EXPRES for a given image motion on the fiber and scrambling gain. The RV error is given by
\begin{equation}
\mathrm{RV}_{\mathrm{error}} = \left( \frac{c}{R_{\mathrm{eff}}} \right) \left( \frac{\delta_\theta}{\theta} \right) \left( \frac{1.0}{\mathrm{scrambling\;gain}} \right)
\label{eq:scrambling}
\end{equation}
where $R_\mathrm{eff}$ is the effective resolution of the spectrograph, $\delta_\theta$ is the RMS of the guiding errors, and $\theta$ is the size of the fiber on the sky \citep{halverson2015}. With data from the FTT showing an image stability of 28 mas and a scrambling gain of at least $5\times 10^{3}$, radial velocity errors from image motion are constrained to be less than 1.4 cm s$^{-1}$. 

\subsection{Modal Noise}
\label{sec:modal_noise}
Multimode optical fibers allow a number of different spatial propagation modes, which interfere at the fiber exit boundary. This causes a speckle pattern in the output---dependent on the coherence of the incident light, wavelength, and geometry of the optical fiber core---that introduces a limit on S/N and radial velocity precision \citep[e.g.,][]{baudrand2001,lemke2011,petersburg2018}. The effects of modal noise are more severe when fewer modes at a given wavelength are propagated through an optical fiber. For this reason, longer wavelengths are more limited in achievable S/N than shorter wavelengths. Additionally, optical fibers with larger core diameters allow more propagation modes and are thus less prone to radial velocity errors from modal noise. However, regardless of the number of propagated modes, modal noise will only average out when the proper mitigation techniques are applied.

Fiber agitation is among the most efficient ways to mitigate the effects of modal noise \citep{petersburg2018}. EXPRES employs a custom fiber agitator that is used during every exposure of calibration and science light. The fiber agitator takes advantage of the fact that chaotic agitation produces optimal mitigation of modal noise \citep{petersburg2018}. The device consists of two disks of diameter 30 cm, with loop attachments at the edge of the disks to hold the fibers. The two disks rotate at different frequencies, typically at 0.5 Hz and 0.45 Hz. This produces vertical fiber motion with an amplitude of 15 cm as well as quasi-chaoticism such that an equivalent fiber configuration is not reached within 10 seconds.

\subsubsection{Calibration Source Modal Noise}
Modal noise is more significant for calibration light than for stellar light due to a higher temporal coherence \citep{mahadevan2014}. Given that the wavelength calibration source for EXPRES is an LFC coupled to the spectrograph via multimode fibers, mitigation of modal noise is critical for achieving a precise wavelength solution. In order to maximize the mitigation of modal noise for these calibration sources, we take wavelength calibration images longer than 10 seconds, thereby allowing the fiber agitator to reach as many quasi-chaotic configurations as possible.

\begin{figure*}
\begin{center}
\includegraphics{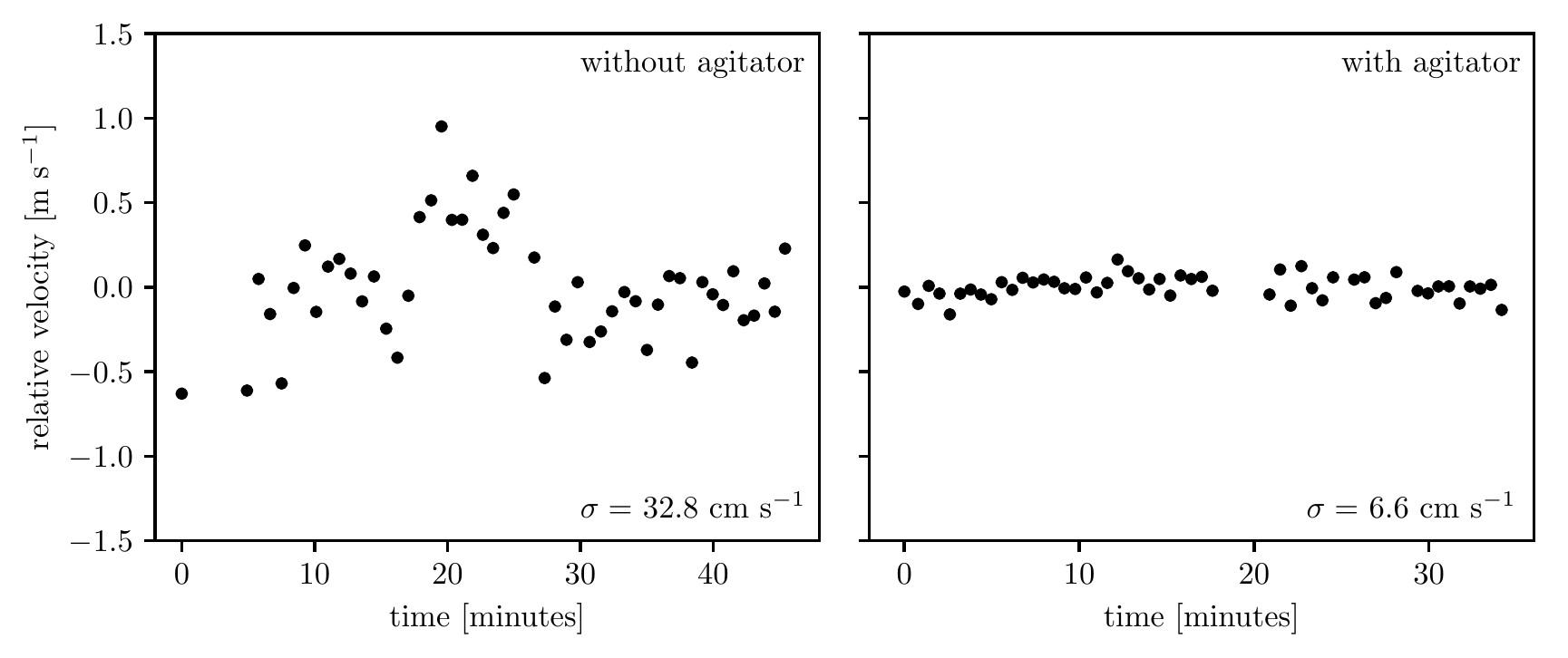} 
\end{center}
\caption{Comparison of instrument stability from cross-correlating LFC exposures without fiber agitation (left) and with fiber agitation (right). Fiber agitation improves the velocity scatter by a factor on the order of 10, reducing this error source to a level below a few cm s$^{-1}$.}
\label{fig:agitator}
\end{figure*}

In Figure \ref{fig:agitator}, we show the relative displacement of the wavelength solution, in terms of velocity, of many consecutive LFC exposures relative to the first exposure with and without fiber agitation. The velocities are solved for by cross-correlating the exposures against an analytic template. Without fiber agitation, the LFC velocity drift has a standard deviation of 32.8 cm s$^{-1}$, compared to 6.6 cm s$^{-1}$ with agitation. The details and rationale behind this test are discussed further in section \ref{sec:lab_tests}. Both sets of data were detrended with a linear function to account for slow, calibratable instrument drift. This improvement of a factor of several illustrates the importance of mitigating modal noise for radial velocity measurements with Doppler spectrographs. The residual impact of modal noise with fiber agitation is limited to less than a few cm s$^{-1}$, accounting for the fact that photon noise and other uncalibratable radial velocity error sources impact this data as well. 

\subsubsection{Continuum Modal Noise}
Continuum modal noise incurred for observations of stars and calibration exposures with the flat-field light source is also mitigated by fiber agitation. Because such continuum sources are highly incoherent, modal noise is expected to be an issue at a much lower level compared to modal noise from coherent calibration source light. Given that the radial velocity error in coherent calibration light has been constrained to a level less than a few cm s $^{-1}$, with a large improvement achieved with the fiber agitator, continuum modal noise is expected to be negligible when the agitator is in operation.

\section{Calibration Sources}
\label{sec:cal_sources}
The absolute accuracy of the wavelength calibration source, as well as the methods used to produce a wavelength solution from it, will directly impact radial velocity precision. The EXPRES LFC features a mode spacing of 14 Ghz, producing an emission line with known wavelength on the detector every 12 pixels in the blue, and 18 pixels in the red.

\subsection{Wavelength Calibration Accuracy}
There is an intrinsic limit on the accuracy of the calibration spectrum. The long-term stability of the Menlo Systems LFC has been measured to approximately 1-2 cm s$^{-1}$ \citep{halverson2016,milakovic2020,probst2020}. Photon noise places an additional limit on the precision of the calibration source at a level of 2 cm s$^{-1}$. The exposure length of LFC frames is typically 10 seconds, and requires the use of a neutral density filter, as the LFC source is naturally very bright. The exposure length needs to be long enough for residual modal noise to average out, as discussed in \ref{sec:modal_noise}. We do not observe any offsets in the wavelength solution when toggling the LFC between on and standby mode throughout each night, indicating robust performance and repeatability of the accuracy of the source. When the LFC is in standby mode, the source comb is active and filtered, but the main amplification and broadening is off and no light is produced. This allows for a quick transition to turn the LFC on.

\subsection{Calibration Injection Repeatability}
\begin{figure}
\begin{center}
\includegraphics{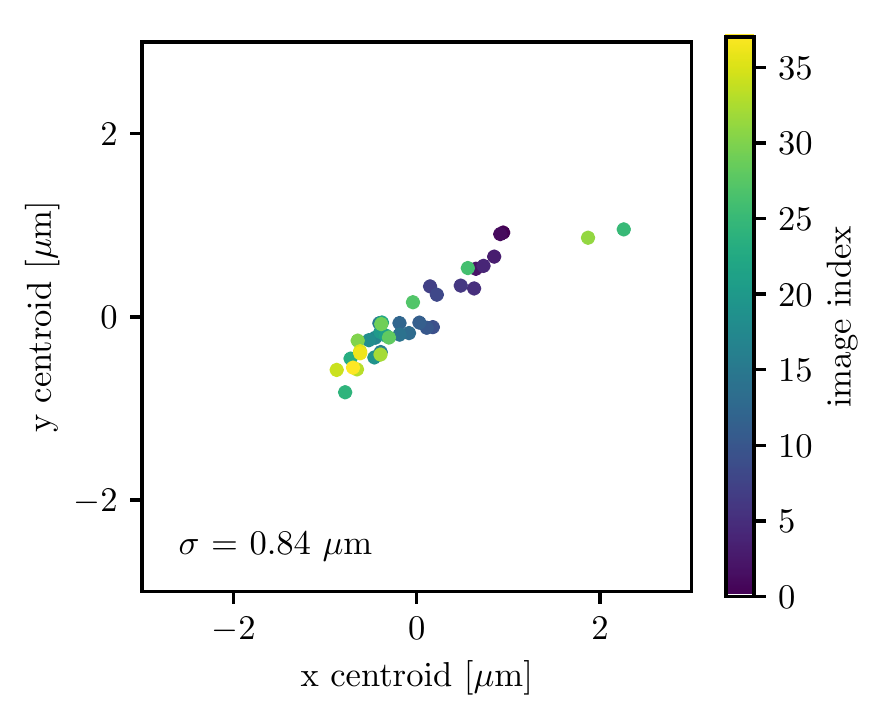} 
\end{center}
\caption{The x and y centroid positions of calibration light being injected into the science fiber, where each exposure was taken after moving the calibration injection mirror in and out of position.}
\label{fig:cim}
\end{figure}

In the current wavelength calibration procedure of EXPRES, an LFC exposure is taken every $\sim$15 minutes. The LFC calibration exposure is generally made while slewing to a new target so there is essentially no overhead (no lost telescope time) for this calibration. To perform wavelength calibration on the same pixels that are used for science, a calibration injection mirror drops into the beam path to inject calibration light into the octagonal science fiber within the FEM. Any significant offsets in the centroid of the LFC light on the fiber would propagate to a shift in the wavelength solution due to imperfect scrambling in the fibers, as in the case of imperfect guiding of starlight on the fiber. To assess this possibility, we measured the calibration spot location on the fiber with the FTT camera between many instances of moving the calibration injection mirror in and out of position. The results of these centroid positions are shown in Figure \ref{fig:cim}. The points are colored by their index in the sequence, which is equivalently their position in time. The standard deviation of the quadrature sum of the x and y centroid positions is 0.84 $\mu$m. However, the change in centroid position is not random, it generally moves from the upper right to the lower left on the detector. This may contribute to a false linear drift of the wavelength solution over time. With the known movement, it is possible to calculate the centroid motion of the output of the fiber via
\begin{equation}
\mathrm{SG} = \frac{d_\mathrm{input}/D_\mathrm{input}}{d_\mathrm{output}/D_\mathrm{output}}
\end{equation}
where SG is the scrambling gain, $d$ is the displacement of the spot centroid at the input or output and $D$ is the diameter of the fiber input or output \cite{halverson2015}. Assuming a scrambling gain of $5\times10^3$ and the appropriate fiber sizes of EXPRES, a 1 $\mu$m displacement at the science fiber input translates to a displacement at the output of the science fiber of 0.1 nm. Such a displacement makes up a fraction of $3\times10^{-6}$ of the fiber output face. This fiber output is dispersed in the spectrograph with a point spread function of 4 pixels, which collectively make up 2400 m s$^{-1}$ in velocity near the center of the spectral format (600 m s$^{-1}$ per pixel). A fractional shift in the dispersion direction of $3\times10^{-6}$ equates to a velocity shift of 0.7 cm s$^{-1}$ in the wavelength solution. This is an upper limit to this source of radial velocity error, as we implicitly assumed that all of the displacement was in the dispersion direction in this calculation. If the displacement was purely in the cross-dispersion direction, there would be no immediate impact on the wavelength solution. In reality, there is probably some combination of displacement in both dimensions, leading to an error somewhere in between 0 cm s$^{-1}$ and 0.7 cm s$^{-1}$. This source of error is uncalibratable in the data reduction, and while small, the positional repeatability of the calibration injection mirror should be monitored over any given epoch. In between some epochs in the EXPRES survey, the FEM was disassembled and removed from the telescope. When reinstalled, the typical calibration injection position may change substantially compared to that of night-to-night changes, leading to a larger radial velocity offset for different epochs. This is accounted for in the EXPRES radial velocity analysis.  

\subsection{Wavelength Calibration Process}
Accuracy of the calibration source is also limited by S/N variations in different comb lines and the finite pixel sampling of the PSF of the spectrograph. \citet{terrien2014} performed simulations for the \textit{Habitable Zone Planet Finder} (HPF) optical frequency comb, finding an error of $<5$ cm s$^{-1}$. \cite{halverson2016} note that this error will be smaller for the NEID instrument compared to HPF, due to a larger sampling and improved spectral flattening with a Menlo Systems LFC. We expect similarly improved performance, owing to the similarities between the NEID instrument and EXPRES. HPF has a spectral sampling of 3 pixels, while the sampling of EXPRES is 4 pixels. The spatial light modulator (SLM) of the Menlo Systems LFC can produce an extremely flat output of the comb. Brightness variations of the spectrum across the detector are then primarily set by efficiency differences across spectral orders (e.g., the middle spectrum in Figure \ref{fig:55Cnc_snr}) and the overall wavelength dependent throughput of the instrument, set primarily by blue photon losses in the optical fibers. While the wavelengths of the calibration lines may be stable, one concern is that variable background influences the quality of the fits to line centroids, which propagates to a radial velocity error. Although this effect can be mitigated by fitting a high order function to the rapidly varying background of the LFC, any such fit is prone to untrackable systematic uncertainties and may cause subtle changes to the resultant radial velocity calculation.

\begin{figure*}
\begin{center}
\includegraphics{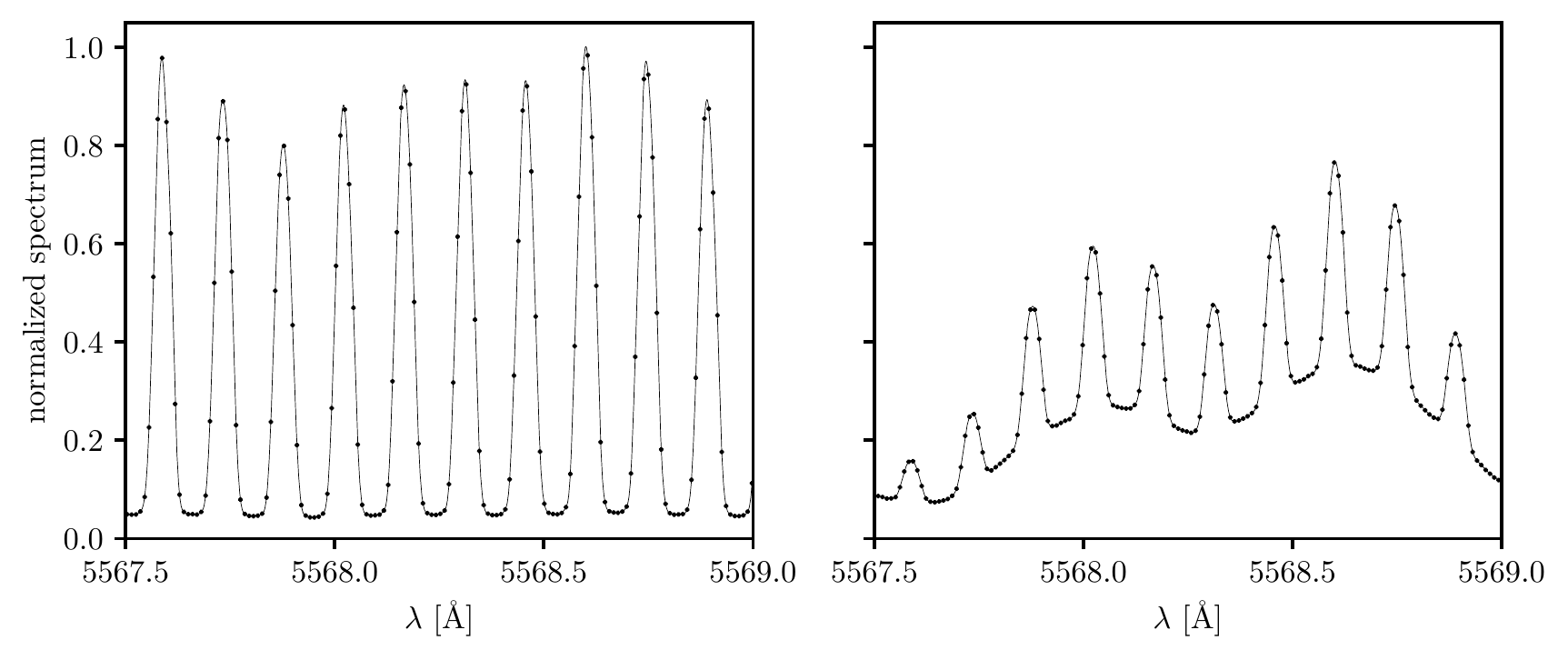} 
\end{center}
\caption{\textit{Left}: a small region of the LFC spectrum after the PCF was replaced, along with other hardware improvements. \textit{Right}: spectrum of the LFC with a degraded PCF. In both panels, a cubic spline interpolation has been applied to help visualize the shapes of the lines.}
\label{fig:lfc_comp}
\end{figure*}

When the PCF of the LFC degrades, the spectral flattening of the comb becomes sub-optimal, and the background exhibits higher amplitude variations. The PCF is responsible for broadening the initial, infrared comb to visible wavelengths, and degrades over time due to damage sustained from high-energy UV and blue photons. These photons heat the fiber, causing imperfections in the material. As more damage is sustained, the transmission and spectral broadening capabilities of the fiber will decrease. It is therefore important to monitor the health of this fiber, and replace it when necessary. In Figure \ref{fig:lfc_comp}, a spectrum of the LFC is shown for part of one order, both when the PCF was degraded (right panel) and shortly after the PCF was replaced (left panel), along with other minor adjustments to the LFC hardware. The spectrum in the left panel exhibits a smoother and smaller background. Cross-correlating many LFC spectra in the poor condition of the PCF showed velocity scatter of up to 30 cm s$^{-1}$ worse than with the fiber in good condition, indicating a significant degradation in calibration precision of the instrument.

The lifetime of the PCF depends primarily on how much it is used, but also the polarization axis used. The lifetime of the fast polarization axis is estimated at 200 hours, while the slow polarization axis has an estimated lifetime of 900 hours. Broadening with the slow axis is more efficient, but has a longer starting wavelength of the spectrum (500 nm compared to 450 nm for the fast axis). In EXPRES, the slow polarization axis is now used to maximize lifetime of the PCF, but the fast polarization axis was used in the first PCF. The nature of the degradation of the PCF is also of high interest. When this starts to occur, noticeable dips in the flattened LFC spectrum appear in the low-resolution diagnostic plot within the operating software. Given the qualities in the spectra shown in Figure \ref{fig:lfc_comp}, we see that the S/N will decrease at a given exposure time when the PCF degrades, with more complex shapes of the background. When this occurs, centroid fitting of the lines becomes less robust, reducing the quality of the wavelength solution. It is also possible to monitor the health of the PCF in the extracted data by running a peak-finding algorithm to track the number of peaks that meet specific quality checks. We have found that the undesirable qualities of the PCF degradation do not appear until late in the PCF lifetime, and that the PCF will degrade quickly following that. For the first EXPRES PCF, this degradation took place over a period of about two months under a normal operating schedule until the PCF was replaced. We will be closely monitoring the behavior of the second PCF when it starts to degrade to compare to the first PCF. However, after seven months of use, we have not yet observed the same signs of degradation that we saw in the first PCF. The first EXPRES PCF was used in the fast polarization axis, thus it is possible that the results of the degradation will be different when using the slow axis.

\subsection{Flat-Field Light Source}
Imperfections in the flat-field light source may propagate to errors in correcting for pixel-to-pixel quantum efficiency variations in the CCD. We assume in the flat-fielding procedure that the spectral energy distribution (SED) of this light source is uniform in power over any resolution element in the spectrum. Even if the SED of the light source was perfectly uniform, we expect the S/N in a given order to vary strongly, but smoothly, due to the blaze function of the spectrograph. However, any sharp peaks in the SED of the light source may propagate to poor flat-fielding, inducing radial velocity errors as different parts of the stellar spectra pass over such peaks due to the barycentric motion of the Earth. A second concern is that a flat-field light source that is not spectrally uniform in power outpout will contribute to errors in continuum normalization. This error is likely very small and has not been observed in reduced EXPRES data.

\begin{figure*}[ht]
\begin{center}
\includegraphics{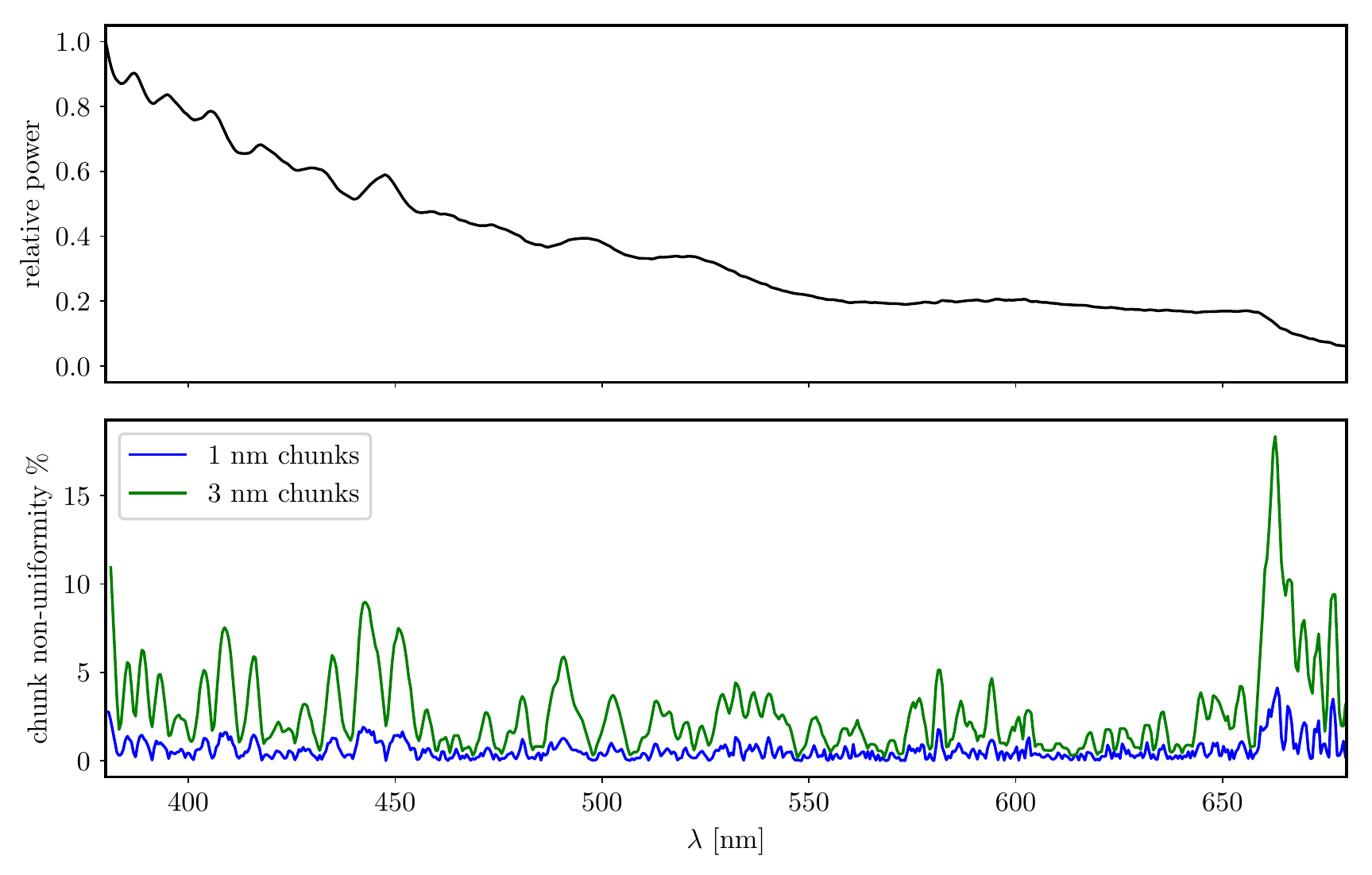} 
\end{center}
\caption{\textit{Top}: The SED of the LED-based flat-field light source. \textit{Bottom}: The variation in power for two wavelength windows across the wavelength range. The spectrum is relatively smooth with no sharp peaks in the SED, and allows for relatively high S/N to be achieved throughout the wavelength range of the instrument in a single exposure.}
\label{fig:led_chunk}
\end{figure*}

In Figure \ref{fig:led_chunk}, we show the normalized, absolute spectral energy distribution of the EXPRES flat-field light source in the top panel, provided by the vendor of the device, FiberTech Optica. The light source is relatively blue to compensate for extra losses of blue photons in the optical fibers. In the bottom panel of Figure \ref{fig:led_chunk}, we show the power amplitude variation within small wavelength chunks of 1 nm and 3 nm. While the source may exhibit SED variations, these are expected to be negligible over the small size of a resolution element of EXPRES, which vary in size from approximately $0.033$ $\mathrm{\AA}$ in blue wavelengths to $0.057$ $\mathrm{\AA}$ in red wavelengths. The SED is relatively smooth, owing the broad emission of LEDs. No sharp peaks are observed, with a power amplitude variability of less than 5\% over 1 nm wavelength windows.

The stability in power output of the source over time is also of interest, as this source can be used to measure detector properties such as gain and charge transfer inefficiency. With a large set of flat-field exposures, we have determined that the variance in power for the EXPRES flat source is at most 0.5\%. This extra variance complicates the typically Poisson-like noise model for an optimal extraction, but it can be folded in once measured. The quartz lamp included in the light source exhibits a negligible spectral variance, which aided in the modeling of the noise characteristics of the flat-field data. 

\section{Detector Effects}
\label{sec:detector}
\subsection{CCD Cryostat Temperature Stability}
\begin{figure*}
\begin{center}
\includegraphics{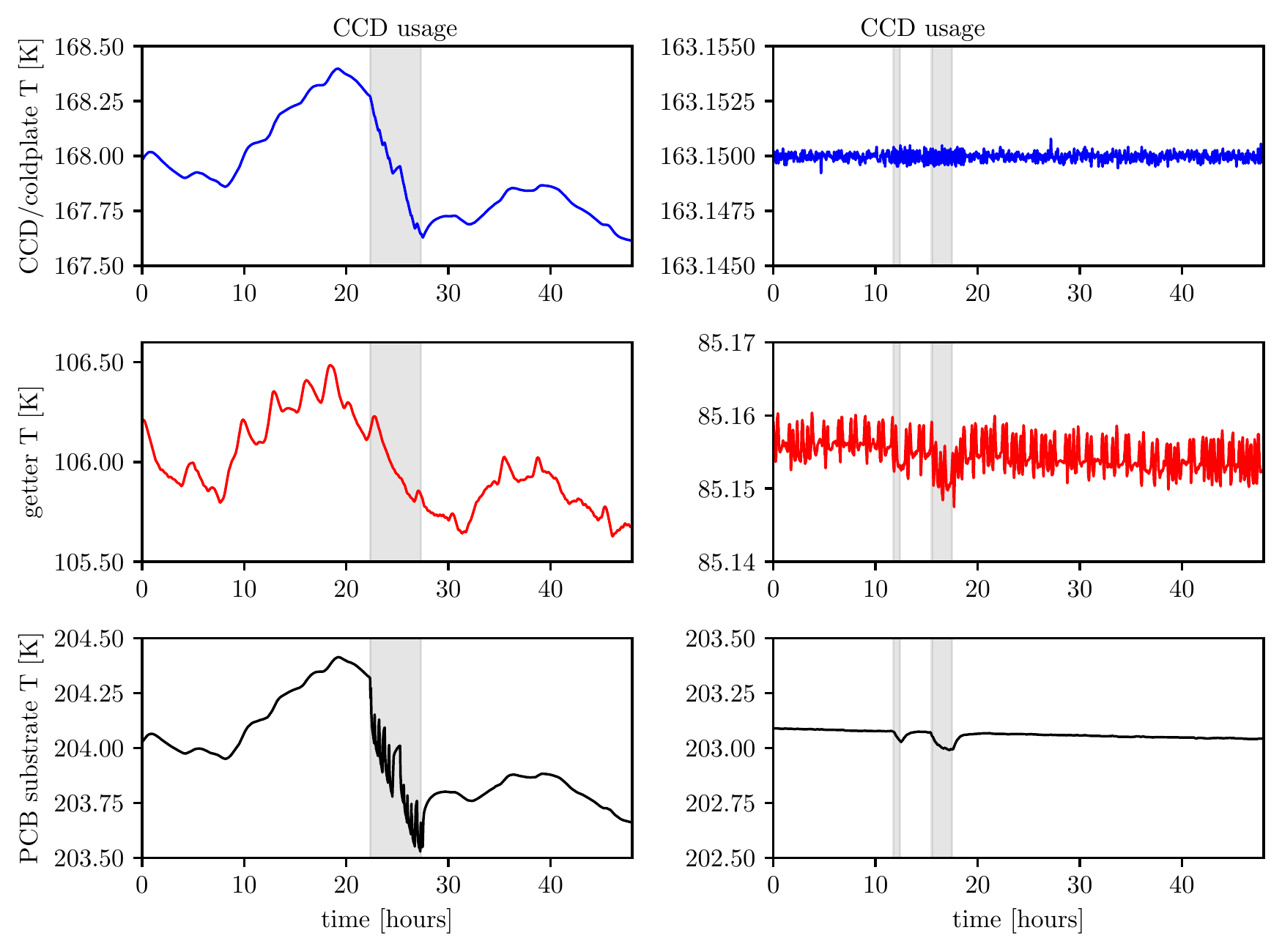} 
\end{center}
\caption{Temperature drifts in the CCD/coldplate, getter, and PCB substrate, for time periods before (\textit{left}) and after (\textit{right}) the constant power mode was implemented. The CCD was in use in the shaded gray regions. Both time periods feature exposures being taken in the middle of the data.}
\label{fig:ccd_temps}
\end{figure*}

Temperature and pressure stability of the CCD are of paramount importance to minimizing spurious radial velocity shifts. Temperature changes lead to thermal expansion and contraction of the detector, and any such changes occurring on timescales between calibration frames and science exposures will be uncalibratable. Of particular interest is the temperature change caused by changing states from standby to integration to readout. We have implemented firmware for the CCD so that the amount of clocking power dissipated in the chip during the exposures is at the same level as during the readout, mitigating these effects. This was to reduce any potential distortions in the CCD caused by changes in the thermal loading. Three temperature readings in the CCD cryostat are shown in Figure \ref{fig:ccd_temps} over a period of 48 hours. The left panels show temperatures before the firmware update was implemented, and the right panels show temperatures after the firmware update. The top row shows the temperature of the coldplate on which the CCD is mounted. This sensor tracks the CCD temperature closely. The middle row shows the temperature of the activated charcoal getter in the cryostat. The bottom row is the temperature of the printed circuit board (PCB) substrate that surrounds the CCD. Times when the CCD was in use are shown by the shaded gray regions, and significant temperature shifts at each reading are apparent when the CCD was in use before the firmware update. Following the update, temperature shifts were reduced by a factor of $10^3$ in the CCD/coldplate, by a factor of 100 in the getter, and by a factor of 5 in the PCB substrate. Before this change, temperature changes in the CCD were present at the 0.1 K level during integration and readout. With the constant power firmware update, temperature changes in the CCD were greatly reduced.

Temperature changes in the CCD/coldplate between readout and integration are now only present at the 1 mK level. However, it is possible that the CCD itself deviates slightly more than the coldplate temperature readings, despite good thermal contact between the two. The coldplate is maintained at a temperature of -110 C$^\circ$ (163.15 K). At this temperature, the coefficient of thermal expansion of silicon $\alpha$ has been measured to be $\alpha(T) = 745.5 \times 10^{-9}$ K$^{-1} $\citep{middelmann2015}. The total size of the CCD is 95.4 mm. The change in size along the dispersion direction can then be found via
\begin{equation}
\frac{\Delta L}{L} = \alpha(T) \Delta T
\end{equation}
where $L$ is the length of the material and $\Delta L$ is the change in length. Even for a larger change in temperature of 10 mK, $\Delta L \approx 0.7$ nm for the whole CCD. Each of the 10.6k pixels is 9 $\mu$m, and will change in physical size by $6.7 \times 10^{-14}$ m. At the center of the spectral format, each pixel covers $\sim$0.01 $\mathrm{\AA}$ in wavelength, equating to a velocity scale of 600 m s$^{-1}$/pixel. A change in pixel size of $6.7 \times 10^{-14}$ m leads to a negligible radial velocity error several orders of magnitude below 1 cm s$^{-1}$. In the constant power mode of the EXPRES CCD, changes in CCD temperature are not significant enough to impact radial velocity precision.

The EXPRES CCD is maintained at cryogenic temperatures via a Stirling cooler. The cooler produces small-scale vibrations in the detector at a frequency of 120 Hz. This effect resulted in a problematic periodicity in the derived wavelength solution over time, caused by a beat frequency with the spatial light modulator (SLM) of the LFC. The SLM is necessary to suppress the naturally large power output variations of the LFC over the full wavelength range, so that the source brightness is relatively constant across the wavelength range of the instrument, and consistent between exposures. The SLM had been performing these flattening corrections of the LFC spectrum at a frequency of 120 Hz during exposures. A small deviation in either the vibrational frequency of the CCD or the SLM correction frequency resulted in biased positions of the LFC emission lines on the detector that varied with a period of six minutes. The amplitude of the deviations in the derived wavelength solutions varied between 3 m s$^{-1}$ and 7.5 m s$^{-1}$ in velocity. This problem was eliminated by changing the SLM correction frequency of the LFC from 120 Hz to 110 Hz, which greatly increased the beat frequency between these two sources to 10 Hz. This effectively averages over the CCD vibrations in any single exposure when forming a wavelength solution with an LFC. Instead of a Stirling cooler, other instruments may use a liquid nitrogen flow to cool CCDs without any added vibration, and may incur errors related to temperature changes due to fill cycles \citep[e.g.,][]{halverson2016}. 

\subsection{CCD Cryostat Pressure Stability}
\begin{figure}
\begin{center}
\includegraphics{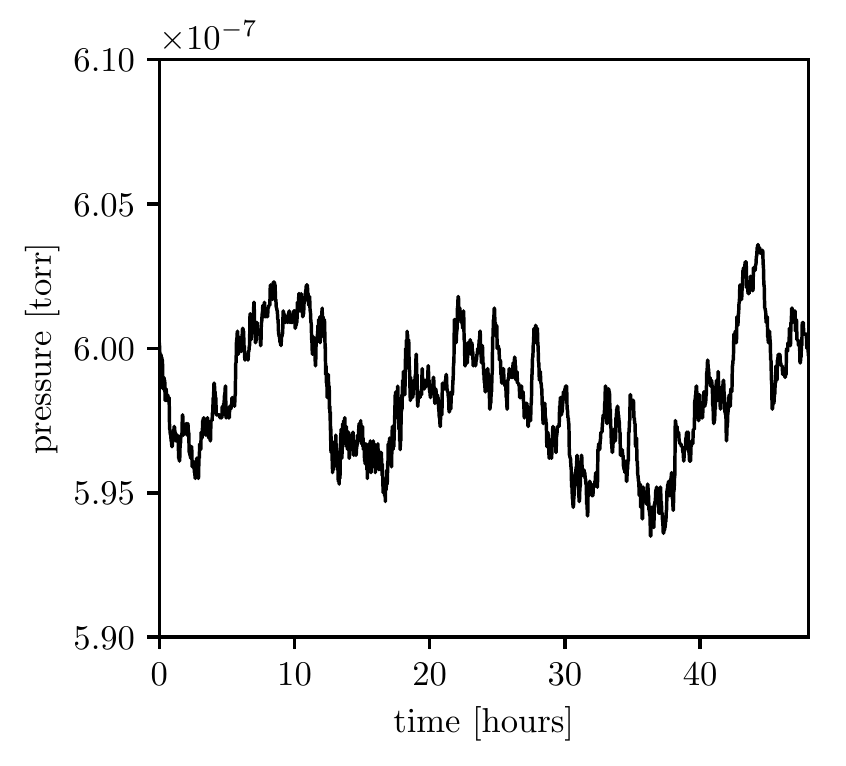} 
\end{center}
\caption{Pressure drift in the CCD cryostat over a 48 hour period. The changes in pressure occur at a level less than $1\times 10^{-8}$ torr, rendering this source of error negligible.}
\label{fig:cryo_pressures}
\end{figure}

The EXPRES CCD is housed in its own cryostat within the vacuum chamber. This cryostat is pumped around once per year, and an activated charcoal getter maintains low pressure in the meantime. Stable pressure in the CCD cryostat is desired to minimize temperature fluctuations. Figure \ref{fig:cryo_pressures} shows the cryostat pressure stability of better than $1\times 10^{-8}$ torr over 48 hours. However, pressure changes in the cryostat that impact the refractive index of the medium will not induce radial velocity errors, because the spectral dispersion has already occurred in the instrument optics. A pressure change in the cryostat will not change the pixel that a photon of a given wavelength hits, even if the wavelength of that photon changes between the instrument chamber and the cryostat.

\subsection{Pixel Position Non-uniformities}
\label{sec:ppnu}
Imperfections in the photolithography process used in fabricating commercial CCDs lead to random positional non-uniformity in pixel response centroid locations, which cannot be sampled by the LFC when obtaining a wavelength solution, unless the LFC is actively tuned to scan across all pixels by changing the mode spacing and offset frequency \citep[e.g.,][]{wilken2010}. With a comb line every 12-18 pixels across the detector, we are left to interpolate the wavelengths of pixels in between comb lines. In this process, we are implicitly assuming that the pixel positions form a uniform grid. However, deviations in pixel position will lead to the incorrect wavelength being assigned to a given pixel, decreasing the fidelity of observed spectral features in these regions. The amplitude of this pixel-positioning non-uniformity (PPNU) in commercial CCDs is typically around 0.02 pixels.

To address this ``stochastic" PPNU, the EXPRES CCD was characterized with the laser fringe interferometric method detailed in \cite{shaklan_high-precision_1995} and \cite{shao_sub-pixel_2013}, which returns measurements of pixel locations with sub-pixel accuracy. However, the results of this experiment were not robust enough to create a complete pixel position map, owing to low S/N, loss of interferometric stability, and regions with dust on the CCD dewar window; due to equipment failure, we were unable to repeat the procedure successfully before the CCD was needed for EXPRES commissioning. Nonetheless, we were able to characterize the statistical properties of this non-uniformity over limited regions of the CCD (which were fortuitously free of dust).

\begin{figure}
\begin{center}
\includegraphics[trim=0 .2cm 0 0, clip, scale=.9]{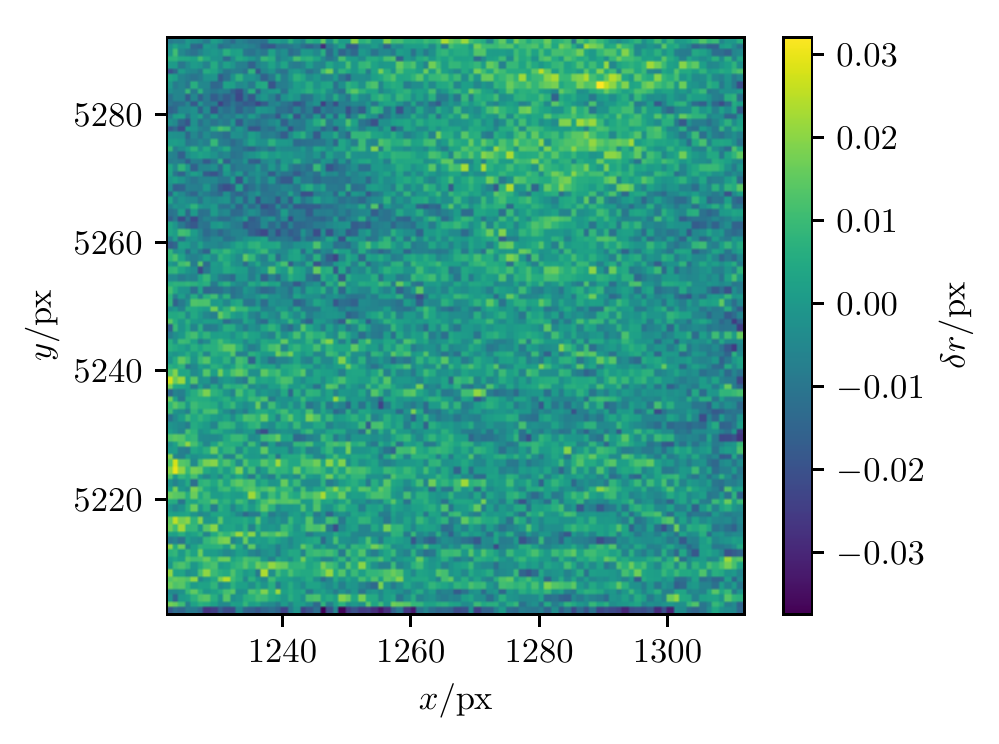} 
\end{center}
\caption{Absolute displacement of pixel response centroids ($\delta r$, in units of pixels) from uniform-pitch pixel grid, over a small (90 px $\times$ 90 px) region of interest of the CCD. Owing to extensive dust and equipment failure preventing interferometric stability, the majority of the CCD was not amenable to the recovery of this information.}
\label{fig:ppnu}
\end{figure}

Based on these results, we estimate that the pointwise standard deviation of the stochastic PPNU is indeed bounded from above by 0.02 pixels, although we are unable to obtain a precise measurement characterizing the entire detector. This upper bound was also returned independently from analysing the distribution of residuals from LFC lines compared to a smooth wavelength solution, over many realisations of photon noise in our calibration exposures. However, both of these analyses also indicated that the positioning errors were correlated between pixels, rather than being independent and identically distributed for each pixel --- this is easily demonstrated by the presence of large-scale features (spanning several pixels) in Figure \ref{fig:ppnu}.

\begin{figure}
\begin{center}
\includegraphics[scale=.9]{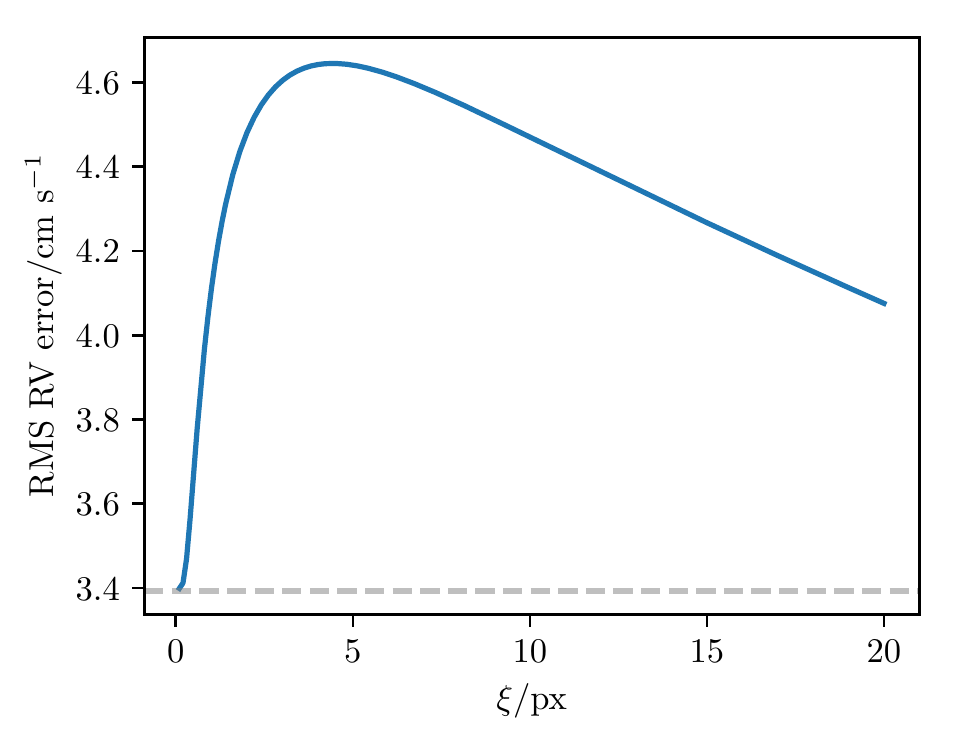} 
\end{center}
\caption{Predicted RMS velocity errors induced by stochastic pixel-positioning non-uniformity as a function of the effective correlation length, using parameters determined from our interferometric and LFC measurements.}
\label{fig:correlations}
\end{figure}

With this in mind, we performed a Monte-Carlo analysis of the distribution of cross-correlation radial velocity errors induced by many realizations of stochastic PPNU for a simulated solar spectrum and wavelength solution, with a simulated extraction aperture 20 pixels high, in the absence of photon noise. This analysis was repeated while varying the effective correlation length of the stochastic PPNU, assuming a red-noise (i.e., isotropic exponentially correlated) model;  we show the results in Figure \ref{fig:correlations}. We estimate the typical single measurement error caused by pixel non-uniformities to be of order 5 cm s$^{-1}$ (RMS). The exact error incurred depends on the pixels being used at the time for determining radial velocities. Due to motion of the Earth about the barycenter of the solar system, spectral features move on the CCD throughout the year at a level of tens of pixels, constituting independent realizations of this pixel-positioning error. However, observations taken at the same time of year (and likely same time of night, owing to changing air mass considerations) will sample essentially the same realization of this measurement error. The effect of stochastic PPNU on the resulting radial velocity error is therefore strongly dependent on the observing strategy and temporal baseline, and will be temporally correlated with a periodicity of one year.

A second type of pixel non-uniformity comes from stitching errors in the photolithography process. For some large format CCDs, a stepper motor is used to position blocks of pixels across the silicon substrate. At the boundaries of these blocks, large positional errors are incurred due to the limited accuracy of the stepper motor. These stitching errors have been observed in the HARPS spectrograph \citep{dumusque2015}, which lead to a spurious radial velocity signal with a one-year period, caused by the stellar spectra moving across the stitching boundaries due to the motion of the Earth about the barycenter of the solar system, in much the same manner as described above. The STA1600 CCD is constructed from a single photolithography mask and thus no stepping errors are incurred.

\subsection{Charge Transfer Inefficiency}
The readout process of the CCD will inherently reduce radial velocity precision, as the transfer of photo-electrons is not perfectly efficient. This charge transfer inefficiency (CTI) may result in charge being read out from a pixel other than the one it originated in, producing asymmetries in line profiles that can manifest as spurious Doppler shifts. This effect has been studied in the context of the Space Telescope Imaging Spectrograph on the \textit{Hubble Space Telescope} \citep{goudfrooij2006}, as well as high-resolution spectrographs SOPHIE and HARPS \citep{bouchy2009,milakovic2020}, and simulated for the case of the NEID spectrograph \citep{blake2017}. These studies have shown that CTI is a function of S/N, that it can change over time, and that it becomes constant in the case of high S/N. For convenience, CTI is defined as CTI = 1 - CTE, where CTE is charge transfer efficiency. Considering the PPNU mapping completed in the previous section, CTI may be the largest uncertainty among detector-related radial velocity errors.

EXPRES employs an STA 10.6k $\times$ 10.6k pixel CCD with 16 individual readout amplifiers. In practice, four amplifiers on the outer edges of the CCD are not used, as no part of the echelle spectral format makes contact with them. Therefore, the effective size of the CCD is 8k $ \times$ 8k pixels, and each amplifier contains a rectangular grid of 5280 $\times$ 1320 pixels. The dispersion of the spectra are in the direction of serial readout, which will minimize the impact of CTI, as fewer pixels are present in that dimension per amplifier. The readout speed is set to 500 kHz. CTI decreases with faster readout speed, with the penalty of higher read noise. CTI measurements of the EXPRES CCD were performed in the laboratory by the vendor for two readout speeds, 100 kHz and 1000 kHz, but not for the 500 kHz speed that is currently being used. Each mode was tested with a signal level of 50k electrons per pixel. We assume that CTI in the 500 kHz mode lies somewhere between those of the two modes that were measured. In the laboratory tests, each amplifier exhibited a slightly different CTI. In the 100 kHz mode, the worst CTI was $4\times 10^{-7}$ and the best CTI was $2\times 10^{-7}$ in the serial direction. In the 1000 kHz mode, a lower limit of $1\times 10^{-7}$ was placed in the serial direction. In the parallel direction, a lower limit of $1\times 10^{-7}$ was placed in both readout modes. These measurements were in the regime of high S/N. The effect of CTI in the parallel readout is expected to be smaller, for two reasons. First, the CTI is better in these pixels owing to their smaller charge capacity. Pixels with larger wells are more susceptible to worse CTI, as a given signal level will take up less capacity in the pixel compared to a smaller pixel. Pixels in the serial readout are generally built this way to accommodate on-chip binning. The second reason is that charge loss in the parallel direction will primarily serve to decrease the overall signal signal level of a given order (at a very small level), rather than contribute to small line asymmetries in the dispersion direction that translate to radial velocity shifts. However, as the orders are curved in 2D across the detector, the pixels on one side of an order within an amplifier will be on different rows than the pixels on the right side of the order. This may contribute some line asymmetry, as these pixels will undergo a slightly larger or smaller number of transfers in the parallel direction. However, this effect should be much smaller than that from serial readout. We note that this is the opposite situation of that presented in \cite{bouchy2009}, as their serial readout is perpendicular to the spectral orders while ours is parallel to them.  

If CTI was constant and not dependent on time or signal level, the effect would be negligible due to the fact that all observations would be impacted equally. The realistic concern is that radial velocity observations of stars may have different S/N depending on observing conditions. It is common to attempt to match the S/N of different observations of the same star to minimize this effect, which can be accomplished by terminating the exposure when an exposure meter has reached a certain level of counts. However, there is also a desire to choose exposure times based on a star's p-mode oscillation timescale \citep{chaplin2019}, the idea being that always exposing for an integer number of such timescales will minimize the radial velocity shift due to this activity feature. Therefore, exposure times cannot be chosen to satisfy both conditions if the required times are different. In the high S/N regime, due to constant CTI, the exposure time could be chosen to first satisfy a minimum acceptable S/N and then chosen for an integral number of p-mode oscillation timescales. Observations with higher S/N than the rest will have a smaller velocity shift due to CTI than observations with lower S/N than the rest of the set. An alternative strategy would be to implement CTI corrections for observations depending on their S/N, as was done in \cite{bouchy2009} in the low-S/N regime. Implementing an algorithm to correct for CTI depending on pixel position and signal level enabled the recovery of photon-noise limited Doppler precision in the SOPHIE instrument. However, this study was limited to studying CTI at the level of a few m s$^{-1}$ due to the calibration source (a ThAr lamp) and stability of the instrument, and covered up to a S/N of about 100. The goal for EXPRES observations is generally to reach a S/N of 250 in the middle of the optical spectrum. With the next generation of Doppler spectrographs using a more ideal LFC for wavelength calibration, the effect may be apparent at a level of tens of cm s$^{-1}$, which would be a significant term in the error budgets. 

\begin{figure}
\begin{center}
\includegraphics{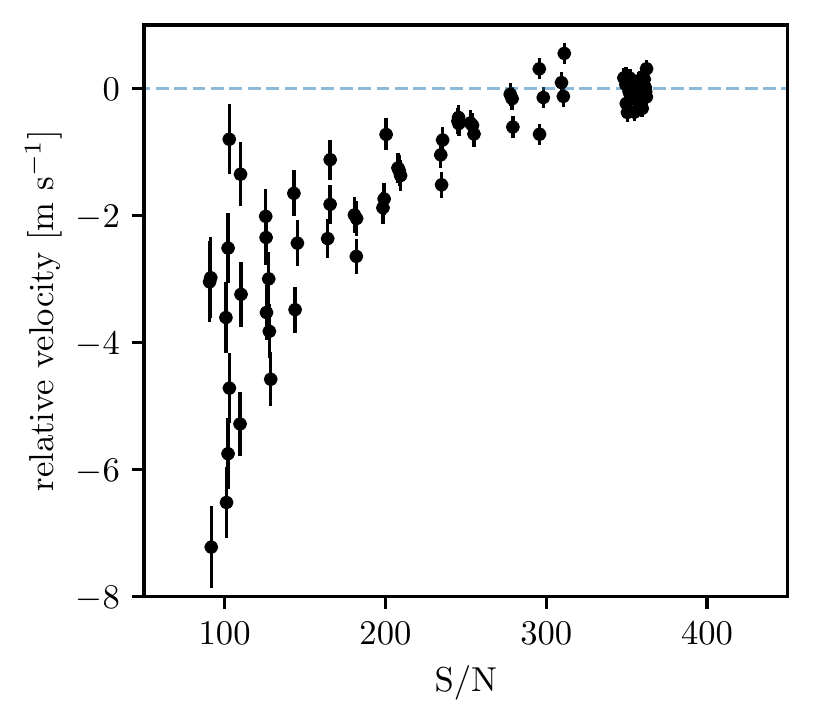} 
\end{center}
\caption{Velocity shifts in the LFC relative to high S/N as a result of the CTI effect. The S/N is measured as the median of the peak S/N in each spectral order.}
\label{fig:cte}
\end{figure}

To assess this possibility, we first repeat the experiment in \cite{bouchy2009} to obtain the systematic velocity drift due to changing S/N. First, a short series of LFC exposures was taken at high S/N, to assess any instrumental drift. Then, LFC exposures with decreasing exposures times were taken, to effectively decrease the S/N. The relative velocities were then computed with the CCF method relative to the first LFC exposure. The S/N at which velocity shifts become apparent is the level that stars and calibration exposures should reach without having to correct for this effect, which is the level that CTI becomes constant. The results of this test are shown in Figure \ref{fig:cte}. The reported S/N is a median of the peak S/N per pixel in each extracted spectral order occupied by LFC light. Given this, the true average signal level of pixels in this test is lower. It is difficult to compare this measure of S/N to the typical continuum S/N in stellar spectra, owing to the very different spectrum of the LFC compared to a star. Still, we are interested in knowing the possible radial velocity error between high S/N and low S/N. The incurred radial velocity error exceeds 10 cm s$^{-1}$ at a S/N of 300 in this test, and reaches several m/s at a S/N of 100. This error can be avoided by matching the S/N of different exposures of the same star, even if the S/N is lower than 300, as the result will be a constant offset that will not impact the ability to disentangle planetary signals. We have revised our observing strategy so that instead of setting a fixed exposure time, the exposure meter terminates an observation at the minimum of S/N or 1200 seconds, as discussed in Section \ref{sec:expm_sn_cal}. Given the brightness of our stars, even with a range of observing conditions, this ensures a constant S/N of 250 per pixel in the V-band for all of our program observations.   

One challenge with this test is that modal noise begins to increase as exposure time is decreased. However, since we expect this error source to be stochastic, the impact should only be to increase the spread in velocities obtained at a given S/N. One way to remedy this effect would be to change the neutral density filter in the LFC to higher values while maintaining the same exposure time, to effectively decrease the S/N. A second problem is that photon noise will naturally increase the scatter in velocity measurements as well, which should be random in nature. In addition, the peak S/N of an LFC exposure may not be directly comparable to the S/N in the continuum of a stellar spectrum. The S/N in stellar absorption lines is inherently lower than in the continuum, amplifying the significance of this effect.

Owing to the relative similarity between the NEID and EXPRES spectral formats, we may compare the results here to those of the simulated results for NEID in \cite{blake2017}. In that work, it was shown that changes in CTI of $10^{-6}$ would lead to radial velocity offsets of several m s$^{-1}$. The effect of parallel CTI was not included, which we expect to be very small with EXPRES. This may be what we are incurring through the LFC exposure test with EXPRES with a change in peak S/N from 400 to 100. 

\subsection{Electronics Noise}
For control and read out of the CCD, EXPRES uses an Archon controller, which was supplied and integrated by STA. In addition to the characteristics of the on-chip amplifiers, the read noise for a CCD system is heavily influenced by the details of the pattern of clocks used to shift the charge packets and operate the gates of the amplifiers. The smallest read noise will be obtained with long integration times at the output nodes, but at a cost of requiring a longer time to read out each image. Originally, EXPRES read out at a 1 MHz pixel rate, but we eventually decided that the optimization of the noise/read out time tradeoff would be better at a 500 kHz rate. After dropping to this rate, STA adjusted the clocking patterns further to reduce systematic offsets in the amplifier pedestal levels. The read noise of the EXPRES detector has been measured to be between 5 and 10 e$^{-}$ (depending on the amplifier region) and this noise is included in the S/N measurements presented in this paper.

\subsection{CCD Fringing}
\begin{figure*}
\begin{center}
\includegraphics{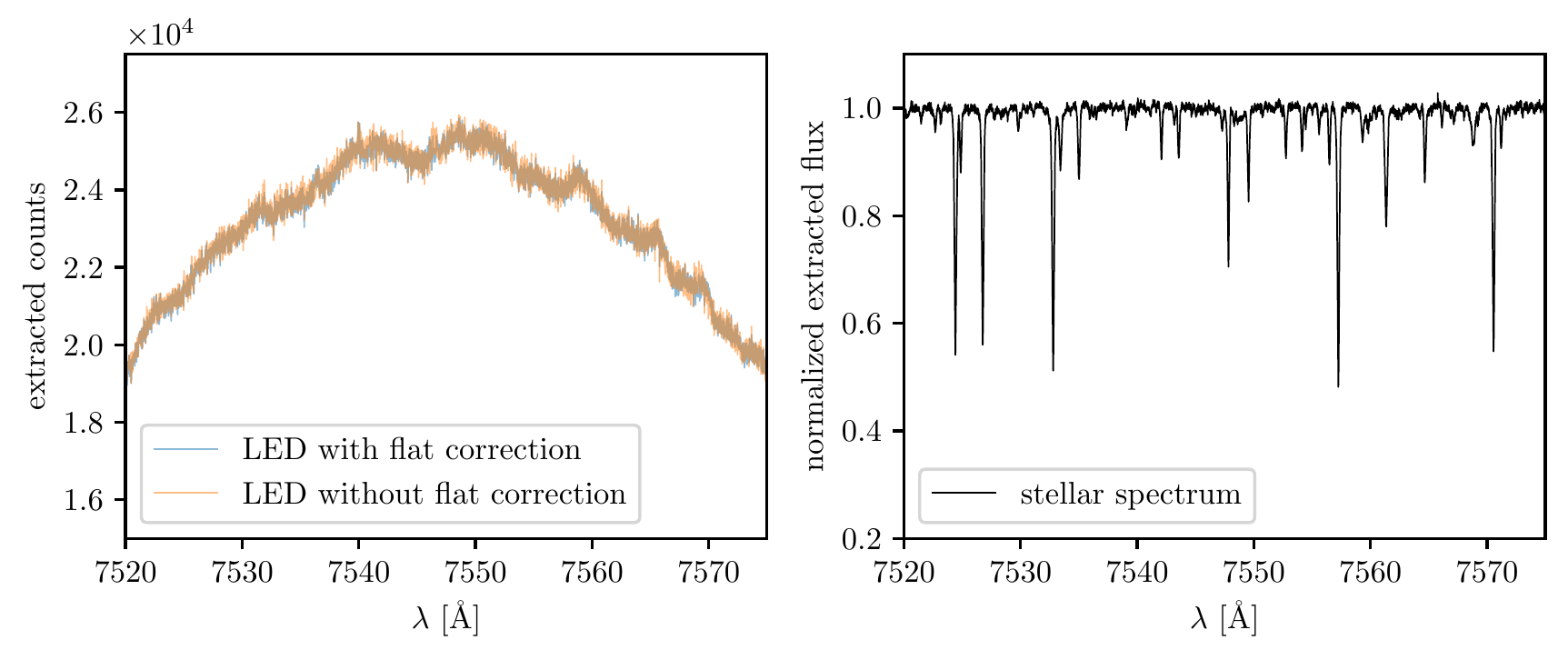} 
\end{center}
\caption{\textit{Left:} Spectrum of the flat-field LED in red wavelengths, with fringing manifested as a wave-like pattern on the CCD. \textit{Right:} A flat-relative optimally extracted stellar spectrum over the same wavelengths shown in the left panel, with no noticeable fringing present.}
\label{fig:red_flat}
\end{figure*}

The EXPRES CCD is 30 $\mu$m thick, and exhibits some fringing in red wavelengths. Fringing occurs when long wavelength photons penetrate the top silicon layer of the CCD, reflecting off of different surfaces with the CCD, and interfere with incoming photons. Constructive and destructive interference is manifested as a wave-like pattern in the spectrum, tracing the small differences in thickness of the CCD. An example of this is shown in Figure \ref{fig:red_flat}. In the left panel, the spectrum of the EXPRES LED source is shown both with and without 2D flat-fielding from the extended fiber. Fringing is present, causing up to a 4\% deviation in the measured counts in the continuum. 2D flat-fielding does not remove the fringing effect initially, as it is performed column-wise and then smoothed over local chunks of $\sim$100 pixels. In the right panel of Figure \ref{fig:red_flat}, an extracted stellar spectrum is shown over the same wavelength range. The EXPRES flat-relative optimal extraction scheme, presented in \cite{petersburg2019}, effectively eliminates the fringing pattern in the extracted science data, as the extraction is performed relative to a science fiber flat spectral profile. The fringing effect is therefore calibratable, and given that it is only significant outside of the wavelengths used for radial velocity analysis (approximately 4850-7150 \AA), we expect a negligible contribution from it to radial velocity measurement error.

\section{Stray Light and Cosmic Ray Removal}
\label{sec:stray_light}
Stray light incident on the detector from unwanted reflections within the instrument may lead to degraded line profiles and radial velocity errors. Stray light is minimized by extensive baffling in the optical bench structure and camera barrel. Light that is reflected off of the center gap in the grating mosaic is absorbed by a dark screen. Still, residual scattered light may be manifested as an unfocused glow across the detector, and is not spatially correlated with the science data, though its intensity does scale with the overall brightness of the spectrum.

The extraction pipeline for EXPRES has a custom implementation of scattered light subtraction \citep{petersburg2019}. Smooth functions are fit to the scattered light that resides between each order. These functions are then interpolated through the orders, approximating the scattered light in the extracted pixels and generating a full two-dimensional scattered light model on the detector. This model is subtracted from the data before extraction.

The optimal extraction algorithm of the EXPRES pipeline also implements robust cosmic ray identification and removal. Cosmic rays present as significant outliers during the optimal extraction process and can be easily masked accordingly \citep{horne1986,zechmeister2013}, as long as the proper noise model is used. One at a time, data within the extraction window with the largest residual above a certain threshold (typically 8$\sigma$) is masked before repeating the optimal extraction. This operation also enables the removal of dark pixels and other highly localized issues of the CCD. This cosmic ray removal is included in the formal uncertainty of the extraction.

\section{Sky Contamination and Telluric Contamination}
\label{sec:sky}
Earth's atmosphere produces emission at a very faint level, in conjunction with scattered light from the Moon, stars, and nearby human activity. The surface brightness of the sky at the LDT site has been measured to be 22.0 mag/arcsec$^2$ in the V band at zenith, and 21.3 mag/arcsec$^2$ at an elevation of 30$^\circ$ \citep{massey_sky}. These measurements are preliminary and will be refined over the next observing cycle.  

\begin{figure}
\begin{center}
\includegraphics{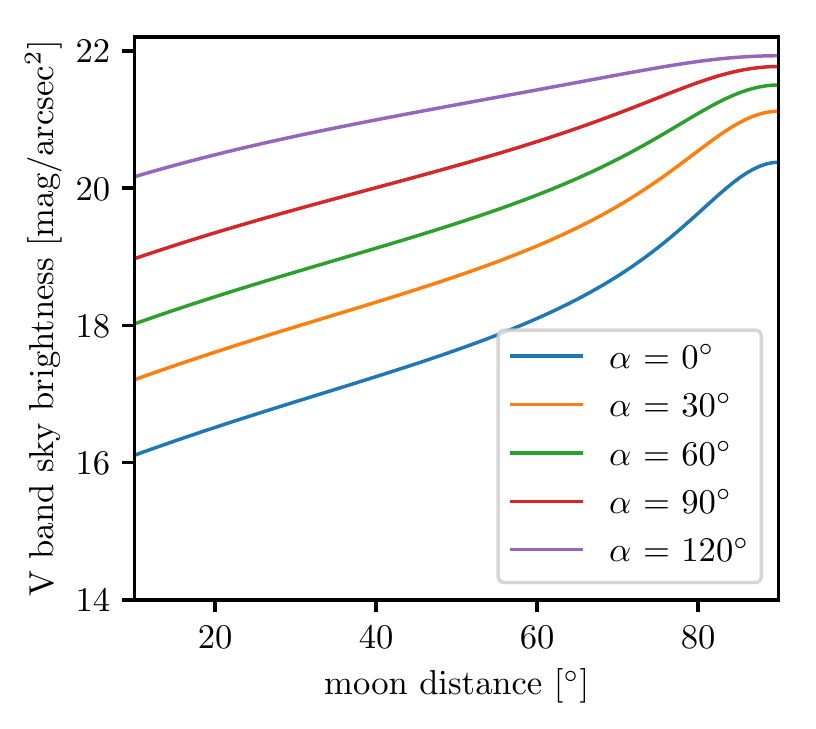} 
\end{center}
\caption{Sky brightness at the LDT as a function of distance and phase of the Moon, assuming a target at the zenith with a nominal sky background magnitude of $V=20$.}
\label{fig:sky}
\end{figure}

Scattered moonlight is the largest contributor of contamination light into the science fiber, as we expect the strength of this effect to be several orders of magnitude larger than that of zodiacal light, scattered starlight, airglow, and sky emission, with the exception of a few bright lines in the optical. Scattered moonlight is essentially a reddened solar spectrum. The concern is that if this imposed solar spectrum lines up closely to that of a star with a spectral type similar to the Sun, the contamination will induce absorption line asymmetries that bias the measured radial velocity. The surface brightness of this emission is dependent on lunar phase and distance from the moon to the target. Generalized expressions for calculating this surface brightness were presented in \cite{krisciunas1991}, however, reflections from clouds can further increase the sky brightness. In Figure \ref{fig:sky}, we show the predicted brightness of the sky including contamination from the Moon, at different phases $\alpha$ and distances in degrees. For simplicity, we assume the target is at zenith, and that the Moon moves closer to the horizon at greater distances. In reality, observing at higher air masses will naturally increase the brightness of the sky in addition to Moon contamination, but this effect is small.

To assess whether or not this contamination will produce significant error, we consider that the faintest stars on which we expect our most precise radial velocity measurements are magnitude $V=7$. For stars up to a magnitude of $V=7$, we are able to reach a S/N of 250 in less than 20 minutes, under reasonable atmospheric conditions. Reaching this S/N reduces the error from photon noise to 30 cm s$^{-1}$. After this point, increasing the S/N results in smaller reductions of the radial velocity error from photon noise. More details regarding this effect are provided in \cite{petersburg2019}. Additionally, it may be important to limit the smearing of stellar spectra across the detector due to barycentric motion, as well as minimize the residual barycentric correction errors discussed in \cite{blackman2019}. 
Generally, the brightest the sky will be at a reasonable distance from the Moon will be 12 magnitudes fainter than our target stars. Based on the results presented in \cite{roy2020}, we then expect a worst case error of 10 cm s$^{-1}$ from moonlight contamination. The average error will be less, and precaution is taken to point at large separations from the Moon.  

For EXPRES, a more significant effect of Earth's atmosphere comes from telluric contamination. Telluric lines are atomic and molecular absorption lines present in the spectrum from Earth's atmosphere, and do not depend on the brightness of the star, warranting a careful treatment. The source particles of the lines can be divided into two categories, water and non-water lines. The depth of water lines depends on the amount of precipitable water vapor along the line of sight, while the depth of non-water lines depends primarily on air mass. Large telluric absorption lines are saturated, and are avoided when computing the radial velocity by masks in the CCF. However, smaller telluric lines known as microtellurics populate a large portion of the spectrum. 

We have implemented the methods of \cite{leet2019} into the EXPRES reduction pipeline, to empirically obtain a telluric line model for every observation over the relevant wavelengths. Because most of the telluruic lines in our program observations are from water and the water column density can vary by more than a factor of two, we check every observation to exclude lines from the CCF masks that are deeper than 10\% of the continuum. We are now working to characterize the impact that this has on the CCF radial velocities, but we have seen improvement between a few cm s$^{-1}$ up to 15 cm s$^{-1}$ in the RMS of the radial velocity residuals to known exoplanet hosting stars. \cite{cunha2014} estimate the radial velocity error contribution of telluric contamination reaches as high as nearly 1 m s$^{-1}$, depending on the spectral type of the star, air mass, and systemic radial velocity of the star system. However, for G-dwarf and K-dwarf type stars in that study, the impact was more commonly 10-30 cm s$^{-1}$.

\section{Chromatic Exposure Meter}
\label{sec:expm}
\subsection{Chromatic Barycentric Correction}
The barycentric correction is an essential component of precision Doppler spectroscopy. This is the step to shift stellar radial velocity observations into the frame of the barycenter of the solar system, which can assumed to be at rest with respect to most stellar systems over the timescales of typical surveys. Any errors in the correction velocity propagate directly to the measured stellar radial velocity. Many effects related to Earth's motion must be taken into account for an accurate barycentric correction, e.g., rotation, precession, nutation, gravitational redshift of photons, the Shapiro delay, and light travel time delay \citep{we2014}. In addition, the 3D location of the observatory must be known to high-accuracy. In principle, all of these things may be known to an accuracy sufficient to render the barycentric correction error to less than 1 cm s$^{-1}$.

For all EXPRES observations, we use the barycentric correction package \textbf{barycorrpy} \citep{kanodia2018}, which has been tested for accuracy against previously standards \textbf{barycorr} \citep{we2014} and \textbf{TEMPO2} \citep{hobbs2006}. Given accurate inputs, this algorithm produces corrections accurate to the mm s$^{-1}$ level. 

Despite an accurate algorithm for barycentric corrections, several effects may contribute to radial velocity errors being incurred at this step. The absolute time of the exposure must be known to 0.25 seconds to constrain the accuracy to less than 1 cm s$^{-1}$ \citep{we2014}. The EXPRES shutters are controlled by a computer that is constantly syncing its clock to atomic sources. Errors on the order of tens of ms are expected in this process. In additional, stellar flux throughout each exposure is measured with an exposure meter to ensure that accurate weights are applied to the barycentric correction \citep{tronsgaard2019}. The integration intervals of the exposure meter need to be short to account for atmospheric changes on short timescales. This leads to a limited S/N being achieved in the exposure meter data, which propagates to errors in the weights being applied to the barycentric corrections. For the bright stars observed with EXPRES, high S/N can be achieved in one-second exposures of the meter. For fainter stars, the exposure meter integration length needs to be increased to achieve good S/N in the weights. This is a trade off between S/N and time resolution. However, \cite{blackman2019} noted that integration lengths up to 30 seconds are typically safe to use in keeping the barycentric correction error constrained to less than 1 cm s$^{-1}$. 

The location of the observatory has been measured with two different GPS units, multiple times, which differed in absolute 3D position in the WGS 84 frame by 54 meters in the worst case. Generally speaking, the observatory location needs to be accurate to 100 meters in order to constrain the barycentric correction error to less than 0.5 cm s$^{-1}$. Assuming that the observatory location is correct to a few tens of meters, then this source of error should contribute less than a 0.2 cm s$^{-1}$ to the error budget. Likewise, stellar coordinates must be known accurately. All stars in our observations have had coordinates, proper motions, and parallaxes measured by \textit{Gaia} \citep{gaia2016}, which far exceeds the accuracy requirements for barycentric corrections good to 1 cm s$^{-1}$. 

EXPRES is one of the first radial velocity instruments to implement a chromatic exposure meter \citep{blackman2019}, enabling wavelength-dependent barycentric corrections, which are needed at the highest precision levels to account for chromatic attenuation of starlight in Earth's atmosphere \citep{blackman2017}. The exposure meter is composed of a low-resolution spectrograph, and the detector is the same model EMCCD used in the FTT system. With one year of chromatic exposure meter data from EXPRES, \cite{blackman2019} demonstrated that unless accounted for, chromatic variability in Earth's atmosphere will regularly lead to radial velocity errors up to 10 cm s$^{-1}$, and larger errors occasionally occur. This error depends on the nightly observing conditions and the method used for solving for the radial velocity, as well as potential instrumental effects. A chromatic dependence in throughput could also be caused by a combination of poor guiding and lack of atmospheric dispersion compensation. A photon-weighted, chromatic barycentric correction is applied to every EXPRES radial velocity measurement as a standard procedure \citep{petersburg2019}. Radial velocity errors incurred at this step stem from S/N as discussed above, and other minor effects such as the accuracy of the wavelength solution of the exposure meter spectrograph. This can be regularly re-calibrated, and a wavelength accurate to a few nanometers per pixel is sufficient to render this error source negligible. 

\subsection{S/N Calibration}
\label{sec:expm_sn_cal}
\begin{figure}
\begin{center}
\includegraphics{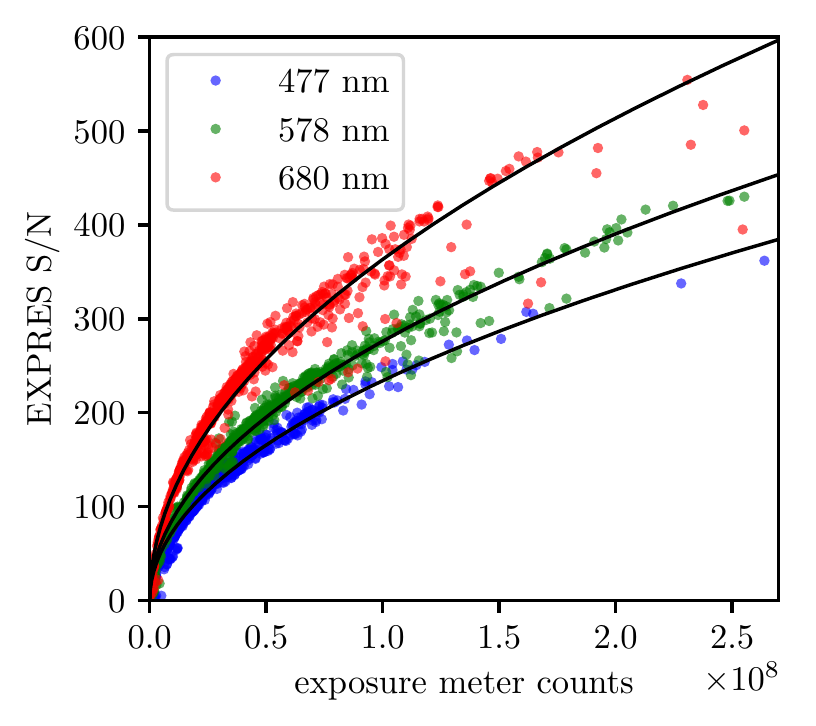} 
\end{center}
\caption{S/N per pixel in three EXPRES orders plotted against summed counts in the same wavelength bins on the exposure meter. The strong correlation allows observers to accurately assess the S/N as a function of wavelength on the EXPRES detector without reducing the data.}
\label{fig:expm_snr_cal}
\end{figure}

One advantage of the exposure meter spectrograph is that it can also be used to chromatically estimate the S/N per pixel on the EXPRES detector in real time. This enables observers to accurately set exposure times to reach the desired S/N under different atmospheric conditions, and is accomplished without any reduction of the EXPRES data. In Figure \ref{fig:expm_snr_cal}, we show the peak S/N per pixel of three EXPRES orders in blue, green, and red wavelengths, plotted against binned exposure meter counts in the same wavelengths. Each set of data is fit with a function of the form 
\begin{equation}
\mathrm{S/N} = C \sqrt{\mathrm{counts}},
\end{equation}
where counts are the summed pixel counts from a region on the exposure meter centered on the given wavelength, and $C$ is a fitted constant.

\section{Stability Tests}
\label{sec:lab_tests}

\subsection{Stability in the Nightly Wavelength Solutions}
\begin{figure*}
\begin{center}
\includegraphics{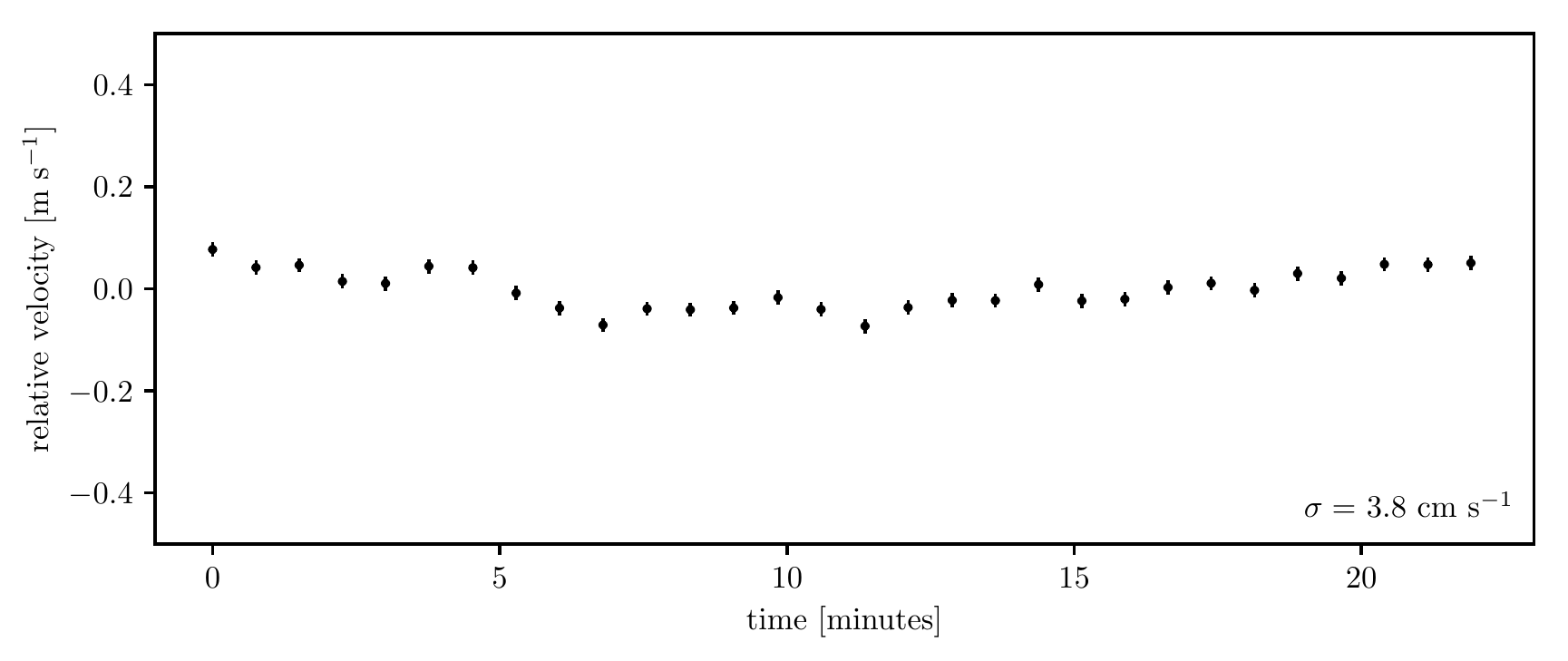} 
\end{center}
\caption{Linear-detrended relative velocities with error bars of a set of LFC exposures. The standard deviation of 3.8 cm s$^{-1}$ is the typical instrumental calibration precision between calibration and science exposures when the system is performing optimally.}
\label{fig:lfc_stability}
\end{figure*}

The first test of spectrograph stability comes from the amount of drift in the wavelength solution over time. To test this, we examine the apparent velocity shifts of LFC exposures that are used to generate wavelength solutions throughout a given night, as was done for the fiber agitator analysis in Section \ref{sec:illumination}. Typically, a new wavelength solution is obtained every $\sim$15 minutes when observing, to account for drifts due to vibrations or thermal variations in the instrument, where the wavelength solution for a given exposure is interpolated from the nearest wavelength solutions in time. Because the LFC is a stable source, any observed drifts are due to instrument instability. In Figure \ref{fig:lfc_stability}, we show the relative velocities of LFC exposures over a 25-minute period. The data have been detrended by a linear polynomial, which is representative of the typical instrument drift. Following this detrending, the remaining scatter represents the typical error between calibration and science exposures. The standard deviation of the data is 3.8 cm s$^{-1}$. This is representative of the instrument contribution to radial velocity errors during observations, excluding on-sky effects such as guiding, barycentric correction errors, calibration injection repeatability, and velocity shifts due to signal-dependent CTI effects. It does represent the errors incurred from environmental instability, other detector effects, photon noise (in the calibration images), and modal noise. Additional error incurred during stellar observations will stem from photon noise limited by the brightness of the star as well as stellar activity and telluric contamination from Earth's atmosphere. With many similar sets of LFC data taken throughout instrument commissioning, the standard deviation tends to vary between 3 cm s$^{-1}$ and 9 cm s$^{-1}$. These data sets were taken over similar lengths of time with the PCF in good condition.

From the slow and linear drift of consecutive LFC exposures, we may draw several conclusions. The instrumental drift is slow enough and linear enough such that simultaneous calibration is not required. When calibrating wavelengths on different pixels than the science data, there is the added possibility of systematic error from pixel effects in the CCD, such as those discussed in Section \ref{sec:ppnu}. This additional source of error is eliminated by calibrating with the science fiber. When simultaneous calibration is performed, it may be preferable to have the calibration source on for the duration of the exposure, such as what naturally comes when using an iodine cell Doppler spectrograph. This helps average out any instrument drift incurred during the exposure, enabling accurate calibration. However, it is not practical to have the LFC on for the duration of all science exposures, as the PCF would quickly degrade and require replacing every few months. Finally, simultaneous calibration on pixels between orders would complicate the extraction of science data. With science orders and simultaneous orders close together on the detector, the optimal extraction model would need to be more complex, there may be additional scattered light between the orders, and there is the potential for cross-talk between the two fibers, as they are separated from only a thin layer of cladding.

\subsection{Chamber Vacuum Pumps}
\begin{figure*}
\begin{center}
\includegraphics{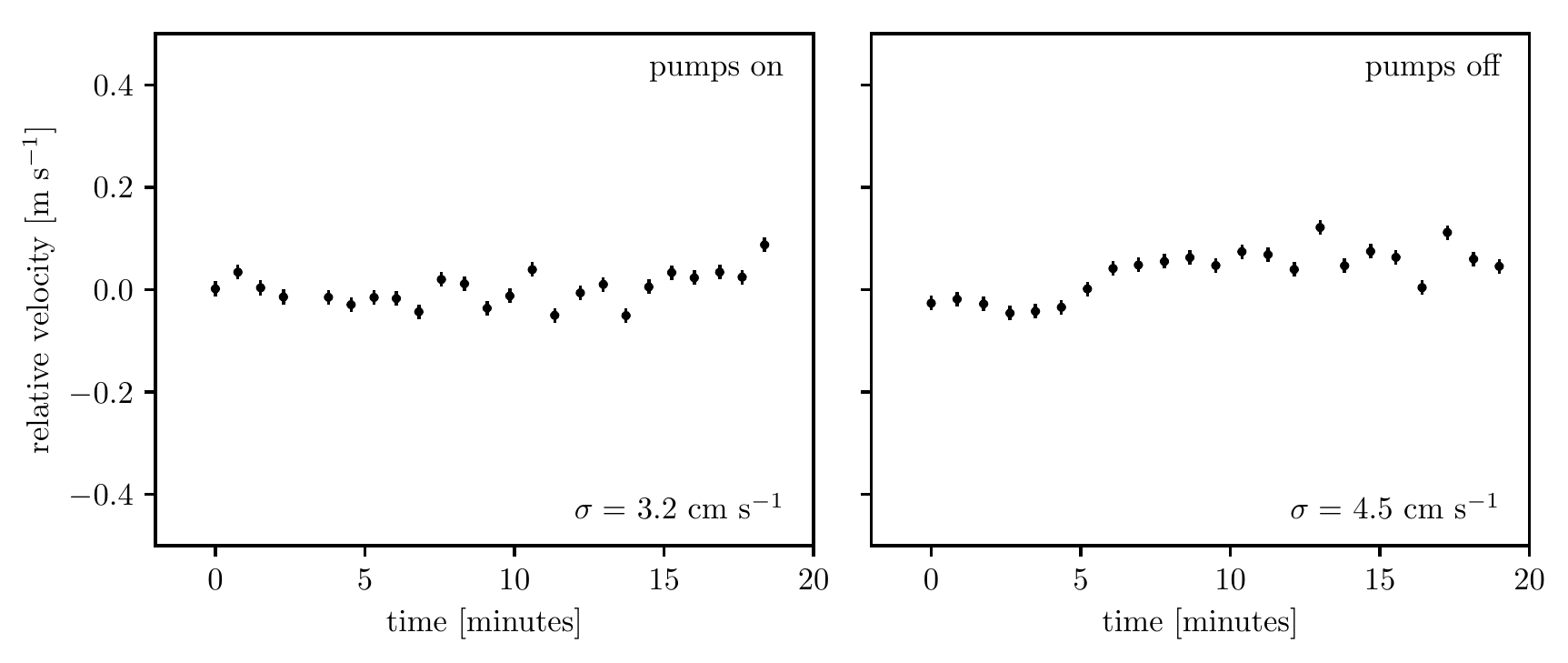} 
\end{center}
\caption{Comparison of instrument stability from cross-correlating LFC exposures with the vacuum pumps on (left) and with the vacuum pumps off (right). There is no significant difference in instrument stability with the vacuum pumps running during exposures.}
\label{fig:pumps}
\end{figure*}

In order to maintain extremely stable pressure in the spectrograph chamber, the vacuum pumps may be run continuously throughout observing. To test if vibrations from the pumps contribute instability in the wavelength solution, we took a set of LFC exposures  with and without the chamber vacuum pumps on. The results of this test are shown in Figure \ref{fig:pumps}. In the left panel, the pumps were on, and the LFC exposures were cross-correlated to assess drift in the wavelength solution, as in the previous subsection. In the right panel, the pumps were left off, and the chamber experienced a rise in pressure similar to that shown in Figure \ref{fig:ror}. The similarly small velocity shifts between the two data sets indicate that running the pumps has a negligible effect on the instrument calibration precision. It is likely that the relatively high frequency and low amplitude of the pump vibrations average out any negative effects. As in previous figures, the data have been detrended with a linear function to remove the slow instrumental drift.

\subsection{Echellogram Stability}
\begin{figure*}
\begin{center}
\includegraphics{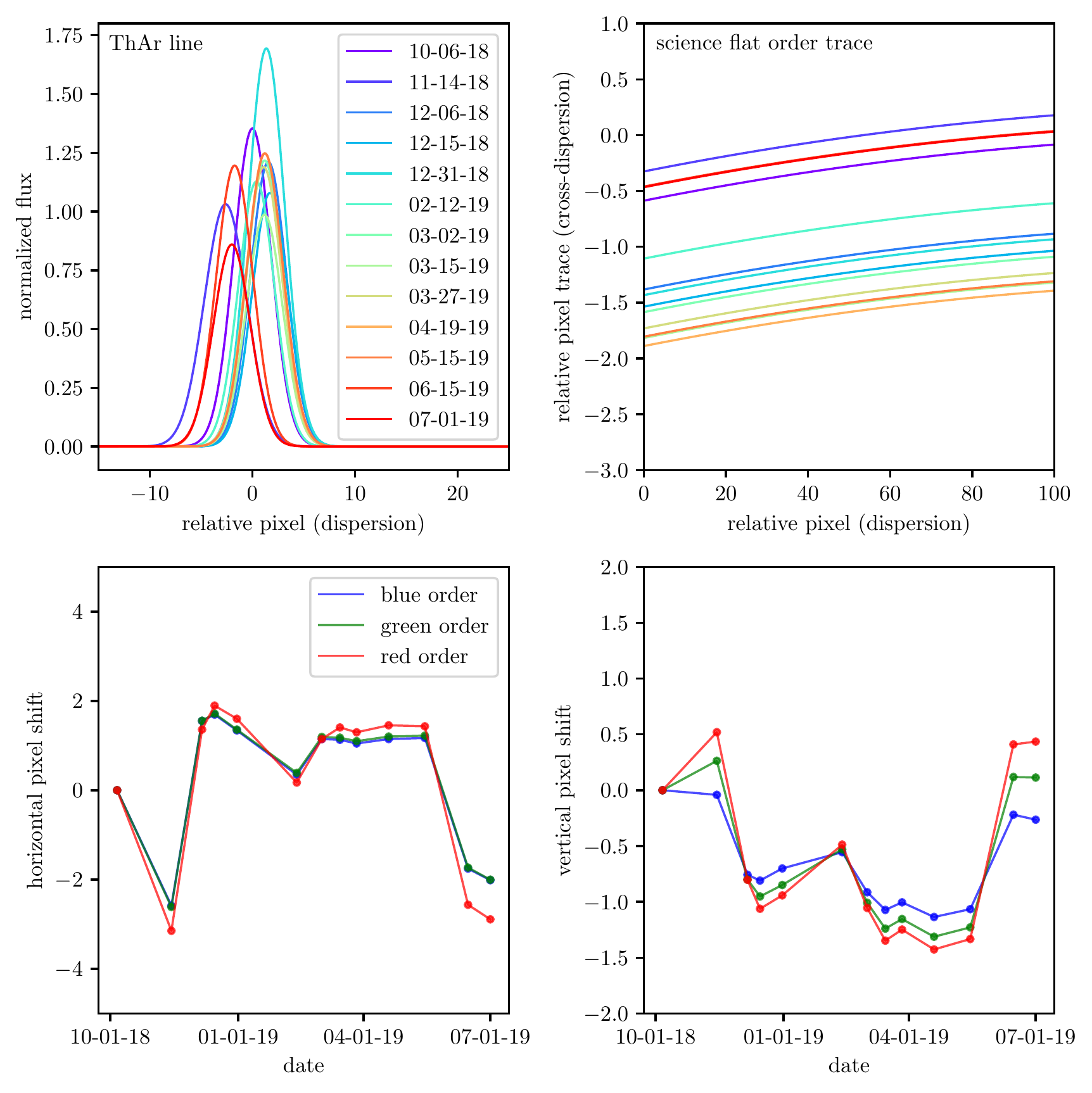} 
\end{center}
\caption{Drift of the echellogram on the CCD in both the dispersion and cross-dispersion dimensions over a 9 month period. \textit{Top left}: Fits of a select ThAr line from a green order, colored by observed date. \textit{Bottom left}: The corresponding drift of the three ThAr line locations. \textit{Top right}: Small region of an order trace from the same order that the previous ThAr line is located in, taken from the same dates and colored accordingly. \textit{Bottom right}: Corresponding vertical drifts of three order traces over time.}
\label{fig:echellogram_drift}
\end{figure*}

Over time, mechanical drifts or changes in focus in the instrument will manifest as changes in position of the echellogram on the detector. On short timescales, this will also manifest as changes in the wavelength solution as discussed previously. Over long timescales, we can assess the stability of the echellogram in terms of pixel displacement based on drifts of the calibration spectra. We have examined the positions of ThAr emission lines to assess drift in the horizontal (dispersion) direction. For the vertical (cross-dispersion) direction, we have tracked the changes in fits of the order traces from flat-fields taken through the science fiber. The drifts have been tracked in three different orders over the echellogram. In Figure \ref{fig:echellogram_drift}, we show these drifts on the left and right sides, respectively. In the top left panel, we show fits to one ThAr line colored by date that the exposure was taken. Variations in the mean line position in the dispersion dimension are noted, and this was performed for three different ThAr lines in blue, green, and red orders. The mean positions of these three lines are then plotted in the bottom left panel over time. The good agreement in line position variations indicates that the whole echellogram is moving across the CCD over time. In the top right panel, order traces of the science flat are plotted for the green order, for the same dates as the ThAr lines and colored accordingly. The variations in the mean vertical position on the CCD are recorded for the same three orders as in the ThAr analysis. These positions are plotted over time in the bottom right panel. Again, good agreement in the position variations is noted, but an additional spread in the data indicates that the vertical motion is somewhat wavelength-dependent. The vertical motion of the echellogram matches the inverse of the horizontal motion, showing that the whole echellogram has moved diagonally, back-and-forth, across the detector over time. We have identified the source of larger shifts in the echellogram to be refocusing of the EXPRES camera. During the more stable period from early March 2019 to the middle of June 2019, the focus of the instrument was not changed. If maximum stability is to be achieved, focusing of the instrument should be avoided if possible. However, these motions are still much smaller than the velocity shifts due to the barycentric motion of the Earth, which is on the order of tens of pixels and cannot be avoided.

\subsection{Long-term Stability Improvements}
One way to assess how instrument stability has improved over time is to examine the nightly drift in velocity of the wavelength solution. Initially, this drift was large, as high as 60 m s$^{-1}$ per night, during the commissioning period in 2018. By the spring of 2019, this drift had decreased to an average of 2.5 m s$^{-1}$ per night, which is a much more stable configuration. This drift has been found to be weakly correlated with the spectrograph temperature, which has exhibited a slow, seasonal drift correlated with the ground temperature, as discussed in Section \ref{sec:environment}. As the temperature warmed in the summer of 2019, the drift became larger with a larger temperature gradient per day. In any case, on timescales of calibrations and observations, the instrument drift appears to be highly calibratable. 

\section{Discussion}
\label{sec:discussion}
\subsection{Recommendations for Further Instrumentation Development}
In the complete error budget of radial velocity instruments such as EXPRES, stellar activity is the single largest term. Without improvement on that front, more improvement in instrumentation will return fewer gains. EXPRES is an instrument built to produce data sets that will enable further development to statistically mitigate the effects of stellar activity. It is also important to note that not every source of radial velocity error for every instrument has been discovered. Therefore, the list of effects described here is probably not complete. Beside this point, we are able to recommend that the community pursue further development in two key areas of instrumentation, which are the wavelength calibration source and sub-pixel effects in CCDs. 

\subsubsection{Calibration source} 
The LFC may be the ultimate wavelength calibration source for optical spectrographs, but there are still two primary concerns with it. The first is that the wavelength range of the LFC currently begins at about 500 nm, while the throughput of EXPRES allows for high S/N in stellar spectra down to 400 nm. This neglects a 100 nm wide window that is an important region of the spectrum, as it is both rich in absorption lines from G-type and K-type stars as well as deficient in telluric absorption lines from Earth's atmosphere. Opening up this window to radial velocity study could greatly improve the information content of a single stellar spectrum. One alternative may be to include a stablized etalon that is able to reach bluer wavelengths, which is able to produce a comb-like emission spectrum similar to the LFC. While such sources are known to drift, if the dispersion does not change, then they could be calibrated using the emission region that overlaps with that of the LFC. One concern with the etalon is that calibration to the LFC in the overlap region may not accurately sample the behavior of the etalon in bluer wavelengths, as a chromatic dependence in instrument drift could limit this potential calibration source. Combining wavelength calibration sources in this manner may be the quickest way to gain complete wavelength calibration over the optical window. The second concern is that the noise characteristics of the LFC change over time, as discussed in Section \ref{sec:cal_sources}. A degraded PCF will have a significant impact on the ability to perform wavelength calibration. It is important to monitor the health of a PCF and replace it before significant degradation impacts the quality of wavelength calibration.  

\subsubsection{PPNU and CTI}
PPNU and CTI are likely the only remaining instrument related errors that could be improved with further calibration in the data reduction stage. As discussed in Section \ref{sec:detector}, it is possible to create a pixel position map to calibrate radial velocity errors from the PPNU effect. We were unable to complete this task for the entire EXPRES CCD. However, based on our evaluation of a sub-region on the detector, it should be possible to realize an improvement of several cm s$^{-1}$ in this error term if such a pixel map could be produced for the entire detector. Exploiting the pixel map to perform a generalized optimal extraction \citep[e.g., ``spectroperfectionism"][]{bolton2010} also increases the computational cost of spectral extractions by an order of magnitude. Regarding CTI, this effect can be mitigated to a negligible level with careful matching of signal between different exposures of the same star. However, radial velocity errors can be significant ($> 10$ cm s$^{-1}$) for only relatively small differences in S/N. A correction algorithm for EXPRES may be implemented in the future to correct for this effect if stars are observed at mismatched S/N. If this is accomplished, this potentially significant source of radial velocity error could be greatly mitigated for such observations.

\section{Conclusion}
\label{sec:conclusion}
EXPRES is a new Doppler spectrograph capable of reaching a 30 cm s$^{-1}$ single-measurement precision on bright stars, based on the instrument error analysis presented in this work as well as the on-sky results presented in \cite{petersburg2019}. This level of precision, along with higher resolution than past instruments, will enable a new survey in the search for Earth-sized exoplanets, potentially in the habitable zones of their host stars. In addition to finding planets, the goal for EXPRES is to provide data to improve observed radial velocities through statistical analysis to mitigate the impact of stellar activity on observed spectra. Under good seeing conditions, the throughput of EXPRES is 10\% to 15\%, enabling an S/N exceeding 300 per pixel on magnitude $V=6$ stars in 10 minute exposures. Cross-correlation of many consecutive LFC exposures demonstrate an instrument calibration precision under 10 cm s$^{-1}$. On-sky, several more instrumental sources of radial velocity error are incurred due to image motion, atmospheric dispersion, and barycentric corrections, making for a total instrument error of approximately 10 cm s$^{-1}$, not including photon noise. Calibratable instrumental drifts persist, potentially due to varying temperature in the spectrograph chamber. Use of an Invar optical bench and frequent wavelength solutions mitigate the impact of these effects. With a characterized CCD, PPNU radial velocity errors have been constrained to be under 5 cm s$^{-1}$, and the impact of CTI has been measured and mitigated by matching the S/N between different exposures of the same star. Atmospheric dispersion compensation, fast tip-tilt guiding, and a chromatic exposure meter ensure that radial velocity errors incurred from Earth's atmosphere are mitigated. A description of the current reduction pipeline, wavelength calibration, spectral extraction, and Doppler analysis of the early data is presented in \citep{petersburg2019}. 

Following what we have learned during the development and commissioning of EXPRES, we are able to recommend further development in several aspects of radial velocity instrumentation. The largest term in the total radial velocity error budget of instruments such as EXPRES is stellar activity. Instrumental errors have been constrained to at least a factor of a few smaller than this term. Regarding instrumentation, the largest errors we find that may still be correctable are PPNU and CTI in CCDs. With the EXPRES detector, PPNU error has been measured to be 5 cm s$^{-1}$, but we were unable to implement a corrective algorithm to mitigate this effect. It should still be possible to obtain the data necessary to complete such an algorithm, with the proper experimental setup. We also note that different detectors may have different pixel non-uniformities resulting in a larger or smaller effect, owing to different construction methods, and that each detector should be individually characterized. The impact of CTI has been measured to contribute up to several m s$^{-1}$ errors in the worst cases, when S/N is not matched between science observations. With a well-measured CTI that is dependent on signal level, this error can be mitigated through software corrections or by matching S/N between observations with the exposure meter. Similarly to PPNU, we note that different detectors will exhibit different CTI characteristics. Finally, a stable wavelength calibration source is still needed in blue wavelengths, where a wealth of stellar absorption lines are present in G-dwarf and K-dwarf stars that are not currently being used in radial velocity analyses.

\software{Barycorrpy \citep{kanodia2018}, Astropy \citep{astropy2013,astropy2018}, Matplotlib \citep{matplotlib2007}, SciPy \citep{scipy2001}, NumPy \citep{numpy2011}}

\acknowledgments
This work was supported by the NSF under grants NSF MRI-1429365 and ATI-1509436. We acknowledge generous support for telescope time provided by the Heising-Simons Foundation and the Yale Astronomy Department. DAF and JMB wish to acknowledge support from an anonymous donation, which has also been used for telescope time. ABD, LLZ, and RRP gratefully acknowledge support from the NSF GRFP. We thank the anonymous referee, whose helpful feedback improved the quality of this manuscript.

\bibliographystyle{aasjournal.bst}

\end{document}